\documentclass[12pt]{article}
\usepackage{amssymb}
\usepackage{amsmath}
\usepackage{amstext}
\usepackage{graphicx,epsfig}
\usepackage{epsfig}
\usepackage{verbatim} 
\usepackage{fancybox}
\usepackage{color}
\usepackage{ulem}
\usepackage{enumitem}
\usepackage{subfigure}
\usepackage{bbm}
\usepackage{parskip}

\newcommand{\Comment}[1]{{}}
\definecolor{MyDarkBlue}{rgb}{0.15,0.15,0.45}
\usepackage[linktocpage=true]{hyperref}
\hypersetup{
colorlinks=true,
citecolor=MyDarkBlue,
linkcolor=MyDarkBlue,
urlcolor=MyDarkBlue,
pdfauthor={Kurt Hinterbichler},
pdftitle={Theoretical Aspects of Massive Gravity},
pdfsubject={hep-th}
}

\newcommand{\be}{\begin{equation}}
\newcommand{\ee}{\end{equation}}
\newcommand{\bea}{\begin{eqnarray}}
\newcommand{\eea}{\end{eqnarray}}
\newcommand{\beas}{\begin{eqnarray*}}
\newcommand{\eeas}{\end{eqnarray*}}
\newcommand{\nn}{\nonumber}

\def\({\left(}
\def\){\right)}

\newcommand{\Tr}{\text{Tr}}

\newcommand{\pb}{\mathbf{p}}
\newcommand{\kb}{\mathbf{k}}
\newcommand{\xb}{\mathbf{x}}
\newcommand{\pd}{(2\pi)^d}
\newcommand{\intx}{\int d^d \mathbf{x}\ }

\newcommand{\la}{\langle}
\newcommand{\ra}{\rangle}
\newcommand{\lb}{\left[}
\newcommand{\rb}{\right]}
\newcommand{\half}{\frac{1}{2}}

\newcommand{\intpd}{\int \frac{d^Dp}{(2\pi)^D}\ }

\newcommand{\intp}{\int d^d \mathbf{p}\ }
\newcommand{\wsp}{\omega_{\mathbf{p}}}
\newcommand{\intpp}{\int \frac{d^d \mathbf{p}}{\sqrt{\pd 2\wsp}}\ }

\title{Theoretical Aspects of Massive Gravity}
\author{Kurt Hinterbichler\footnote{kurthi@physics.upenn.edu}}

\numberwithin{equation}{section}

\begin{document}


\begin{center}
{\LARGE \bf{\sc Theoretical Aspects of Massive Gravity}}
\end{center} 
 \vspace{1truecm}
\thispagestyle{empty} \centerline{
{\large \bf {\sc Kurt Hinterbichler${}^{a,}$}}\footnote{E-mail address: \Comment{\href{mailto:kurthi@physics.upenn.edu}}{\tt kurthi@physics.upenn.edu}}
                                                          }

\vspace{1cm}

\centerline{{\it ${}^a$ 
Center for Particle Cosmology, Department of Physics and Astronomy,}}
 \centerline{{\it University of Pennsylvania, 209 South 33rd Street, }} \centerline{{\it Philadelphia, PA 19104, USA}}

\begin{abstract}
Massive gravity has seen a resurgence of interest due to recent progress which has overcome its traditional problems, yielding an avenue for addressing important open questions such as the cosmological constant naturalness problem.  The possibility of a massive graviton has been studied on and off for the past 70 years.  During this time, curiosities such as the vDVZ discontinuity and the Boulware-Deser ghost were uncovered.  We re-derive these results in a pedagogical manner, and develop the St\"ukelberg formalism to discuss them from the modern effective field theory viewpoint.  We review recent progress of the last decade, including the dissolution of the vDVZ discontinuity via the Vainshtein screening mechanism, the existence of a consistent effective field theory with a stable hierarchy between the graviton mass and the cutoff, and the existence of particular interactions which raise the maximal effective field theory cutoff and remove the ghosts.  In addition, we review some peculiarities of massive gravitons on curved space, novel theories in three dimensions, and examples of the emergence of a massive graviton from extra-dimensions and brane worlds.
\end{abstract}

\newpage

\tableofcontents
\newpage

\section{Introduction}
\parskip=5pt
\normalsize
Our goal is to explore what happens when one tries to give the graviton a mass.  This is a modification of gravity, so we first discuss what gravity is and what it means to modify it.

\subsection{General relativity is massless spin 2}

General relativity (GR) \cite{Einstein:1916vd} is by now widely accepted as the correct theory of gravity at low energies or large distances.  The discovery of GR was in many ways ahead of its time.  It was a leap of insight, from the equivalence principle and general coordinate invariance, to a fully non-linear theory governing the dynamics of spacetime itself.  It provided a solution, one more elaborate than necessary, to the problem of reconciling the insights of special relativity with the non-relativistic action at a distance of newtonian gravity.

Had it not been for Einstein's intuition and years of hard work, general relativity would likely have been discovered anyway, but its discovery may have had to wait several more decades, until developments in field theory in the 1940's and 50's primed the culture.  But in this hypothetical world without Einstein, the path of discovery would likely have been very different, and in many ways more logical.

This logical path starts with the approach to field theory espoused in the first volume of Weinberg's field theory text \cite{Weinberg:1995mt}.  Degrees of freedom in flat four dimensional spacetime are particles, classified by their spin.  These degrees of freedom are carried by fields.  If we wish to describe long range macroscopic forces, only bosonic fields will do, since fermionic fields cannot build up classical coherent states.  By the spin statistics theorem, these bosonic fields must be of integer spin $s=0,1,2,3,$ etc.   A field, $\psi$, which carries a particle of mass $m$, will satisfy the Klein-Gordon equation $(\square-m^2)\psi=0$, whose solution a distance $r$ from a localized source goes like $\sim {1\over r}e^{-mr}$.  Long range forces, those without exponential suppression, must therefore be described by massless fields, $m=0$. 

 Massless particles are characterized by how they transform under rotations transverse to their direction of motion.  The transformation rule for bosons is characterized by an integer $h\geq 0$, which we call the helicity.  For $h= 0$, such massless particles can be carried most simply by a scalar field $\phi$.   For a scalar field, any sort of interaction terms consistent with Lorentz invariance can be added, and so there are a plethora of possible self-consistent interacting theories of spin 0 particles.
 
 For helicities $s\geq1$, the field must carry a gauge symmetry if we are to write interactions with manifest Lorentz symmetry and locality.  For helicity 1, if we choose a vector field $A_\mu$ to carry the particle, its action is fixed to be the Maxwell action, so even without Maxwell, we could have discovered electromagnetism via these arguments.  If we now ask for consistent self interactions of such massless particles, we are led to the problem of deforming the action (and possibly the form of the gauge transformations), in such a way that  the linear form of the gauge transformations is preserved.   These requirements are enough to lead us essentially uniquely to the non-abelian gauge theories, two of which describe the strong and weak forces \cite{Henneaux:1997bm}.
 
 Moving on to helicity 2, the required gauge symmetry is linearized general coordinate invariance.  Asking for consistent self interactions leads essentially uniquely to GR and full general coordinate invariance \cite{Gupta:1954zz,Kraichnan:1955zz,Weinberg:1965rz,Deser:1969wk,Boulware:1974sr,Fang:1978rc,Wald:1986bj} (see also chapter 13 of  \cite{Weinberg:1995mt}, which shows how helicity 2 implies the equivalence principle).  For helicity $\geq 3$, the story ends, because there are no self interactions that can be written \cite{Berends:1984wp} (see also chapter 13 of  \cite{Weinberg:1995mt}, which shows that the scattering amplitudes for helicity $\geq 3$ particles vanish).
 
 This path is straightforward, starting from the principles of special relativity (Lorentz invariance), to the classification of particles and fields that describe them, and finally to their possible interactions.  The path Einstein followed, on the other hand, is a leap of insight and has logical gaps; the equivalence principle and general coordinate invariance, though they suggest GR, do not lead uniquely to GR.  
 
 General coordinate invariance is a gauge symmetry, and gauge symmetries are redundancies of description, not fundamental properties.  In any system with gauge symmetry, one can always fix the gauge and eliminate the gauge symmetry, without breaking the physical global symmetries (such as Lorentz invariance) or changing the physics of the system in any way.  One often hears that gauge symmetry is fundamental, in electromagnetism for example, but the more correct statement is that gauge symmetry in electromagnetism is necessary only if one demands the convenience of linearly realized Lorentz symmetry and locality.  Fixing a gauge will not change the physics, but the price paid is that the Lorentz symmetries and locality are not manifest.   
 
 On the other hand, starting from a system without gauge invariance, it is always possible to introduce gauge symmetry by putting in redundant variables.  Often this can be very useful for studying a system, and can elucidate properties which are otherwise difficult to see.  This is the essence of the St\"ukelberg trick, which we will make use of extensively in our study of massive gravity.  In fact, as we will see, this trick can be used to make \textit{any} lagrangian invariant under general coordinate diffeomorhpisms, the same group under which GR is invariant.  Thus general coordinate invariance cannot be the defining feature of GR. 
 
 Similarly, the principle of equivalence, which demands that all mass and energy gravitate with the same strength, is not unique to GR.   It can be satisfied even in scalar field theories, if one chooses the interactions properly.  For example, this can be achieved by iteratively coupling a canonical massless scalar to its own energy momentum tensor.  Such a theory in fact solves all the problems Einstein set out to solve; it provides a universally attractive force which conforms to the principles of special relativity, reduces to newtonian gravity in the non-relativistic limit, and satisfies the equivalence principle\footnote{This theory is sometimes known as the Einstein-Fokker theory, first introduced in 1913 by Nordstr\"om \cite{Nordstrom1, Nordstrom2}, and later in a different form \cite{Freund:1969hh, deser}.  It was even studied by Einstein when he was searching for a relativistic theory of gravity that embodied the equivalence principle \cite{einsteinfokker}.}.  By introducing diffeomorphism invariance via the St\"ukelberg trick, it can even be made to satisfy the principle of general coordinate invariance.
 
The real underlying principle of GR has nothing to do with coordinate invariance or equivalence principles or geometry, rather it is the statement: \textit{general relativity is the theory of a non-trivially interacting massless helicity 2 particle}.  The other properties are consequences of this statement, and the implication cannot be reversed.

As a quantum theory, GR is not UV complete.  It must be treated as an effective field theory valid at energies up to a cutoff at the Planck mass, $M_P$, beyond which unknown high energy effects will correct the Einstein-Hilbert action.  As we will discuss in Section \ref{GRsection}, for a given background such as the spherical solution around a heavy source of mass $M$ such as the Sun, GR has three distinct regimes.  There is a classical linear regime, where both non-linear effects and quantum effects can be ignored.  This is the regime in which $r$ is greater than the Schwarzschild radius, $r>r_S\sim {M\over M_P^2}$.  For $M$ the mass of the Sun, we have $r_S\sim 1\ \text{km}$, so the classical linear approximation is good nearly everywhere in the solar system.  There is the quantum regime $r<{1\over M_P}$, very near the singularity of the black hole, where the effective field theory description breaks down.   Most importantly, there is a well separated middle ground, a classical non-linear regime, ${1\over M_P}<r<r_S$,  where non-linearities can be summed up without worrying about quantum corrections, the regime which can be used to make controlled statements about what is going on inside a black hole.  One of the challenges of adding a mass to the graviton, or any modification of gravity, is to retain calculable yet interesting regimes such as this.

\subsection{Modifying general relativity}

A theory of massive gravity is a theory which propagates a massive spin 2 particle.  The most straightforward way to construct such a theory is to simply add a mass term to the Einstein-Hilbert action, giving the graviton a mass, $m$, in such a way that GR is recovered as $m\rightarrow 0$.  This is a modification of  gravity, a deformation away from the elegant theory of Einstein.  Since GR is the essentially unique theory of a massless spin 2 degree of freedom, it should be remembered that \textit{modifying gravity means changing its degrees of freedom}.  

Despite the universal consensus that GR is a beautiful and accurate theory, there has in recent years arisen a small industry of physicists working to modify it and test these modifications.  When asked to cite their motivation, they more often than not point to supernova data \cite{Perlmutter:1998np,Riess:1998cb} which shows that the universe has recently started accelerating in its expansion.  If GR is correct, there must exist some dark energy density, $\rho\sim 10^{-29} \ {\rm g}/{\rm cm}^3$.  The simplest interpretation is that there is a constant term, $\Lambda$, in the Einstein-Hilbert action, which would give $\rho\sim M_P^2\Lambda$.  To give the correct vacuum energy, this constant has to take the small value $\Lambda/M_P^2\sim 10^{-65}$, whereas arguments from quantum field theory suggest a value much larger, up to order unity \cite{Weinberg:1988cp}.  
It is therefore tempting to speculate that perhaps GR is wrong, and instead of a dark energy component, gravity is modified in the infrared \cite{Deffayet:2000uy,Deffayet:2001pu}, in such a way as to produce an accelerating universe from nothing.  Indeed may modifications can be cooked up which produce these so-called self accelerating solutions.  For example, one well-studied modification is to replace the Einstein-Hilbert lagrangian with $F(R)$, a general function of the Ricci scalar \cite{DeFelice:2010aj,Sotiriou:2008rp}, which can lead to self-accelerating solutions \cite{Carroll:2003wy,Carroll:2004de}.  This modification is equivalent to adding an additional scalar degree of freedom.

These cosmological reasons for studying modifications to gravity are often criticized on the grounds that they can only take us so far; the small value of the cosmological acceleration relative to the Planck mass must come from somewhere, and the best these modifications can do is to shift the fine-tuning into other parameters (see \cite{Batra:2008cc} for an illustration in the $F(R)$/scalar-tensor case). 

While it is true the small number must come from somewhere, there remains the hope that it can be put somewhere which is technically natural, i.e. stable to quantum corrections.  Some small parameters, such as the ratio of the Higgs mass to the Planck mass in the standard model, are not technically natural, whereas others, like small fermion masses, are technically natural, because their small values are stable under quantum corrections.  A rule of thumb is that a small parameter is technically natural if there is a symmetry that appears as the small parameter is set to zero.   When this is the case, symmetry protects a zero value of the small parameter from quantum corrections.  This means corrections due to the small parameter must be proportional to the parameter itself.  In the case of small fermion masses, it is chiral symmetry that appears, whereas in the case of the Higgs mass and the cosmological constant, there is no obvious symmetry that appears.

Of course there is no logical inconsistency with having small parameters, technically natural or not, and nature may explain them anthropically \cite{barrow:book}, or may just employ them without reason.  But as practical working physicists, we hope that it is the case that a small parameter is technically natural, because then there is a hope that perhaps some classical mechanism can be found that drives the parameter towards zero, or otherwise explains its small value.  If it is not technically natural, any such mechanism will be much harder to find because it must know about the quantum corrections in order to compensate them.

One does not need a cosmological constant problem, however, to justify studying modifications to GR.  There are few better ways to learn about a structure, whether it's a sports car, a computer program or a theory, than to attempt to modify it.  With a rigid theory such as GR, there is a level of appreciation that can only be achieved by witnessing how easily things can go badly with the slightest modification.  In addition, deforming a known structure is one of the best ways to go about discovering new structures, structures which may have unforeseen applications.  

One principle that comes to play is the continuity of physical predictions of a theory in the parameters of the theory.  Surely, we should not be able to say experimentally, given our finite experimental precision, that a parameter of nature is exactly mathematically zero and not just very small.  If we deform GR by a small parameter, the predictions of the deformed theory should be very close to GR, to the extent that the deformation parameter is small.  It follows that any undesirable pathologies associated with the deformation should cure themselves as the parameter is set to zero.  Thus, we uncover a mechanism by which such pathologies can be cured, a mechanism which may have applications in other areas.

Massive gravity is a well developed case study in the infrared modification of gravity, where all of these points are nicely illustrated.  Purely from the consideration of degrees of freedom, it is a natural modification to consider, since it amounts to simply giving a mass to the particle which is already present in GR.   In another sense, it is less minimal than $F(R)$ or scalar tensor theory, which adds a single scalar degree of freedom, because to reach the 5 polarizations of the massive graviton we must add at least 3 degrees of freedom beyond the 2 of the massless graviton.  

With regard to the cosmological constant problem, there is the possibility of a technically natural explanation.  The deformation parameter is $m$, the graviton mass, and GR should be restored as $m\rightarrow 0$.    The force mediated by a massive graviton has a Yukawa profile $\sim {1\over r}e^{-mr}$, which drops off from that of a massless graviton at distances $r \gtrsim {1\over m}$, so one could hope to explain the acceleration of the universe without dark energy by choosing the graviton mass to be of order the Hubble constant $m\sim H$.  Of course, this does not eliminate the small cosmological constant, which reappears as the ratio $m/ M_P$.  But there is now hope that this is a technically natural choice, because deformation by a mass term breaks the gauge symmetry of GR, which is restored in the limit $m\rightarrow 0$.  As we will see, a small $m$ is indeed protected from quantum corrections (though as we will see, there are other issues that prevent this, at our current stage of understanding, from being a completely satisfactory realization of a technically natural cosmological constant).

There are also interesting lessons to be learned regarding the continuity of physical predictions.  The addition of a mass term is a brutality upon the beautiful structure of GR, and does not go unpunished.  Various pathologies appear, which are representative of common pathologies associated with any infrared modification of gravity.  These include strong classical non-linearities, ghost-like instabilities, and a very low cutoff, or region of trustability, for the resulting quantum effective theory.  In short, modifying the infrared often messes up the UV.  New mechanisms also come into play, because the extra degrees of freedom carried by the massive graviton must somehow decouple themselves as $m\rightarrow 0$ to restore the physics of GR.

The study of the massless limit leads to the discovery of the \textit{Vainshtein mechanism}, by which these extra degrees of freedom hide themselves at short distances using non-linearities.  This mechanism has already proven useful for model builders who have long-range scalars, such as moduli from the extra dimensions of string theory, that they want to shield from local experiments that would otherwise rule them out. 

\subsection{History and outline}

The possibility of a graviton mass has been studied off and on since 1939, when Fierz and Pauli \cite{Fierz:1939ix} first wrote the action describing a free massive graviton.  Following this, not much occurred until the early 1970's, when there was a flurry of renewed interest in quantum field theory.  The linear theory coupled to a source was studied by van Dam, Veltman, and Zakharov \cite{vanDam:1970vg,zakharov}, who discovered the curious fact that the theory makes predictions different from those of linear GR even in the limit as the graviton mass goes to zero.  For example, massive gravity in the $m\rightarrow 0$ limit gives a prediction for light bending that is off by 25 percent from the GR prediction.   The linear theory violates the principle of continuity of the physics in the parameters of the theory.  This is known as the \textit{vDVZ discontinuity}.  The discontinuity was soon traced to the fact that not all of the degrees of freedom introduced by the graviton mass decouple as the mass goes to zero.  The massive graviton has 5 spin states, which in the massless limit become the 2 helicity states of a massless graviton, 2 helicity states of a massless vector, and a single massless scalar.  The scalar is essentially the longitudinal graviton, and it maintains a finite coupling to the trace of the source stress tensor even in the massless limit.  In other words, the massless limit of a massive graviton is not a massless graviton, but rather a massless graviton plus a coupled scalar, and the scalar is responsible for the vDVZ discontinuity.

If the linear theory is accurate, then the vDVZ discontinuity represents a true physical discontinuity in predictions, violating our intuition that physics should be continuous in the parameters.  Measuring the light bending in this theory would be a way to show that the graviton mass is mathematically zero rather than just very small.  However, the linear theory is only the start of a complete non-linear theory, coupled to all the particles of the standard model.   The possible non-linearities of a real theory were studied several years later by Vainshtein \cite{Vainshtein:1972sx}, who found that the nonlinearities of the theory become stronger and stronger as the mass of the graviton shrinks.  What he found was that around any massive source of mass $M$, such as the Sun,  there is a new length scale known as the \textit{Vainshtein radius}, $r_V\sim \left(M\over m^4M_P^2\right)^{1/5}$.  At distances $r\lesssim r_V$, non-linearities begin to dominate and the predictions of the linear theory cannot be trusted.   The Vainshtein radius goes to infinity as $m\rightarrow 0$, so there is no radius at which the linear approximation tells us something trustworthy about the massless limit. This opens the possibility that the non-linear effects cure the discontinuity.  To have some values in mind, if we take $M$ the mass of the Sun and $m$ a very small value, say the Hubble constant $m\sim 10^{-33}\ \text{eV}$, the scale at which we might want to modify gravity to explain the cosmological constant, we have $r_V\sim 10^{18}\ \text{km}$, about the size of the Milky Way. 

Later the same year, Boulware and Deser \cite{Boulware:1973my} studied some specific fully non-linear massive gravity theories and showed that they possess a ghost-like instability.  Whereas the linear theory has 5 degrees of freedom, the non-linear theories they studied turned out to have 6, and the extra degree of freedom manifests itself around non-trivial backgrounds as a scalar field with a wrong sign kinetic term, known as the \textit{Boulware-Deser ghost}.

Meanwhile, the ideas of effective field theory were being developed, and it was realized that a non-renormalizable theory, even one with apparent instabilities such as massive gravity, can be made sense of as an effective field theory, valid only at energies below some ultraviolet cutoff scale $\Lambda$.  In 2003, Arkani-Hamed, Georgi and Schwartz \cite{ArkaniHamed:2002sp} brought to attention a method of restoring gauge invariance to massive gravity in a way which makes it very simple to see what the effective field theory properties are.  They showed that massive gravity generically has a maximum UV cutoff of $\Lambda_5=(M_P m^4)^{1/5}$.  For Hubble scale graviton mass, this is a length scale $\Lambda_5^{-1}\sim 10^{11}\ \text{km}$.  This is a very small cutoff, parametrically smaller than the Planck mass, and goes to zero as $m\rightarrow 0$.   Around a massive source, the quantum effects become important at the radius $r_Q= \left(\frac{M}{M_{Pl}}\right)^{1/3}{1\over \Lambda_5}$, which is parametrically larger than the Vainshtein radius at which non-linearities enter.  For the Sun, $r_Q\sim 10^{24}\ \text{km}$. Without finding a UV completion or some other re-summation, there is no sense in which we can trust the solution inside this radius, and the usefulness of massive gravity is limited.  In particular, since the whole non-linear regime is below this radius, there is no hope to examine the continuity of physical quantities in $m$ and explore the Vainshtein mechanism in a controlled way.  On the other hand, it can be seen that the mass of the Boulware-Deser ghost drops below the cutoff only when $r\lesssim r_Q$, so the ghost is not really in the effective theory at all and can be consistently excluded.

Putting aside the issue of quantum corrections, there has been continued study of the Vainshtein mechanism in a purely classical context.  It has been shown that classical non-linearities do indeed restore continuity with GR in certain circumstances.  In fact, the ghost degree of freedom can play an essential role in this, by providing a repulsive force in the non-linear region to counteract the attractive force of the longitudinal scalar mode.

 By adding higher order graviton self-interactions with appropriately tuned coefficients, it is in fact possible to raise the UV cutoff of the theory to $\Lambda_3=(M_P m^2)^{1/3}$, corresponding to roughly $\Lambda_3^{-1}\sim 10^{3}\ \text{km}$.  In 2010, the complete action of this theory in a certain decoupling limit was worked out by de Rham and Gabadadze \cite{deRham:2010ik}, and they show that, remarkably, it is free of the Boulware-Deser ghost.  Recently, it was shown that the complete theory is free of the Boulware-Deser ghost.  This $\Lambda_3$ theory is the best hope of realizing a useful and interesting massive gravity theory.

The subject of massive gravity also naturally arises in extra-dimensional setups.  In a Kaluza-Klein scenario such as GR in 5d compactified on a circle, the higher Kaluza-Klein modes are massive gravitons.  Brane world setups such as the Dvali-Gabadadze-Porrati (DGP) model \cite{Dvali:2000hr} give more intricate gravitons with resonance masses.  The study of such models has complemented the study of pure 4d massive gravity and has pointed towards new research directions.  

The major outstanding question is whether it is possible to UV extend the effective field theory of massive gravity to the Planck scale, and what this UV extension may look like.  This would provide a solution to the problem of making the small cosmological constant technically natural, and is bound to be an interesting theory its own right (the analogous question applied to massive vector bosons leads to the discovery of the Higgs mechanism and spontaneous symmetry breaking).  In the case of massive gravity, there are indications that a UV completion may not have a local Lorentz invariant form.  Another long shot, if UV completion can be found, would be to take the $m\rightarrow 0$ limit of the completion and hope to obtain a UV completion to ordinary GR. 

As this review is focused on the theoretical aspects of Lorentz invariant massive gravity, we will not have much to say about the large literature on Lorentz-violating massive gravity.  We will also not say much about the experimental search for a graviton mass, or what the most likely signals and search modes would be.  There has been much work on these areas, and each could be the topic of a separate review. 

\bigskip
{\bf Conventions}:
Often we will work in an arbitrary number of dimensions, just because it is easy to do so.  In this case, $D$ signifies the number of spacetime dimension and we stick to $D\geq3$. $d$ signifies the number of space dimensions, $d=D-1$.  We use the mostly plus metric signature convention, $\eta_{\mu\nu}=(-,+,+,+,\cdots)$.  Tensors are symmetrized and anti-symmetrized with unit weight, i.e $T_{(\mu\nu)}=\half \left(T_{\mu\nu}+T_{\nu\mu}\right)$,   $T_{[\mu\nu]}=\half \left(T_{\mu\nu}-T_{\nu\mu}\right)$.  The reduced 4d Planck mass is $M_P={1\over (8\pi G)^{1/2}}\approx 2.43\times 10^{18}\ \rm{GeV}$.   Conventions for the curvature tensors, covariant derivatives and Lie derivatives are those of Carroll \cite{Carroll:2004st}.

\section{\label{Massive}The free Fierz-Pauli action}

We start by displaying an action for a single massive spin 2 particle in flat space, carried by a symmetric tensor field $h_{\mu\nu}$,
\be \label{massivefreeaction} S=\int d^D x -\frac{1}{2}\partial_\lambda h_{\mu\nu}\partial^\lambda h^{\mu\nu}+\partial_\mu h_{\nu\lambda}\partial^\nu h^{\mu\lambda}-\partial_\mu h^{\mu\nu}\partial_\nu h+\frac{1}{2}\partial_\lambda h\partial^\lambda h-\frac{1}{2}m^2(h_{\mu\nu}h^{\mu\nu}-h^2).
\ee
This is known as the \textit{Fierz-Pauli action} \cite{Fierz:1939ix}.  Our point of view will be to take this action as given and then show that it describes a massive spin 2.  There are, however, some (less than thorough) ways of motivating this action.  To start with, the action above contains all possible contractions of two powers of $h$, with up to two derivatives.  The two derivative terms, those which survive when $m=0$, are chosen to match exactly those obtained by linearizing the Einstein-Hilbert action.  The $m=0$ terms describe a massless helicity 2 graviton and have the gauge symmetry 
\be\label{gaugesymorig} \delta h_{\mu\nu}=\partial_{\mu}\xi_\nu+\partial_{\nu}\xi_\mu,\ee
for a spacetime dependent gauge parameter $\xi_\mu(x)$.  This symmetry fixes all the coefficients of the two-derivative part of (\ref{massivefreeaction}), up to an overall coefficient.  The mass term, however, violates this gauge symmetry.  The relative coefficient of $-1$ between the $h^2$ and $h_{\mu\nu}h^{\mu\nu}$ contractions is called the \textit{Fierz-Pauli tuning}, and it not enforced by any known symmetry.

However, the only thing that needs to be said about this action is that it describes a single massive spin 2 degree of freedom of mass $m$.  We will show this explicitly in what follows.  Any deviation from the form (\ref{massivefreeaction}) and the action will no longer describe a single massive spin 2.  For example, violating the Fierz-Pauli tuning in the mass term by changing to $-\frac{1}{2}m^2(h_{\mu\nu}h^{\mu\nu}-(1-a)h^2)$ for $a\not=0$ gives an action describing a scalar ghost (a scalar with negative kinetic energy) of mass $m_g^2={3-4a\over 2a}m^2$, in addition to the massive spin 2.  For small $a$, the ghost mass squared goes like $\sim {1\over a}$.  It goes to infinity as the Fierz-Pauli tuning is approached, rendering it non-dynamical.  Violating the tuning in the kinetic terms will similarly alter the number of degrees of freedom, see  \cite{VanNieuwenhuizen:1973fi} for a general analysis.

There is a method of constructing lagrangians such as (\ref{massivefreeaction}) to describe any given spin.  See for example the first few chapters of \cite{Weinberg:1995mt}, the classic papers on higher spin lagrangians \cite{Fronsdal:1978rb,Singh:1974qz}, and the reviews \cite{Bouatta:2004kk,Sorokin:2004ie}.

\subsection{\label{canonicalanalysis}Hamiltonian and degree of freedom count}

We will begin our study of the Fierz-Pauli action (\ref{massivefreeaction}) by casting it into hamiltonian form and counting the number of degrees of freedom.  We will show that it propagates ${D(D-1)\over 2}-1$ degrees of freedom in $D$ dimensions (5 degrees of freedom for $D=4$), the right number for a massive spin 2 particle.

We start by Legendre transforming (\ref{massivefreeaction}) only with respect to the spatial components $h_{ij}$.  The canonical momenta are\footnote{Note that canonical momenta can change under integrations by parts of the time derivatives.  We have fixed this ambiguity by integrating by parts such as to remove all derivatives from $h_{0i}$ and $h_{00}$.}
\be \pi_{ij}={\partial {\cal L}\over \partial \dot h_{ij}}=\dot h_{ij}-\dot h_{kk}\delta_{ij}-2\partial_{(i}h_{j)0}+2\partial_k h_{0k}\delta_{ij}.\ee
Inverting for the velocities, we have
\be \dot h_{ij}=\pi_{ij}-{1\over D-2}\pi_{kk} \delta_{ij}+2\partial_{(i}h_{j)0}.\ee

In terms of these hamiltonian variables, the Fierz-Pauli action (\ref{massivefreeaction}) becomes
\be \label{hamlinearaction} S=\int d^Dx\ \pi_{ij}\dot h_{ij}-{\cal H}+2h_{0i}\left(\partial_j\pi_{ij}\right)+m^2h_{0i}^2+h_{00}\left(\vec\nabla^2h_{ii}-\partial_i\partial_j h_{ij}-m^2 h_{ii}\right),\ee
where 
\bea\label{hamlinear} {\cal H}=&&\half \pi_{ij}^2-\half{1\over D-2}\pi_{ii}^2 \nn \\
&&+\frac{1}{2}\partial_k h_{ij}\partial_k h_{ij}-\partial_i h_{jk}\partial_j h_{ik}+\partial_i h_{ij}\partial_j h_{kk}-\frac{1}{2}\partial_i h_{jj}\partial_i h_{kk}+\frac{1}{2}m^2(h_{ij}h_{ij}-h_{ii}^2) .\nn \\ 
\eea

First consider the case $m=0$.  The time-like components $h_{0i}$ and $h_{00}$ appear linearly multiplied by terms with no time derivatives.  We can interpret them as Lagrange multipliers enforcing the constraints $\partial_j\pi_{ij}=0$ and $\vec\nabla^2h_{ii}-\partial_i\partial_j h_{ij}=0$.  It is straightforward to check that these are first class constraints, and that the hamiltonian (\ref{hamlinear}) is first class.  Thus (\ref{hamlinearaction}) is a first class gauge system.  For $D=4$, the $h_{ij}$ and $\pi_{ij}$ each have 6 components, because they are $3\times 3$ symmetric tensors, so together they span a 12 dimensional (for each space point) phase space.  We have 4 constraints (at each space point), leaving an 8 dimensional constraint surface.  The constraints then generate 4 gauge invariances, so the gauge orbits are 4 dimensional, and the gauge invariant quotient by the orbits is 4 dimensional (see \cite{Henneaux:1992ig} for an introduction to constrained hamiltonian systems, gauge theories, and the terminology used here).  These are the two polarizations of the massless graviton, along with their conjugate momenta.

In the case $m\not=0$, the $h_{0i}$ are no longer Lagrange multipliers.  Instead, they appear quadratically and are auxiliary variables.  Their equations of motion yield 
\be h_{0i}=-{1\over m^2}\partial_j\pi_{ij},\ee
which can be plugged back into the action (\ref{hamlinearaction}) to give
\be \label{hamlinearaction2} S=\int d^Dx\ \pi_{ij}\dot h_{ij}-{\cal H}+h_{00}\left(\vec\nabla^2h_{ii}-\partial_i\partial_j h_{ij}-m^2 h_{ii}\right),\ee
where 
\bea\label{hamlinear2} {\cal H}=&&\half \pi_{ij}^2-\half{1\over D-2}\pi_{ii}^2 +\frac{1}{2}\partial_k h_{ij}\partial_k h_{ij}-\partial_i h_{jk}\partial_j h_{ik}+\partial_i h_{ij}\partial_j h_{kk}-\frac{1}{2}\partial_i h_{jj}\partial_i h_{kk}\nn \\
&&+\frac{1}{2}m^2(h_{ij}h_{ij}-h_{ii}^2) +{1\over m^2}\left(\partial_j \pi_{ij}\right)^2.\nn \\ 
\eea

The component $h_{00}$ remains a Lagrange multiplier enforcing a single constraint ${\cal C}=-\vec\nabla^2h_{ii}+\partial_i\partial_j h_{ij}+m^2 h_{ii}=0$, but the hamiltonian is no longer first class.  One secondary constraint arises from the Poisson bracket with the hamiltonian $H=\int d^dx\ {\cal H}$, namely $\{ H,{\cal C}\}_{\rm PB}={1\over D-2}m^2\pi_{ii}+\partial_i\partial_j\pi_{ij}$.  The resulting set of two constraints is second class, so there is no longer any gauge freedom.  For $D=4$ the 12 dimensional phase space has 2 constraints for a total of $10$ degrees of freedom, corresponding to the 5 polarizations of the massive graviton and their conjugate momenta.  

Note that the Fierz-Pauli tuning is crucial to the appearance of $h_{00}$ as a Lagrange multiplier.  If the tuning is violated, then $h_{00}$ appears quadratically and is an auxiliary variable, and no longer enforces a constraint.  There are then no constraints, and the full 12 degrees of freedom in the phase space are active.  The extra 2 degrees of freedom are the scalar ghost and its conjugate momentum.

\subsection{\label{modesolutions}Free solutions and graviton mode functions}

We now proceed to find the space of solutions of (\ref{massivefreeaction}), and show that it transforms as a massive spin 2 representation of the Lorentz group, showing that the action propagates precisely one massive graviton.  The equations of motion from (\ref{massivefreeaction}) are 
\bea \label{freeeom} \frac{\delta S}{\delta h^{\mu\nu}}=
\square h_{\mu\nu}-\partial_\lambda\partial_\mu h^\lambda_{\ \nu}-\partial_\lambda\partial_\nu h^\lambda_{\ \mu}+\eta_{\mu\nu}\partial_\lambda\partial_\sigma h^{\lambda\sigma}+\partial_\mu\partial_\nu h-\eta_{\mu\nu}\square h-m^2(h_{\mu\nu}-\eta_{\mu\nu}h)=0.\nn \\
\eea

Acting on (\ref{freeeom}) with $\partial^\mu$, we find, assuming $m^2\not=0$, the constraint $\partial^\mu h_{\mu\nu}-\partial_\nu h$. Plugging this back into the equations of motion, we find $\square h_{\mu\nu}-\partial_\mu\partial_\nu h-m^2(h_{\mu\nu}-\eta_{\mu\nu}h)=0$.  Taking the trace of this we find $h=0$, which in turn implies $\partial^\mu h_{\mu\nu}$=0.  
This, along with $h=0$ applied to the equations of motion (\ref{freeeom}), gives $(\square-m^2)h_{\mu\nu}=0$.

Thus the equations of motion (\ref{freeeom}) imply the three equations,
\be \label{free3eq} (\square -m^2)h_{\mu\nu}=0,\ \ \ \partial^\mu h_{\mu\nu}=0,\ \ \  h=0.\ee
Conversely, it is straightforward to see that these three equations imply the equations of motion (\ref{freeeom}), so (\ref{free3eq}) and (\ref{freeeom}) are equivalent.  The form (\ref{free3eq}) makes it easy to count the degrees of freedom as well.  For $D=4$, the first of (\ref{free3eq}) is an evolution equation for the $10$ components of the symmetric tensor $h_{\mu\nu}$, and the last two are constraint equations on the initial conditions and velocities of $h_{\mu\nu}$.  The last determines the trace completely, killing one real space degree of freedom.  The second gives 4 initial value constraints, and the vanishing of its time derivative, i.e. demanding that it be preserved in time, implies 4 more initial value constraints, thus killing 4 real space degrees of freedom.  In total, we are left with the $5$ real space degrees of freedom of a 4 dimensional spin 2 particle, in agreement with the hamiltonian analysis of Section \ref{canonicalanalysis}.  

The first equation in (\ref{free3eq}) is the standard Klein-Gordon equation, with the general solution
\begin{equation}\label{gen1massivegravity}
h^{\mu\nu}(x)=\intpp \left( h^{\mu\nu}(\pb)e^{ip\cdot x}+h^{\mu\nu\ast}(\pb)e^{-ip\cdot x}\right).
\end{equation}
Here $\pb$ are the spatial momenta, $\omega_{\pb}=\sqrt{\pb^2+m^2}$, and the $D$-momenta $p^\mu$ are on shell so that $p^\mu=(\omega_\pb,\pb)$.  

Next we expand the Fourier coefficients $h^{\mu\nu}(\pb)$ over some basis of symmetric tensors, indexed by $\lambda$,
\be h^{\mu\nu}(\pb)=a_{\pb,\lambda}\bar\epsilon^{\mu\nu}(\pb,\lambda).\ee

We will fix the momentum dependence of the basis elements $\bar\epsilon^{\mu\nu}(\pb,\lambda)$ by choosing some basis $\bar\epsilon^{\mu\nu}(\kb,\lambda)$ at the standard momentum $k^\mu=(m,0,0,0,\ldots)$ and then acting with some standard boost\footnote{We choose the standard boost to be 
\bea\label{standard1}
&&L^i_{\ j}(p)=\delta_{ij}+\frac{1}{\left|\pb\right|^2}(\gamma-1)\pb^i \pb^j,\nn\\
&& L^i_{\ 0}(p)=L^0_{\ i}(\pb)=\frac{\pb^i}{\left|\pb\right|}\sqrt{\gamma^2-1},\nn\\
&&L^0_{\ 0}(p)=\gamma,
\eea
where 
\begin{equation}
\gamma=\frac{p^0}{m}=\frac{\sqrt{\left|\pb\right|^2+m^2}}{m}.\nn
\end{equation}
is the usual relativistic $\gamma$.  See chapter 2 of \cite{Weinberg:1995mt} for discussions of this standard boost and general representation theory of the Poincare group.}  $L(p)$, which takes $k$ into $p$, $p^\mu=L(p)^\mu_{\ \nu}k^\nu$.  This standard boost will choose for us the basis at $\pb$, relative to that at $\kb$.  Thus we have
\begin{equation}
\bar\epsilon^{\mu\nu}(\pb,\lambda)=L(p)^\mu_{\ \alpha}L(p)^\nu_{\ \beta}\bar\epsilon^{\alpha\beta}(\kb,\lambda).
\end{equation}

Imposing the conditions $\partial_\mu h^{\mu\nu}=0$ and $h=0$ on (\ref{gen1massivegravity}) then reduces to imposing
\be k_\mu \bar\epsilon^{\mu\nu}(\kb,\lambda)=0,\ \ \ \eta_{\mu\nu}\bar\epsilon^{\mu\nu}(\kb,\lambda)=0.\ee
The first says that $\bar\epsilon^{\mu\nu}(\kb,\lambda)$ is purely spatial, i.e. $\bar\epsilon^{0\mu}(\kb,\lambda)=0$.  The second says that it is traceless, so that $\bar\epsilon^{i}_{\ i}(\kb,\lambda)=0$ also.  Thus the basis need only be a basis of symmetric traceless spatial tensors, $\lambda=1,\ldots,{d(d+1)\over 2}-1$.  We demand that the basis be orthonormal,
\be \bar\epsilon^{\mu\nu}(\kb,\lambda)\bar\epsilon^\ast_{\mu\nu}(\kb,\lambda')=\delta_{\lambda\lambda'}.\ee

This basis forms the symmetric traceless representation of the rotation group $SO(d)$, which is the little group for a massive particle in $D$ dimensions.  If $R^{\mu}_{\ \nu}$ is a spatial rotation, we have
\be R^{\mu}_{\ \mu'}R^{\nu}_{\ \nu'}\bar\epsilon^{\mu\nu}(\kb,\lambda')=R^{\lambda'}_{\ \lambda}\bar\epsilon^{\mu\nu}(\kb,\lambda'),\ee
where $R^{\lambda'}_{\ \lambda}$ is the symmetric traceless tensor representation of $R^{\mu}_{\ \mu'}$.  We are free to use any other basis $\epsilon^{\mu\nu}(\kb,\lambda)$, related to the $\bar\epsilon^{\mu\nu}(\kb,\lambda)$ by 
\be \epsilon^{\mu\nu}(\kb,\lambda)=B^{\lambda'}_{\ \lambda} \bar\epsilon^{\mu\nu}(\kb,\lambda'),\ee
where $B$ is any unitary matrix.  

Given a particular spatial direction, with unit vector $\hat k^i$, there is an $SO(d-1)$ subgroup of the little group $SO(d)$ which leaves $\hat k^i$ invariant, and the symmetric traceless rep of $SO(d)$ breaks up into three reps of $SO(d-1)$, a scalar, a vector, and a symmetric traceless tensor.  The scalar mode is called the longitudinal graviton, and has spatial components
\be \label{longitudinalmode}\epsilon^{ij}_{L}=\sqrt{d\over d-1}\( \hat k^i\hat k^j-{1\over d}\delta^{ij}\).\ee
After a large boost in the $\hat k^i$ direction, it goes like $\epsilon_L\sim p^2/ m^2$.  As we will see later, in the massless limit, or large boost limit, this mode is carried by a scalar field, which generally becomes strongly coupled once interactions are taken into account. The vector modes have spatial components
\be \label{vectormodes} \epsilon^{ij}_{V,k}=\sqrt{2}\hat k^{(i}\delta^{j)}_k,\ee
and after a large boost in the $\hat k^i$ direction, they go like $\epsilon_L\sim p/ m$.  In the massless limit, these modes are carried by a vector field, which decouples from conserved sources.
The remaining linearly independent modes are symmetric traceless tensors with no components in the $\hat k^i$ directions, and form the symmetric traceless mode of $SO(d-1)$.   They are invariant under a boost in the $\hat k^i$ direction, and in the massless limit, they are carried by a massless graviton.  In the massless limit, we should therefore expect that the extra degrees of freedom of the massive graviton should organize themselves into a massless vector and a massless scalar.  We will see later explicitly how this comes about at the lagrangian level.

Upon boosting to $\pb$, the polarization tensors satisfy the following properties: they are transverse to $p^\mu$ and traceless,
\be p_\mu \epsilon^{\mu\nu}(\pb,\lambda)=0,\ \ \ \eta_{\mu\nu} \epsilon^{\mu\nu}(\pb,\lambda)=0,\ee
and they satisfy orthogonality and completeness relations
\bea &&\epsilon^{\mu\nu}(\pb,\lambda)\epsilon^\ast_{\mu\nu}(\pb,\lambda')=\delta_{\lambda\lambda'}, \\
&& \sum_\lambda  \epsilon^{\mu\nu}(\pb,\lambda)\epsilon^{\ast\alpha\beta}(\pb,\lambda)={1\over 2}(P^{\mu\alpha}P^{\nu\beta}+P^{\mu\beta}P^{\nu\alpha})-\frac{1}{D-1}P^{\mu\nu}P^{\alpha\beta},\label{completenessmode}
\eea
where $P^{\alpha\beta}\equiv \eta^{\alpha\beta}+\frac{p^\alpha p^\beta}{m^2}$.  The right hand side of the completeness relation (\ref{completenessmode}) is the projector onto the symmetric and transverse traceless subspace of tensors, i.e. the identity on this space.  We also have the following symmetry properties in $\pb$, which can be deduced from the form of the standard boost (\ref{standard1}),
\bea
\epsilon^{ij}(-\pb,\lambda)&=&\epsilon^{ij}(\pb,\lambda),\ \ \ i,j=1,2,\ldots,d \\
\epsilon^{0i}(-\pb,\lambda)&=&-\epsilon^{0i}(\pb,\lambda),\ \ \ i=1,2,\ldots,d \\
\epsilon^{00}(-\pb,\lambda)&=&\epsilon^{00}(\pb,\lambda).
\eea
The general solution to (\ref{freeeom}) thus reads 
\begin{equation}\label{hmassivegravityexpansion}{
h^{\mu\nu}(x)=\intpp\sum_{\lambda} a_{\pb,\lambda}\epsilon^{\mu\nu}(\pb,\lambda)e^{ip\cdot x}+a^\ast_{\pb,\lambda}\epsilon^{\ast \mu\nu}(\pb,\lambda)e^{-ip\cdot x}.}
\ee

The solution is a general linear combination of the following mode functions and their conjugates
\begin{equation}\label{modefuntns}
u^{\mu\nu}_{\pb,\lambda}(x)\equiv\frac{1}{\sqrt{\pd 2\wsp}}\epsilon^{\mu\nu}(\pb,\lambda)e^{ip\cdot x},\ \ \ \lambda=1,2,\ldots,d.
\end{equation}
These are the solutions representing gravitons, and they have the following Poincare transformation properties
\begin{equation}
u^{\mu\nu}_{\pb,\lambda}(x-a)=u^{\mu\nu}_{\pb,\lambda}(x)e^{-ip\cdot a},
\end{equation}
\begin{equation}
\Lambda^\mu_{\ \mu'}\Lambda^\nu_{\ \nu'}u^{\mu'\nu'}_{\pb,\lambda}(\Lambda^{-1}x)=\sqrt{\frac{\omega_{\Lambda\pb}}{\wsp}}W(\Lambda,p)_{\lambda' \lambda}u^{\mu\nu}_{\Lambda\pb,\lambda'}(x),
\end{equation}
where $W(\Lambda,p)=L^{-1}(\Lambda p)\Lambda L(p)$ is the Wigner rotation, and $W(\Lambda,p)_{\lambda' \lambda}$ is its spin 2 rep, $R^\mu_{\ \nu}\rightarrow (B^{-1}RB)_{\lambda' \lambda}$.\footnote{We show the Lorentz transformation property as follows
\[ \Lambda^\mu_{\ \mu'} \Lambda^\nu_{\ \nu'}\epsilon^{\mu'\nu'}(p,\lambda)e^{i p\cdot \Lambda^{-1} x} =\left[\Lambda L(p)\right]^\mu_{\ \mu'}\left[\Lambda L(p)\right]^\nu_{\ \nu'}\epsilon^{\mu'\nu'}(k,\lambda)e^{i \Lambda p\cdot  x}\]
\[=\left[L(\Lambda p)\left(L^{-1}(\Lambda p)\Lambda L(p)\right)\right]^{\mu}_{\ \mu'}\left[L(\Lambda p)\left(L^{-1}(\Lambda p)\Lambda L(p)\right)\right]^{\nu}_{\ \nu'}\epsilon(k,\lambda)^{\mu'\nu'} e^{i \Lambda p\cdot  x}\]\[= \left[L(\Lambda p)W\left(\Lambda,p\right)\right]^{\mu}_{\ \mu'}\left[L(\Lambda p)W\left(\Lambda,p\right)\right]^{\nu}_{\ \nu'}\epsilon(k,\lambda)^{\mu'\nu'} e^{i \Lambda p\cdot  x}.\]
The little group element is a spatial rotation.  For any spatial rotation $R^\mu_{\ \nu}$, we have
\[ R^\mu_{\ \mu'}R^\nu_{\ \nu'}\epsilon^{\mu'\nu'}(k,\lambda)= R^\mu_{\ \mu'}R^\nu_{\ \nu'}B^{\lambda'}_{\ \lambda}\bar\epsilon^{\mu'\nu'}(\kb,\lambda')=B^{\lambda'}_{\ \lambda} R^{\lambda''}_{\ \lambda'}\bar\epsilon^{\mu\nu}(k,\lambda'')\]
\[=\left(B^{-1}RB\right)^{\lambda'}_{\ \lambda}\epsilon^{\mu\nu}(k,\lambda').\]
Plugging back into the above,
\[ \Lambda^\mu_{\ \mu'} \Lambda^\nu_{\ \nu'}\epsilon^{\mu'\nu'}(p,\lambda)e^{i p\cdot \Lambda^{-1} x} =L(\Lambda p)^\mu_{\ \mu'} L(\Lambda p)^\nu_{\ \nu'} W(\Lambda,p)^{\lambda'}_{\ \lambda}\epsilon^{\mu'\nu'}(k,\lambda')e^{i \Lambda p\cdot  x}\]
\[=W(\Lambda,p)^{\lambda'}_{\ \lambda}\epsilon^{\mu\nu}(\Lambda p,\lambda')e^{i \Lambda p\cdot  x},\]
where $W$ is the spin 2 representation of the little group in a basis rotated by $B$, $W=B^{-1}RB$.}  Thus the gravitons are spin 2 solutions.

In terms of the modes, the general solution reads
\begin{equation}\label{amodes}
h^{\mu\nu}(x)=\intp\sum_{\lambda} \left(a_{\pb,\lambda}u^{\mu\nu}_{\pb,\lambda}(x)+a^\ast_{\pb,\lambda}u^{\mu\nu\ast}_{\pb,\lambda}(x)\right).
\end{equation}
The inner (symplectic) product on the space of solutions to the equations of motion is,
\begin{equation}
(h,h')=\left.i\intx h^{\mu\nu\ast}(x)\overset{\leftrightarrow}{\partial_0}h'_{\mu\nu}(x)\right|_{t=0}.
\end{equation}
The $u$ functions are orthonormal with respect to this product,
\begin{align}
&(u_{\pb,\lambda},u_{\pb',\lambda'})=\delta^d(\pb-\pb')\delta_{\lambda\lambda'}, \\
&(u^\ast_{\pb,\lambda},u^\ast_{\pb',\lambda'})=-\delta^d(\pb-\pb')\delta_{\lambda\lambda'}, \\
&(u_{\pb,\lambda},u^\ast_{\pb',\lambda'})=0,
\end{align}
and we can use the product to extract the $a$ and $a^\ast$ coefficients from any solution $h_{\mu\nu}(x)$,
\begin{align}
&a_{\pb,\lambda}=(u_{\pb,\lambda},h), \\
&a^\ast_{\pb,\lambda}=-(u^\ast_{\pb,\lambda},h).
\end{align}

In the quantum theory, the  $a$ and $a^\ast$ become creation and annihilation operators which satisfy the usual commutations relations and produce massive spin 2 states.  The fields $h_{ij}$ and their canonical momenta $\pi_{ij}$, constructed from the  $a$ and $a^\ast$, will then automatically satisfy the Dirac algebra and constraints of the hamiltonian analysis of Section \ref{canonicalanalysis}, providing a quantization of the system.  Once interactions are taken into account, external lines of Feynman diagrams will get a factor of the mode functions (\ref{modefuntns}).

\subsection{Propagator}

Integrating by parts, we can rewrite the Fierz-Pauli action (\ref{massivefreeaction}) as 
\begin{equation}
 S=\int d^D x\ \frac{1}{2}h_{\mu\nu}{\cal O}^{\mu\nu,\alpha\beta}h_{\alpha\beta},
\end{equation}
where
\be  {\cal O}^{\mu\nu}_{\ \ \alpha\beta}=\left(\eta^{(\mu}_{\ \alpha}\eta^{\nu)}_{\ \beta}-\eta^{\mu\nu}\eta_{\alpha\beta}\right)(\square-m^2)-2\partial^{(\mu}\partial_{(\alpha}\eta^{\nu)}_{\ \beta)} 
+\partial^\mu\partial^\nu\eta_{\alpha\beta}+\partial_\alpha\partial_\beta\eta^{\mu\nu},
\ee
is a second order differential operator satisfying
\be\label{osyms} {\cal O}^{\mu\nu,\alpha\beta}={\cal O}^{\nu\mu,\alpha\beta}={\cal O}^{\mu\nu,\beta \alpha}= {\cal O}^{\alpha\beta,\mu\nu}.\ee
In terms of this operator, the equations of motion (\ref{freeeom}) can be written simply as $\frac{\delta S}{\delta h_{\mu\nu}}= {\cal O}^{\mu\nu,\alpha\beta}h_{\alpha\beta}=0.$

To derive the propagator, we go to momentum space, 
\be \label{Ooperator}  {\cal O}^{\mu\nu}_{\ \ \alpha\beta}(\partial\rightarrow ip)=-\left(\eta^{(\mu}_{\ \alpha}\eta^{\nu)}_{\ \beta}-\eta^{\mu\nu}\eta_{\alpha\beta}\right)(p^2+m^2)+2p^{(\mu}p_{(\alpha}\eta^{\nu)}_{\ \beta)} 
-p^\mu p^\nu\eta_{\alpha\beta}-p_\alpha p_\beta\eta^{\mu\nu}.
\ee
The propagator is the operator ${\cal D}_{\alpha\beta,\sigma\lambda}$ with the same symmetries (\ref{osyms}) which satisfies
\begin{equation}\label{propidentrelation}
{\cal O}^{\mu\nu,\alpha\beta}{\cal D}_{\alpha\beta,\sigma\lambda}= \frac{i}{2}(\delta^\mu_\sigma\delta^\nu_\lambda+\delta^\nu_\sigma\delta^\mu_\lambda).
\end{equation}
The right side is the identity operator on the space of symmetric tensors.

Solving (\ref{propidentrelation}), we find
\begin{equation}{
{\cal D}_{\alpha\beta,\sigma\lambda}=\frac{-i}{p^2+m^2}\left[\half\left(P_{\alpha\sigma}P_{\beta\lambda}+P_{\alpha\lambda}P_{\beta\sigma}\right)-\frac{1}{D-1}P_{\alpha\beta}P_{\sigma\lambda}\right],}\label{massmprop}
\end{equation}
where $P_{\alpha\beta}\equiv \eta_{\alpha\beta}+\frac{p_\alpha p_\beta}{m^2}.$

In the interacting quantum theory, internal lines with momentum $p$ will be assigned this propagator, which for large $p$ behaves as $\sim {p^2\over m^4}$.  This growth with $p$ means we cannot apply standard power counting arguments (like those of chapter 12 of \cite{Weinberg:1995mt}) to deduce the renormalizability properties or strong coupling scales of a theory.  We will see later how to overcome this difficulty by rewriting the theory in a way in which all propagators go like $\sim {1\over p^2}$ at high energy.

\subsubsection*{Massless propagator}

The massive graviton propagator (\ref{massmprop}) may be compared to the propagator for the case $m=0$.  For $m=0$, the action becomes
\begin{equation}\label{masslessoform}
 S_{m=0}=\int d^D x\ \frac{1}{2}h_{\mu\nu}{\cal E}^{\mu\nu,\alpha\beta}h_{\alpha\beta},
\end{equation}
where the kinetic operator is 
\begin{equation}\label{Eoper}
  {\cal E}^{\mu\nu}_{\ \ \alpha\beta}= \left.{\cal O}^{\mu\nu}_{\ \ \alpha\beta}\right|_{m=0}=\left(\eta^{(\mu}_{\ \alpha}\eta^{\nu)}_{\ \beta}-\eta^{\mu\nu}\eta_{\alpha\beta}\right)\square-2\partial^{(\mu}\partial_{(\alpha}\eta^{\nu)}_{\ \beta)} 
+\partial^\mu\partial^\nu\eta_{\alpha\beta}+\partial_\alpha\partial_\beta\eta^{\mu\nu}.
\end{equation}
This operator has the symmetries (\ref{osyms}).  Acting on a symmetric tensor $Z_{\mu\nu}$ it reads
\be \epsilon^{\mu\nu,\alpha\beta}Z_{\alpha\beta}=\square Z^{\mu\nu}-\eta^{\mu\nu}\square Z-2\partial^{(\mu}\partial_\alpha Z^{\nu)\alpha}+\partial^\mu\partial^\nu Z+\eta^{\mu\nu}\partial_\alpha\partial_\beta Z^{\alpha\beta}.\ee
The $m=0$ action has the gauge symmetry (\ref{gaugesymorig}), and the operator (\ref{Eoper}) is not invertible.  Acting with it results in a tensor which is automatically transverse, and it annihilates anything which is pure gauge
\be \partial_\mu\left(\epsilon^{\mu\nu,\alpha\beta}Z_{\alpha\beta}\right)=0,\ \ \ \epsilon^{\mu\nu,\alpha\beta}\left(\partial_\alpha\xi_\beta+\partial_\beta\xi_\alpha\right)=0.\ee 

To find a propagator, we must fix the gauge freedom.  We choose the Lorenz gauge (also called harmonic, or de Donder gauge),
\be\label{lorenzgaugecon} \partial^\mu h_{\mu\nu}-\half \partial_\nu h=0.\ee
We can reach this gauge by making a gauge transformation with $\xi_\mu$ chosen to satisfy $\square\xi_\mu=-\left(\partial^\nu h_{\mu\nu}-{1\over 2}\partial_\mu h\right).$
This condition fixes the gauge only up to gauge transformations with parameter $\xi_\mu$ satisfying $\square\xi_\mu=0$.  
In this gauge, the equations of motion simplify to 
\be\label{grav1free}\square h_{\mu\nu}-\half \eta_{\mu\nu}\square h=0.\ee
The solutions to this equation which also satisfy the gauge condition (\ref{lorenzgaugecon}) are the Lorenz gauge solutions to the original equations of motion.

To the lagrangian of (\ref{masslessoform}) we add the following gauge fixing term
\be \mathcal{L}_{\rm GF}=-\left(\partial^\nu h_{\mu\nu}-{1\over 2}\partial_\mu h\right)^2.\ee
Quantum mechanically, this results from the Fadeev-Popov gauge fixing procedure.  Classically, we may justify it on the grounds that the equations of motion obtained from the action plus the gauge fixing term are the same as the gauge fixed equations of motion (\ref{grav1free}).  The gauge condition itself, however, is not obtained as an equation of motion, and must be imposed separately.  We have
\be \mathcal{L}+ \mathcal{L}_{\rm GF}=\frac{1}{2} h_{\mu\nu}\square h^{\mu\nu}-\frac{1}{4} h\square h,\ee
whose equations of motion are (\ref{grav1free}).

We can write the gauge fixed lagrangian as $\mathcal{L}+\mathcal{L}_{\rm GF}=\frac{1}{2}h_{\mu\nu}\tilde {\cal O}^{\mu\nu,\alpha\beta}h_{\alpha\beta},$
where
\be \tilde {\cal O}^{\mu\nu,\alpha\beta}=\square\left[ {1\over 2}\left(\eta^{\mu\alpha}\eta^{\nu\beta}+\eta^{\mu\beta}\eta^{\nu\alpha}\right)-{1\over 2} \eta^{\mu\nu}\eta^{\alpha\beta}\right].\ee
Going to momentum space and inverting, we obtain the propagator, 
\begin{equation}\label{masszeroprop} {
{\cal D}_{\alpha\beta,\sigma\lambda}=\frac{-i}{p^2}\left[\half\left(\eta_{\alpha\sigma} \eta_{\beta\lambda}+ \eta_{\alpha\lambda} \eta_{\beta\sigma}\right)-\frac{1}{D-2} \eta_{\alpha\beta} \eta_{\sigma\lambda}\right],}
\end{equation}
which satisfies the relation (\ref{propidentrelation}) with $\tilde {\cal O}$ in place of ${\cal O}$.
This propagator grows like $\sim {1\over p^2}$ at high energy.  Comparing the massive and massless propagators, (\ref{masszeroprop}) and (\ref{massmprop}), and ignoring for a second the terms in (\ref{massmprop}) which are singular as $m\rightarrow 0$, there is a difference in coefficient for the last term, even as $m\rightarrow 0$.  For $D=4$, it is $1/2$ vs. $1/3$.  This is the first sign of a discontinuity in the $m\rightarrow 0$ limit.

\section{Linear response to sources}

We now add a fixed external symmetric source $T^{\mu\nu}(x)$ to the action (\ref{massivefreeaction}) , 
\be \label{linearsources} S=\int d^D x -\frac{1}{2}\partial_\lambda h_{\mu\nu}\partial^\lambda h^{\mu\nu}+\partial_\mu h_{\nu\lambda}\partial^\nu h^{\mu\lambda}-\partial_\mu h^{\mu\nu}\partial_\nu h+\frac{1}{2}\partial_\lambda h\partial^\lambda h-\frac{1}{2}m^2(h_{\mu\nu}h^{\mu\nu}-h^2)+\kappa h_{\mu\nu}T^{\mu\nu}.
\ee
Here $\kappa=M_P^{-{D-2\over 2}}$ is the coupling strength to the source\footnote{The normalizations chosen here are in accord with the general relativity definition $T^{\mu\nu}={2\over\sqrt{-g}}{\delta{\cal L}\over \delta g_{\mu\nu}}$, as well as the normalization $\delta g_{\mu\nu}=2\kappa h_{\mu\nu}$.}.

The equations of motion are now sourced by $T_{\mu\nu}$,
\begin{equation}\label{sourcedeq}{
\square h_{\mu\nu}-\partial_\lambda\partial_\mu h^\lambda_{\ \nu}-\partial_\lambda\partial_\nu h^\lambda_{\ \mu}+\eta_{\mu\nu}\partial_\lambda\partial_\sigma h^{\lambda\sigma}+\partial_\mu\partial_\nu h-\eta_{\mu\nu}\square h-m^2(h_{\mu\nu}-\eta_{\mu\nu}h)=-\kappa T_{\mu\nu}.}
\end{equation}
In the case $m=0$, acting on the left with $\partial^\mu$ gives identically zero, so we must have the conservation condition $\partial^\mu T_{\mu\nu}=0$ if there is to be a solution.  
For $m\not=0$, there is no such condition. 

\subsection{General solution to the sourced equations}

We now find the retarded solution of (\ref{sourcedeq}), to which the homogeneous solutions of (\ref{modesolutions}) can be added to obtain the general solution.
Acting on the equations of motion (\ref{sourcedeq}) with $\partial^\mu$, we find,
\begin{equation}\label{lorentzg}
\partial^\mu h_{\mu\nu}-\partial_\nu h={\kappa\over m^2}\partial^\mu T_{\mu\nu}.
\end{equation}
Plugging this back into (\ref{sourcedeq}), we find
\[ \square h_{\mu\nu}-\partial_\mu\partial_\nu h-m^2(h_{\mu\nu}-\eta_{\mu\nu}h)=-\kappa T_{\mu\nu}+{\kappa\over m^2}\left[\partial^\lambda\partial_\mu T_{\nu\lambda}+\partial^\lambda\partial_\nu T_{\mu\lambda}-\eta_{\mu\nu}\partial\partial T\right].\]
Where $\partial\partial T$ is short for the double divergence $\partial_\mu\partial_\nu T^{\mu\nu}$.  Taking the trace of this we find 
\begin{equation}\label{traceless}
h=-{\kappa\over m^2 (D-1)} T-{\kappa\over m^4}{D-2\over D-1}\partial\partial T.
\end{equation}
Applying this to (\ref{lorentzg}), we find 
\begin{equation}
\partial^\mu h_{\mu\nu}=-{\kappa\over m^2 (D-1)}\partial_\nu T+{\kappa\over m^2}\partial^\mu T_{\mu\nu}-{\kappa\over m^4}{D-2\over D-1}\partial_\nu \partial\partial T, 
\end{equation}
which when applied along with \eqref{traceless} to the equations of motion, gives
\bea 
(\partial^2-m^2)h_{\mu\nu}=&&-\kappa\left[T_{\mu\nu}-{1\over D-1}\left(\eta_{\mu\nu}-{\partial_\mu\partial_\nu\over m^2}\right)T\right] \\ \nn
&&+{\kappa\over m^2}\left[\partial^\lambda\partial_\mu T_{\nu\lambda}+\partial^\lambda\partial_\nu T_{\mu\lambda}-{1\over D-1}\left(\eta_{\mu\nu}+(D-2){\partial_\mu\partial_\nu\over m^2}\right)\partial\partial T\right].
\eea
Thus we have seen that the equations of motion (\ref{sourcedeq}) imply the following three equations,
\bea \nn
(\square-m^2)h_{\mu\nu}&=&-\kappa\left[T_{\mu\nu}-{1\over D-1}\left(\eta_{\mu\nu}-{\partial_\mu\partial_\nu\over m^2}\right)T\right]\\ \nn
    &&+{\kappa\over m^2}\left[\partial^\lambda\partial_\mu T_{\nu\lambda}+\partial^\lambda\partial_\nu T_{\mu\lambda}-{1\over D-1}\left(\eta_{\mu\nu}+(D-2){\partial_\mu\partial_\nu\over m^2}\right)\partial\partial T\right],\\ \nn 
      \partial^\mu h_{\mu\nu}&=&-{\kappa\over m^2 (D-1)}\partial_\nu T+{\kappa\over m^2}\partial^\mu T_{\mu\nu}-{\kappa\over m^4}{D-2\over D-1}\partial_\nu \partial\partial T, \\
        h&=&-{\kappa\over m^2 (D-1)}T-{\kappa\over m^4}{D-2\over D-1}\partial\partial T. \label{massiveeqns}
 \eea
Conversely, it is straightforward to see that these three equations imply the equations of motion (\ref{sourcedeq}).

Taking the first equation of (\ref{massiveeqns}) and tracing, we find $(\square-m^2)\left[h+{\kappa\over m^2 (D-1)}T+{\kappa\over m^4}{D-2\over D-1}\partial\partial T\right]=0$.
Under the assumption that $(\partial^2-m^2) f=0 \Rightarrow f=0$ for any function $f$, the third equation is implied.  This will be the case with good boundary conditions, such as the retarded boundary conditions we impose when we are interested in the classical response to sources. The second equation of (\ref{massiveeqns}) can also be shown to follow under this assumption, so that we may obtain the solution by Fourier transforming only the first equation of (\ref{massiveeqns}).  This solution can also be obtained by applying the propagator (\ref{massmprop}) to the Fourier transform of the source.

Despite the absence of gauge symmetry, we will often be interested in sources which are conserved anyway, $\partial_\mu T^{\mu\nu}=0.$  When the source is conserved, and under the assumptions in the paragraph above, we are left with just the equation, 
\be (\partial^2-m^2)h_{\mu\nu}=-\kappa\left[T_{\mu\nu}-{1\over D-1}\left(\eta_{\mu\nu}-{\partial_\mu\partial_\nu\over m^2}\right)T\right].
\ee
The general solution for a conserved source is then,
\be \label{sourcegen} h_{\mu\nu}(x)=\kappa\intpd e^{ipx}{1\over p^2+m^2} \left[ T_{\mu\nu}(p)-{1\over D-1}\left(\eta_{\mu\nu}+{p_\mu p_\nu\over m^2}\right)T(p)\right],\ee
where $T^{\mu\nu}(p)$ is the Fourier transform of the source, $T^{\mu\nu}(p)=\int d^Dx\ e^{-ipx}\ T^{\mu\nu}(x).$  To get the retarded field, we should integrate above the poles in the $p^0$ plane.

\subsection{Solution for a point source}

We now specialize to four dimensions so that $\kappa={1/ M_P}$, and we consider as source the stress tensor of a mass $M$ point particle at rest at the origin
\be\label{particlesource} T^{\mu\nu}(x)=M\delta^{\mu}_0\delta^{\nu}_0\delta^3(\xb),\ \ T^{\mu\nu}(p)=2\pi M\delta^{\mu}_0\delta^{\nu}_0 \delta(p^0).  \ee
Note that this source is conserved.  
For this source, the general solution (\ref{sourcegen}) gives  
\bea \nn h_{00}(x)&=&{2 M\over 3M_P}\int {d^3 \pb\over (2\pi)^3} e^{i\pb \xb}{1\over \pb^2+m^2},\\ \nn
h_{0i}(x)&=&0, \\ h_{ij}(x)&=&{M\over 3M_P}\int {d^3 \pb\over (2\pi)^3} e^{i\pb \xb}{1\over \pb^2+m^2}\left(\delta_{ij}+{p_i p_j \over m^2}\right). \label{metric1masiv} 
\eea
Using the formulae
\bea \nn \int {d^3 \pb\over (2\pi)^3} e^{i\pb \xb}{1\over \pb^2+m^2}&=&{1\over 4\pi}{e^{-mr}\over r},\\\nn
\int {d^3 \pb\over (2\pi)^3} e^{i\pb \xb}{p_ip_j\over \pb^2+m^2}&=&-\partial_i\partial_j  \int {d^3 \pb\over (2\pi)^3} e^{i\pb \xb}{1\over \pb^2+m^2} \\
&=&{1\over 4\pi}{e^{-mr}\over r}\left[{1\over r^2}(1+mr)\delta_{ij}-{1\over  r^4}(3+3mr+m^2r^2)x_ix_j\right],\nn\\
\eea
where $r\equiv \sqrt{x_i x_i}$, we have
\bea \nn h_{00}(x)&=&{2 M\over 3M_P}{1\over 4\pi}{e^{-mr}\over r},\\ \nn
h_{0i}(x)&=&0, \\  h_{ij}(x)&=&{M\over 3M_P}{1\over 4\pi}{e^{-mr}\over r}\left[{1+mr+m^2r^2\over m^2r^2}\delta_{ij}-{1\over m^2 r^4}(3+3mr+m^2r^2)x_ix_j\right].\nn\\ \label{massivehform}
\eea
Note the Yukawa suppression factors $e^{-mr}$, characteristic of a massive field.

For future reference, it will be convenient to record these expressions in spherical coordinates for the spatial variables.  Using the formula 
$\left[F(r)\delta_{ij}+G(r)x_ix_j\right]dx^i dx^j=\left(F(r)+r^2 G(r)\right)dr^2+F(r)r^2 d\Omega^2$ to get to spherical coordinates
we find
\be h_{\mu\nu}dx^\mu dx^\nu= -B(r)dt^2+C(r)dr^2+A(r)r^2d\Omega^2 ,\ee
where
\bea B(r)&=&-{2 M\over 3M_P}{1\over 4\pi}{e^{-mr}\over r},\nn \\
C(r)&=&-{2 M\over 3M_P}{1\over 4\pi}{e^{-mr}\over r}{1+mr\over m^2r^2},\nn \\
A(r)&=&{M\over 3M_P}{1\over 4\pi}{e^{-mr}\over r}{1+mr+m^2r^2\over m^2r^2}. \label{masssphersol}
\eea
In the limit $r\ll 1/m$ these reduce to 
\bea B(r)&=&-{2 M\over 3M_P}{1\over 4\pi r},\nn \\
C(r)&=&-{2 M\over 3M_P}{1\over 4\pi m^2 r^3},\nn\\
A(r)&=&{M\over 3M_P}{1\over 4\pi m^2 r^3}.
\eea
Corrections are suppressed by powers of $mr$.  

\subsubsection*{Solution for the massless graviton}

For the purposes of comparison, we will compute the point source solution for the massless case as well.  We choose the Lorenz gauge (\ref{lorenzgaugecon}).
In this gauge, the equations of motion simplify to 
\be\label{grav1}\square h_{\mu\nu}-\half \eta_{\mu\nu}\square h=-\kappa T_{\mu\nu}.\ee
Taking the trace, we find $\square h={2\over D-2} \kappa T,$
and upon substituting back, we get
\be\label{grav1b} \square h_{\mu\nu}=-\kappa\left[ T_{\mu\nu} -{1\over D-2}\eta_{\mu\nu} T\right].\ee
This equation, along with the Lorenz gauge condition (\ref{lorenzgaugecon}), is equivalent to the original equation of motion in Lorenz gauge.  

Taking $\partial^\mu$ on (\ref{grav1}) and on its trace, using conservation of $T_{\mu\nu}$ and comparing, we have $\square(\partial^\mu h_{\mu\nu}-\half \partial_\nu h)=0$, so that the Lorentz condition is automatically satisfied when boundary conditions are satisfied with the property that $\square f=0 \Rightarrow f=0$ for any function $f$, as is the case when we impose retarded boundary conditions.  We can then solve  \ref{grav1} by Fourier transforming.  
\be\label{masslgenso} h_{\mu\nu}(x)=\kappa\intpd e^{ip\cdot x}{1\over p^2} \left[ T_{\mu\nu}(p) -{1\over D-2}\eta_{\mu\nu} T(p)\right],\ee
where $T^{\mu\nu}(p)=\int d^Dx\ e^{-ip\cdot x}\ T^{\mu\nu}(x),$ is the Fourier transform of the source.
To get the retarded field, we should integrate above the poles in the $p^0$ plane.  

Now we specialize to $D=4$, and we consider as a source the point particle of mass $M$ at the origin (\ref{particlesource}).
For this source, the general solution (\ref{masslgenso}) gives 
\bea \nn h_{00}(x)&=&{M\over 2M_P}\int {d^3 \pb\over (2\pi)^3} e^{i\pb \xb}{1\over \pb^2}={M\over 2M_P}{1\over 4\pi r},\\ \nn
h_{0i}(x)&=&0, \\ h_{ij}(x)&=&{M\over 2M_P}\int {d^3 \pb\over (2\pi)^3} e^{i\pb \xb}{1\over \pb^2}\delta_{ij}={M\over 2M_P}{1\over 4\pi r}\delta_{ij}. \label{metric1}
\eea

For later reference, we record this result in spherical spatial coordinates as well.  Using the formula 
$\left[F(r)\delta_{ij}+G(r)x_ix_j\right]dx^i dx^j=\left(F(r)+r^2 G(r)\right)dr^2+F(r)r^2 d\Omega^2$ to get to spherical coordinates
we find
\be h_{\mu\nu}dx^\mu dx^\nu= -B(r)dt^2+C(r)dr^2+A(r)r^2d\Omega^2 ,\ee
where
\bea B(r)&=&-{ M\over 2M_P}{1\over 4\pi r},\nn \\
C(r)&=&{ M\over 2M_P}{1\over 4\pi r},\nn \\
A(r)&=&{M\over 2M_P}{1\over 4\pi r}. \label{masslesspoinspher}
\eea

\subsection{The vDVZ discontinuity}

We would now like to extract some physical predictions from the point source solution.  Let's assume we have a test particle moving in this field, and that this test particle responds to $h_{\mu\nu}$ in the same way that a test particle in general relativity responds to the metric deviation $\delta g_{\mu\nu}={2\over M_P}h_{\mu\nu}$.  We know from the textbooks (see for example chapter 7 of \cite{Carroll:2004st}) that if $h_{\mu\nu}$ takes the form $2 h_{00}/M_P=-2\phi$, $2 h_{ij}/M_P=-2\psi \delta_{ij}$, $h_{0i}=0$ for some functions $\phi(r)$ and $\psi(r)$, then the newtonian potential experienced by the particle is given by $\phi(r)$.  Furthermore, if $\psi(r)=\gamma \phi(r)$ for some constant $\gamma$, called the PPN parameter, and if $\phi(r)=-{k\over r}$ for some constant $k$, then the angle for the bending of light at impact parameter $b$ around the heavy source is given by $\alpha=2(1+\gamma)/b$.  Looking at (\ref{metric1}), the massless graviton gives us the values 
\bea \phi =-{GM\over r},\ \ \ \  \psi =-{GM\over r}, \ \ \ {\rm massless\ graviton,}
\eea
using ${1\over M_P^2}=8\pi G$. The PPN parameter is therefore $\gamma=1$ and the magnitude of the light bending angle for light incident at impact parameter $b$ is
\be \alpha={4GM\over b},\ \ \ \ {\rm massless\ graviton.}\ee

For the massive case, the metric (\ref{massivehform}) is not quite in the right form to read off the newtonian potential and light bending.  To simplify things, we notice that while the massive gravity action is not gauge invariant, we have assumed that the coupling to the test particle is that of GR, so this coupling is gauge invariant.  Thus we are free to make a gauge transformation on the solution $h_{\mu\nu}$, and there will be no effect on the test particle.  To simplify the metric (\ref{massivehform}), we go back to (\ref{metric1masiv}) and notice that the $p_ip_j\over m^2$ term in $h_{ij}$ is pure gauge, so we can ignore this term.  Thus our metric is gauge equivalent to the metric
\bea h_{00}(x)&=&{2 M\over 3M_P}{1\over 4\pi}{e^{-mr}\over r},\nn \\
h_{0i}(x)&=&0, \nn \\ 
h_{ij}(x)&=&{M\over 3M_P}{1\over 4\pi}{e^{-mr}\over r}\delta_{ij}.
\eea
We then have, in the small mass limit, 
\be \phi =-{4\over 3}{GM\over r},\ \ \ \  \psi =-{2\over 3}{GM\over r}\delta_{ij}.  \ \ \ {\rm massive\ graviton,}
\ee
These are the same values as obtained for $\omega=0$ Brans-Dicke theory.  The newtonian potential is larger than for the massless case.  The PPN parameter is $\gamma=\half$, and the magnitude of the light bending angle for light incident at impact parameter $b$ is the same as in the massless case,
\be \alpha={4GM\over b},  \ \ \ {\rm massive\ graviton.}\ee
If we like, we can make the newtonian potential agree with GR by scaling $G\rightarrow {3\over 4}G$.  Then the light bending would then change to $\alpha={3GM\over b}$, off by 25 percent from GR.  

What this all means is that linearized massive gravity, even in the limit of zero mass, gives predictions which are order one different from linearized GR.  If nature were described by either one or the other of these theories, we would, by making a finite measurement, be able to tell whether the graviton mass is mathematically zero or not, in violation of our intuition that the physics of nature should be continuous in its parameters. This is the vDVZ (van Dam, Veltman, Zakharov) discontinuity \cite{vanDam:1970vg, zakharov} (see also \cite{PhysRevD.2.2255,Carrera:2001pj}).  It is present in other physical predictions as well, such as the emission of gravitational radiation \cite{VanNieuwenhuizen:1973qf}. 

\section{\label{stukelbergtrick}The St\"uckelberg trick}

We have seen that there is a discontinuity in the physical predictions of linear massless gravity and the massless limit of linear massive gravity, known as the vDVZ discontinuity.  In this section, we will expose the origin of this discontinuity.  We will see explicitly that the correct massless limit of massive gravity is not massless gravity, but rather massless gravity plus extra degrees of freedom, as expected since the gauge symmetry which kills the extra degrees of freedom only appears when the mass is strictly zero.  The extra degrees of freedom are a massless vector, and a massless scalar which couples to the trace of the energy momentum tensor.  This extra scalar coupling is responsible for the vDVZ discontinuity.

Taking $m\rightarrow 0$ straight away in the lagrangian (\ref{linearsources}) does not yield not a smooth limit, because degrees of freedom are lost.  To find the correct limit, the trick is to introduce new fields and gauge symmetries into the massive theory in a way that does not alter the theory.  This is the St\"ukelberg trick.  Once this is done, a limit can be found in which no degrees of freedom are gained or lost. 

\subsection{\label{vectorstukelberg}Vector example}

To introduce the idea, we consider a simpler case, the theory of a massive photon $A_{\mu}$ coupled to a (not necessarily conserved) source $J_\mu$,  
\begin{equation}\label{massivevect}{
S=\int d^Dx\ -\frac{1}{4} F_{\mu\nu}F^{\mu\nu}-\half m^2 A_\mu A^\mu+A_\mu J^\mu,}
\end{equation}
where $F_{\mu\nu}\equiv\partial_\mu A_\nu-\partial_\nu A_\mu.$  The mass term breaks the would-be gauge invariance, $\delta A_\mu=\partial_\mu \Lambda$, and for $D=4$ this theory describes the 3 degrees of freedom of a massive spin 1 particle.  Recall that the propagator for a massive vector is ${-i\over p^2+m^2}\(\eta_{\mu\nu}+{p_\mu p_\nu\over m^2}\)$, which goes like $\sim {1\over m^2}$ for large momenta, invalidating the usual power counting arguments.

As it stands, the limit $m\rightarrow 0$ of the lagrangian (\ref{massivevect}) is not a smooth limit because we lose a degree of freedom -- for $m=0$ we have Maxwell electromagnetism which in $D=4$ propagates only 2 degrees of freedom, the two polarizations of a massless helicity 1 particle.  Also, the limit does not exist unless the source is conserved, as this is a consistency requirement in the massless case.

The St\"uckelberg trick consists of introducing a new scalar field $\phi$, in such a way that the new action has gauge symmetry but is still dynamically equivalent to the original action.  It will expose a different $m\rightarrow 0$ limit which is smooth, in that no degrees of freedom are gained or lost.
We introduce a field, $\phi$, by making the replacement 
\be A_\mu\rightarrow A_\mu+\partial_\mu\phi,\ee
following the pattern of the gauge symmetry we want to introduce \cite{stukelberg}.  This is emphatically \textit{not} a change of field variables.  It is \textit{not} a decomposition of $A_{\mu}$ into transverse and longitudinal parts ($A_\mu$ is not meant to identically satisfy $\partial_\mu A^\mu=0$ after the replacement), and it is \textit{not} a gauge transformation (the lagrangian (\ref{massivevect}) isn't gauge invariant).  Rather, this is creating a new lagrangian from the old one, by the addition of a new field $\phi$.  $F_{\mu\nu}$ is invariant under this replacement, since the replacement looks like a gauge transformation and $F_{\mu\nu}$ is gauge invariant.   The only thing that changes is the mass term and the coupling to the source, 
\begin{equation}\label{massivestuk}{
S=\int d^Dx\ -\frac{1}{4} F_{\mu\nu}F^{\mu\nu}-\half m^2 (A_\mu+\partial_\mu\phi)^2+A_\mu J^\mu-\phi\partial_\mu J^\mu.}
\end{equation}
We have integrated by parts in the coupling to the source.  The new action now has the gauge symmetry 
\be \delta A_\mu=\partial_\mu \Lambda,\ \ \ \delta\phi=-\Lambda.\ee
By fixing the gauge $\phi=0$, called the unitary gauge (a gauge condition for which it is permissible to substitute back into the action, because the potentially lost $\phi$ equation is implied by the divergence of the $A_{\mu}$ equation) we recover the original massive lagrangian (\ref{massivevect}), which means (\ref{massivestuk}) and (\ref{massivevect}) are equivalent theories.  They both describe the three degrees of freedom of a massive spin 1 in $D=4$.  The new lagrangian (\ref{massivestuk}) does the job using more fields and gauge symmetry.

The St\"ukelberg trick is a terrific illustration of the fact that gauge symmetry is a complete sham.  It represents nothing more than a redundancy of description.  We can take any theory and make it a gauge theory by introducing redundant variables.  Conversely, given any gauge theory, we can always eliminate the gauge symmetry by eliminating the redundant degrees of freedom.  The catch is that removing the redundancy is not always a smart thing to do.  For example, in Maxwell electromagnetism it is impossible to remove the redundancy and at the same time preserve manifest Lorentz invariance and locality.  Of course, electromagnetism with gauge redundancy removed is still electromagnetism, so it is still Lorentz invariant and local, it is just not manifestly so.  With the St\"ukelberg trick presented here, on the other hand, we are adding and removing extra gauge symmetry in a rather simple way, which does not mess with the manifest Lorentz invariance and locality. 

We see from (\ref{massivestuk}) that $\phi$ has a kinetic term, in addition to cross terms.  Rescaling $\phi\rightarrow {1\over m}\phi$ in order to normalize the kinetic term, we have 
\begin{equation}\label{massivephoton}{
S=\int d^Dx\ -\frac{1}{4} F_{\mu\nu}F^{\mu\nu}-\half m^2 A_\mu A^\mu -m A_\mu\partial^\mu\phi -\half\partial_\mu\phi\partial^\mu\phi+A_\mu J^\mu-{1\over m}\phi \partial_\mu J^\mu,}
\end{equation}
and the gauge symmetry reads 
\be \delta A_\mu=\partial_\mu \Lambda,\ \ \ \delta\phi=-m\Lambda.\ee

Consider now the $m\rightarrow 0$ limit.  Note that if the current is not conserved (or its divergence does not go to zero with at least a power of $m$ \cite{Fronsdal:1979yn}), then the scalar becomes strongly coupled to the divergence of the source and the limit does not exist.  Assuming a conserved source, the lagrangian becomes in the limit 
\begin{equation}{
\mathcal{L}= -\frac{1}{4} F_{\mu\nu}F^{\mu\nu} -\half\partial_\mu\phi\partial^\mu\phi+A_\mu J^\mu,}
\end{equation}
and the gauge symmetry is 
\be \delta A_\mu=\partial_\mu \Lambda,\ \ \ \delta\phi=0.\ee
It is now clear that the number of degrees of freedom is preserved in the limit.  For $D=4$ two of the three degrees of freedom go into the massless vector, and one goes into the scalar. 

In the limit, the vector decouples from the scalar, and we are left with a massless gauge vector interacting with the source, as well as a completely decoupled free scalar.  This $m\rightarrow 0$ limit is a different limit than the non-smooth limit we would have obtained by taking $m\rightarrow 0$ straight away in (\ref{massivevect}).  We have scaled $\phi\rightarrow {1\over m}\phi$ in order to canonically normalize the scalar kinetic term, so we are actually using a new scalar $\phi_{new}=m\phi_{old}$ which does not scale with $m$, so the smooth limit we are taking is to scale the old scalar degree of freedom up as we scale $m$ down, in such a way that the new scalar degree of freedom remains preserved.  

Rather than unitary gauge, we can instead fix a Lorentz-like gauge for the action (\ref{massivestuk}),
\be \partial_\mu A^\mu+m\phi=0.\ee
This gauge fixes the gauge freedom up to a residual gauge parameter satisfying $(\square-m^2)\Lambda=0$.  We can add the gauge fixing term
\be S_{\rm GF}=\int d^Dx\ -\half\left(\partial_\mu A^\mu+m\phi\right)^2.\ee
As in the massless case, quantum mechanically this term results from the Fadeev-Popov gauge fixing procedure.  Classically, we may justify it on the grounds that the equations of motion obtained from the action plus the gauge fixing term are the same as the gauge fixed equations of motion (the gauge condition itself, however, is not obtained as an equation of motion, and must be imposed separately).
Adding the gauge fixing term diagonalized the lagrangian,
\begin{equation}\label{massivephotongauge fix}
S+S_{\rm GF}=\int d^Dx\ \frac{1}{2} A_\mu (\square-m^2) A^\mu +\half\phi(\square-m^2)\phi+A_\mu J^\mu-{1\over m}\phi \partial_\mu J^\mu,
\end{equation}
and the propagators for $A_\mu$ and $\phi$ are respectively
\be {-i\eta_{\mu\nu}\over p^2+m^2},\ \ \ {-i\over p^2+m^2},\ee
which go like $\sim{1\over p^2}$ at high momenta.  Thus we have managed to restore the good high energy behavior of the propagators.

It is possible to find the gauge invariant mode functions for $A_\mu$ and $\phi$, which can then be compared to the unitary gauge mode functions of the massive photon.  In the massless limit, there is a direct correspondence; $\phi$ is gauge invariant and becomes the longitudinal photon, the $A_{\mu}$ has the usual Maxwell gauge symmetry and its gauge invariant transverse modes are exactly the transverse modes of the massive photon.

\subsection{\label{gravitonstukelsec}Graviton St\"ukelberg and origin of the vDVZ discontinuity}

Now consider massive gravity,
\be S=\int d^Dx\ \mathcal{L}_{m=0}-\frac{1}{2}m^2(h_{\mu\nu}h^{\mu\nu}-h^2)+\kappa h_{\mu\nu}T^{\mu\nu},\ee
where $\mathcal{L}_{m=0}$ is the lagrangian of the massless graviton.  We want to preserve the gauge symmetry $ \delta h_{\mu\nu}=\partial_\mu\xi_\nu+\partial_\nu\xi_\mu$ present in the $m=0$ case, so we introduce a St\"uckelberg field $A_\mu$ patterned after the gauge symmetry,
\be\label{stuk1ha} h_{\mu\nu}\rightarrow h_{\mu\nu}+\partial_\mu A_\nu+\partial_\nu A_\mu.\ee
The $\mathcal{L}_{m=0}$ term remains invariant because it is gauge invariant and (\ref{stuk1ha}) looks like a gauge transformation, so all that changes is the mass term, 
\bea S=\int d^Dx\ &&\mathcal{L}_{m=0}-\frac{1}{2}m^2(h_{\mu\nu}h^{\mu\nu}-h^2)-\half m^2 F_{\mu\nu}F^{\mu\nu}-2m ^2\left(h_{\mu\nu}\partial^\mu A^\nu-h\partial_\mu A^\mu\right) \nn\\ 
&&+\kappa h_{\mu\nu}T^{\mu\nu}-2\kappa A_\mu\partial_\nu T^{\mu\nu}, \label{astukelberg}\eea
 where we have integrated by parts in the last term, and where $F_{\mu\nu}\equiv\partial_\mu A_\nu-\partial_\nu A_\mu.$

There is now a gauge symmetry 
\be  \delta h_{\mu\nu}=\partial_\mu \xi_\nu+\partial_\nu \xi_\mu,\ \ \delta A_\mu=-\xi_\mu, \ee
and fixing the gauge $\xi_\mu=0$ recovers the original massive gravity action (as in the vector case, this is a gauge condition for which it is permissible to substitute back into the action, because the potentially lost $A_\mu$ equation is implied by the divergence of the $h_{\mu\nu}$ equation).  At this point, we might consider scaling $A_\mu\rightarrow {1\over m}A_\mu$ to normalize the vector kinetic term, then take the $m\rightarrow 0$ limit.  In this limit, we would end up with a massless graviton and a massless photon, for a total of 4 degrees of freedom (in 4 dimensions).  So at this point, $m\rightarrow 0$ is still not a smooth limit, since we would be losing one of the original 5 degrees of freedom.

We have to go one step further and introduce a scalar gauge symmetry, by introducing another St\"uckelberg field $\phi$,
\be A_\mu\rightarrow A_\mu+\partial_\mu\phi.\ee
The action (\ref{astukelberg}) now becomes
\bea \nn S=\int d^Dx\ \mathcal{L}_{m=0}&-&\frac{1}{2}m^2(h_{\mu\nu}h^{\mu\nu}-h^2)-\half m^2 F_{\mu\nu}F^{\mu\nu}\\ \nn &-&2m ^2\left(h_{\mu\nu}\partial^\mu A^\nu-h\partial_\mu A^\mu\right)-2m^2\left(h_{\mu\nu}\partial^\mu\partial^\nu\phi-h\partial^2\phi \right)+\kappa h_{\mu\nu}T^{\mu\nu}\\ \label{massivenolimit}
&-&2\kappa A_\mu\partial_\nu T^{\mu\nu}+2\kappa \phi \partial\partial T,\eea
where $\partial\partial T\equiv \partial_\mu\partial_\nu T^{\mu\nu}$ and we have integrated by parts in the last term.  

There are now two gauge symmetries 
\bea \delta h_{\mu\nu}&=&\partial_\mu \xi_\nu+\partial_\nu \xi_\mu,\ \ \delta A_\mu=-\xi_\mu \\ 
\delta A_\mu&=&\partial_\mu\Lambda,\ \ \ \delta\phi=-\Lambda. \eea
By fixing the gauge $\phi=0$ we recover the lagrangian (\ref{astukelberg}) .  

Suppose we now rescale $A_\mu\rightarrow {1\over m}A_\mu$, $\phi\rightarrow {1\over m^2}\phi$, under which the action becomes 
\bea \nn S=\int d^Dx\ \mathcal{L}_{m=0}&-&\frac{1}{2}m^2(h_{\mu\nu}h^{\mu\nu}-h^2)-\half  F_{\mu\nu}F^{\mu\nu}\\ \nn &-&2m \left(h_{\mu\nu}\partial^\mu A^\nu-h\partial_\mu A^\mu\right)-2\left(h_{\mu\nu}\partial^\mu\partial^\nu\phi-h\partial^2\phi \right)+\kappa h_{\mu\nu}T^{\mu\nu}\\ \label{massivenolimit}
&-&{2\over m}\kappa A_\mu\partial_\nu T^{\mu\nu}+{2\over m^2}\kappa \phi \partial\partial T,\eea
 and the gauge transformations become 
\bea \delta h_{\mu\nu}&=&\partial_\mu \xi_\nu+\partial_\nu \xi_\mu,\ \ \delta A_\mu=-m\xi_\mu \nn \\ \label{massivenolimitstuk}
\delta A_\mu&=&\partial_\mu\Lambda,\ \ \ \delta\phi=-m\Lambda, \eea
where we have absorbed one factor on $m$ into the gauge parameter $\Lambda$.

Now take the $m\rightarrow 0$ limit. (If the source is not conserved and the divergences do not go to zero fast enough with $m$ \cite{Fronsdal:1979yn}, then $\phi$ and $A_\mu$ become strongly coupled to the divergence of the source, so we now assume the source is conserved.)  In this limit, the theory now takes the form 
\be \label{m0actionprediag} S=\int d^Dx\ \mathcal{L}_{m=0}-{1\over 2} F_{\mu\nu}F^{\mu\nu}-2\left(h_{\mu\nu}\partial^\mu\partial^\nu\phi-h\partial^2\phi \right)+\kappa h_{\mu\nu}T^{\mu\nu},\ee
we will see that this has all 5 degrees of freedom; a scalar tensor vector theory where the vector is completely decoupled but the scalar is kinetically mixed with the tensor.  

To see this, we will un-mix the scalar and tensor, at the expense of the minimal coupling to $T^{\mu\nu}$, by a field redefinition.  Consider the change 
\be h_{\mu\nu}= h^\prime_{\mu\nu}+\pi\eta_{\mu\nu}\label{conformalt},\ee
where $\pi$ is any scalar.  This is the linearization of a conformal transformation.   The change in the massless spin 2 part is (no integration by parts here)
\be \mathcal{L}_{m=0}(h)=\mathcal{L}_{m=0}(h^\prime)+(D-2)\left[\partial_\mu\pi\partial^\mu h^\prime-\partial_\mu\pi\partial_\nu h^{\prime\mu\nu}+\half(D-1)\partial_\mu\pi\partial^\mu\pi\right].\ee
This is simply the linearization of the effect of a conformal transformation on the Einstein-Hilbert action.  

By taking $\pi= {2\over D-2}\phi$ in the transformation (\ref{conformalt}), we can arrange to cancel all the off-diagonal $h\phi$ terms in the lagrangian (\ref{m0actionprediag}), trading them in for a $\phi$ kinetic term.  The lagrangian (\ref{m0actionprediag}) now takes the form, 
\be S=\int d^Dx\ \mathcal{L}_{m=0}(h^\prime)-{1\over 2} F_{\mu\nu} F^{\mu\nu}-2{D-1\over D-2}\partial_\mu \phi\partial^\mu \phi+\kappa h^\prime_{\mu\nu}T^{\mu\nu}+ {2\over D-2}\kappa \phi T,\ee
and the gauge transformations read
\bea \delta h^\prime_{\mu\nu}&=&\partial_\mu \xi_\nu+\partial_\nu \xi_\mu,\ \ \delta A_\mu=0 \\ 
\delta A_\mu&=&\partial_\mu\Lambda,\ \ \ \delta \phi=0. \eea
There are now (for $D=4$) manifestly five degrees of freedom, two in a canonical massless graviton, two in a canonical massless vector, and one in a canonical massless scalar\footnote{Ordinarily the Maxwell term should come with a $1/4$ and the scalar kinetic term with a $1/2$, but we leave different factors here just to avoid unwieldiness.}.   

Note however, that the coupling of the scalar to the trace of the stress tensor survives the $m=0$ limit.  We have exposed the origin of the vDVZ discontinuity.  The extra scalar degree of freedom, since it couples to the trace of the stress tensor, does not affect the bending of light (for which $T=0$), but it does affect the newtonian potential.  This extra scalar potential exactly accounts for the discrepancy between the massless limit of massive gravity and massless gravity.  

As a side note, one can see from this St\"uckelberg trick that violating the Fierz-Pauli tuning for the mass term leads to a ghost.  Any deviation from this form, and the St\"uckelberg scalar will acquire a kinetic term with four derivatives $\sim (\square\phi)^2$, indicating that it carries two degrees of freedom, one of which is ghostlike \cite{deUrries:1995ty,deUrries:1998bi}.  The Fierz-Pauli tuning is required to exactly cancel these terms, up to total derivative.  

Returning to the action for $m\not=0$ (and a not necessarily conserved source), we now know to apply the transformation $h_{\mu\nu}= h^\prime_{\mu\nu}+ {2\over D-2}\phi\eta_{\mu\nu}$, which yields,

\bea \nn S=\int d^Dx\ \mathcal{L}_{m=0}(h^\prime)&-&\frac{1}{2}m^2(h^\prime_{\mu\nu}h^{\prime\mu\nu}-h^{\prime 2})-\half F_{\mu\nu}F^{\mu\nu}+2{D-1\over D-2}\phi \left(\square+{D\over D-2}m^2\right)\phi \\ \nn &-&2m \left(h^\prime_{\mu\nu}\partial^\mu A^\nu-h^\prime\partial_\mu A^\mu\right)+2{D-1\over D-2}\left(m^2 h^\prime\phi +2m\phi\partial_\mu A^\mu\right)\\ \label{massivenolimit}
&+&\kappa h^\prime_{\mu\nu}T^{\mu\nu}+{2\over D-2}\kappa \phi T-{2\over m}\kappa A_\mu\partial_\nu T^{\mu\nu}+{2\over m^2}\kappa \phi \partial\partial T.\eea
The gauge symmetry reads
\bea \delta h^\prime_{\mu\nu}&=&\partial_\mu \xi_\nu+\partial_\nu \xi_\mu+{2\over D-2}m\Lambda\eta_{\mu\nu},\ \ \delta A_\mu=-m\xi_\mu \\ 
\delta A_\mu&=&\partial_\mu\Lambda,\ \ \ \delta\phi=-m\Lambda. \eea

We can go to a Lorentz-like gauge, by imposing the gauge conditions \cite{Nibbelink:2006sz,Huang:2007xf}
\bea &&\partial^\nu h'_{\mu\nu}-\half \partial_\mu h'+m A_\mu=0,\\
&& \partial_\mu A^\mu+m\left(\half h'+2{D-1\over D-2}\phi\right)=0.
\eea
The first condition fixes the $\xi_\mu$ symmetry up to a residual transformation satisfying $(\square-m^2)\xi_\mu=0$.  It is invariant under $\Lambda$ transformations, so it fixes none of this symmetry.  The second condition fixes the $\Lambda$ symmetry up to a residual transformation satisfying $(\square-m^2)\Lambda=0$.  It is invariant under $\xi_\mu$ transformations, so it fixes none of this symmetry.  We add two corresponding gauge fixing terms to the action, resulting from either Fadeev-Popov gauge fixing or classical gauge fixing,
\bea S_{\rm GF1}&=&\int d^Dx\ -\left(\partial^\nu h'_{\mu\nu}-\half \partial_\mu h'+m A_\mu\right)^2,\\
 S_{\rm GF2}&=&\int d^Dx\ -\left( \partial_\mu A^\mu+m\left(\half h'+2{D-1\over D-2}\phi\right)\right)^2=0.
\eea
These have the effect of diagonalizing the action,
\bea \nn S+S_{\rm GF1}+ S_{\rm GF2}&=&\int d^Dx\ \half h'_{\mu\nu}\(\square-m^2\)h'^{\mu\nu}-{1\over 4} h'\(\square-m^2\)h' \nn \\ 
&& +A_\mu\(\square-m^2\) A^\mu+2{D-1\over D-2}\phi \left(\square-m^2\right)\phi \nn \\ \label{massivenolimitdiag}
&&+\kappa h^\prime_{\mu\nu}T^{\mu\nu}+{2\over D-2}\kappa \phi T-{2\over m}\kappa A_\mu\partial_\nu T^{\mu\nu}+{2\over m^2}\kappa \phi \partial\partial T.\nn \\
\eea
The propagators of $h'_{\mu\nu}$, $A_{\mu}$ and $\phi$ are now, respectively, 
\be \label{gravstukelprops} \frac{-i}{p^2+m^2}\left[\half\left(\eta_{\alpha\sigma} \eta_{\beta\lambda}+ \eta_{\alpha\lambda} \eta_{\beta\sigma}\right)-\frac{1}{D-2} \eta_{\alpha\beta} \eta_{\sigma\lambda}\right],\ \ \  \half{-i\eta_{\mu\nu}\over p^2+m^2},\ \ \ {D-2\over 4(D-1)}{-i\over p^2+m^2},\ee
which all behave as $\sim {1\over p^2}$ for high momenta, so we may now apply standard power counting arguments.

With some amount of work, it is possible to find the gauge invariant mode functions for $h'_{\mu\nu}$, $A_\mu$ and $\phi$, which can then be compared to the unitary gauge mode functions of Section \ref{modesolutions}.  In the massless limit, there is a direct correspondence; $\phi$ is gauge invariant and its one degree of freedom is exactly the longitudinal mode (\ref{longitudinalmode}), the $A_{\mu}$ has the usual Maxwell gauge symmetry and its gauge invariant transverse modes are exactly the vector modes (\ref{vectormodes}), and finally the $h'_{\mu\nu}$ has the usual massless gravity gauge symmetry and its gauge invariant transverse modes are exactly the transverse modes of the massive graviton.

\subsection{\label{filtersection}Mass terms as filters and degravitation}

There is a way of interpreting the graviton mass as a kind of high pass filter, through which sources must pass before the graviton sees them.  For a short wavelength source, the mass term does not have much effect, but for a long wavelength source (such as the cosmological constant), the mass term acts to screen it, potentially explaining how the observed cosmic acceleration could be small despite a large underlying cosmological constant \cite{Dvali:2007kt}. 

First we will see how this works in the case of the massive vector.  Return to the action (\ref{massivephoton}), with a conserved source, before taking the $m\rightarrow 0$ limit,
\begin{equation}\label{massivephoton2}{
S=\int d^Dx\  -\frac{1}{4} F_{\mu\nu}F^{\mu\nu}-\half m^2 A_\mu A^\mu -m A_\mu\partial^\mu\phi -\half\partial_\mu\phi\partial^\mu\phi+A_\mu J^\mu.}
\end{equation}
The $\phi$ equation of motion is 
\be \square\phi +m\ \partial\cdot A=0.\ee
We would now like to integrate out $\phi$.  Quantum mechanically we would integrate it out of the path integral.  Classically we would eliminate it with its own equation of motion. Solving the equation of motion involves solving a differential equation, so the result is non-local,
\be \phi=-{m\over \square}\partial\cdot A.\ee
Plugging back into (\ref{massivephoton2}), we obtain a non-local lagrangian
\begin{equation}\label{vectorfilter}{
S=\int d^Dx\  -\frac{1}{4} F_{\mu\nu}\left(1-{m^2\over \square}\right)F^{\mu\nu} +A_\mu J^\mu,}
\end{equation}
where we have used $F_{\mu\nu}{1\over \square}F^{\mu\nu}=-2A_\mu{1\over \square}A^\mu-2 \partial\cdot A{1\over \square}\partial \cdot A,$
arrived at after integration by parts.  The lagrangian (\ref{vectorfilter}) is now a manifestly gauge invariant but non-local lagrangian for a massive vector.  The non-locality results from having integrated out the dynamical scalar mode.  The equation of motion from (\ref{vectorfilter}) is 
\be \left(1-{m^2\over \square}\right)\partial_\mu F^{\mu\nu} =- J^\nu.\ee
This is simply Maxwell electromagnetism, where the source is seen thorough a filter $\left(1-{m^2\over \square}\right)^{-1}$.  For high momenta $p\gg m$, the filter is $\sim 1$ so the theory looks like ordinary electromagnetism.   But for $p\ll m$, the filter becomes very small, so the source appears weakened.  We can think of this as a high-pass filter, where $m$ is the filter scale.   

Applied to gravity, the hope is to explain the small observed value of the cosmological constant.  The cosmological constant, being a constant, is essentially a very long wavelength source.  Gravity equipped with a high pass filter would not respond to a large bare cosmological constant, making the observed effective value appear much smaller, while leaving smaller wavelength sources unsuppressed.  This mechanism is known as \textit{degravitation} \cite{Dvali:2002pe,ArkaniHamed:2002fu,Dvali:2007kt,Patil:2010mq}. 

This filtering is essentially just the Yukawa suppression $e^{-mr}$ that comes in with massive particles, so we should be able to cast the massive graviton into a filtered form.  Look again at the action (\ref{astukelberg}) with a conserved source, before introducing the St\"ukelberg scalar, 
\be \label{astukelberg2}S=\int d^Dx\ \mathcal{L}_{m=0}-\frac{1}{2}m^2(h_{\mu\nu}h^{\mu\nu}-h^2)-\half m^2 F_{\mu\nu}F^{\mu\nu}-2m ^2\left(h_{\mu\nu}\partial^\mu A^\nu-h\partial_\mu A^\mu\right)+\kappa h_{\mu\nu}T^{\mu\nu}.\ee
Now consider the following action containing an additional scalar field $N$, 
\bea \nn S=\int d^Dx\ \mathcal{L}_{m=0}&+&m^2\bigg[-\frac{1}{2}h_{\mu\nu}h^{\mu\nu}+{1\over 4}h^2+A_\mu\square A^\mu+N(h-N) \\
&-&A^\mu\left(\partial_\mu h-2\partial^\nu h_{\mu\nu}+2\partial_\mu N\right)\bigg] +\kappa h_{\mu\nu}T^{\mu\nu}.  \label{auxiliaryaction} \eea
The field $N$ is an auxiliary field.  Its equation of motion is
\be N={1\over 2}h+\partial_\mu A^\mu,\ee
which when plugged into (\ref{auxiliaryaction}) yields (\ref{astukelberg2}). Thus the two actions are equivalent, and (\ref{auxiliaryaction}) is another action describing the massive graviton.  Here, however, there is no gauge symmetry acting on the scalar; $N$ is gauge invariant\footnote{For another form of the massive gravity action, we can take
$N'=N-\partial_\mu A^\mu$ in (\ref{astukelberg}), which gives
\bea \nn S=\int d^Dx\ \mathcal{L}_{m=0}&+&m^2\left[-\frac{1}{2}h_{\mu\nu}h^{\mu\nu}+{1\over 4}h^2-\half F_{\mu\nu}F^{\mu\nu}+N'(h-N')-2A^\mu\left(\partial_\mu h-\partial^\nu h_{\mu\nu}\right)\right] \\ \label{auxiliaryaction2} &+&\kappa h_{\mu\nu}T^{\mu\nu}-2\kappa A_\mu\partial_\nu T^{\mu\nu}.\eea
The field $N'$ now takes the value $N'={1\over 2}h$ and is no longer gauge invariant.}.  

Instead of eliminating the scalar, we can eliminate the vector $A_\mu$ using its equations of motion,
\be A_\mu={1\over \square}\left(\half \partial_\mu h-\partial^\nu h_{\mu\nu}+\partial_\mu N\right).\ee
Plugging back into (\ref{auxiliaryaction}) gives 
\begin{equation}\label{filtintermediate}
S=\int d^Dx\ \frac{1}{2}h_{\mu\nu}\left(1-{m^2\over \square}\right){\cal E}^{\mu\nu,\alpha\beta}h_{\alpha\beta}-2N{1\over \square}\left(\partial_\mu\partial_\nu h^{\mu\nu}-\square h\right)+\kappa h_{\mu\nu}T^{\mu\nu},
\end{equation}
where ${\cal E}^{\mu\nu}_{\ \ \alpha\beta}$ is the second order differential operator for the massless graviton (\ref{Eoper}).
Now, to diagonalize the action, make a conformal transformation
\be h_{\mu\nu}=h^\prime_{\mu\nu}+{2\over D-2}{1\over \square-m^2}N\eta_{\mu\nu},\ee
after which (\ref{filtintermediate}) becomes
\begin{equation}
S=\int d^Dx\ \frac{1}{2}h_{\mu\nu}^\prime \left(1-{m^2\over \square}\right){\cal E}^{\mu\nu,\alpha\beta}h^\prime_{\alpha\beta}+2{D-1\over D-2}N{1\over \square-m^2}N+\kappa h^\prime_{\mu\nu}T^{\mu\nu}+{2\over D-2}\kappa {1\over \square-m^2}NT.
\end{equation}
Finally, making the field redefinition $N^\prime={1\over \square-m^2}N$ to render the coupling to the source local, 
\begin{equation}
S=\int d^Dx\ \frac{1}{2}h_{\mu\nu}^\prime \left(1-{m^2\over \square}\right){\cal E}^{\mu\nu,\alpha\beta}h^\prime_{\alpha\beta}+2{D-1\over D-2}N^\prime (\square-m^2)N^\prime+\kappa h^\prime_{\mu\nu}T^{\mu\nu}+{2\over D-2}\kappa N^\prime T.
\end{equation}
Thus a massive graviton is equivalent to a filtered graviton coupled to $T_{\mu\nu}$ and a scalar with mass $m$ coupled with gravitational strength to the trace $T$.  The scalar is the longitudinal mode responsible for the vDVZ discontinuity.

 It is not hard to see that a linear massive graviton screens a constant source.  Looking at the equations of motion (\ref{sourcedeq}) where the source is a cosmological constant $T_{\mu\nu}\propto \eta_{\mu\nu}$, and taking the double divergence, we find $\partial^\nu\partial^\mu h_{\mu\nu}-\square h=0$, which is the statement that the linearized Ricci scalar vanishes, so a cosmological constant produces no curvature.  If degravitation can be made to work cosmologically, then this provides an interesting take on the cosmological constant problem.  Of course the smallness of the cosmological constant reappears in the ratio $m/M_P$, but as we will see, in massive gravity a small mass is technically natural.  There are other obstacles as well, and promising avenues towards overcoming them, and we will have more to say about these things while studying the non-linear theory. 

\section{\label{massivecurvedspacesection}Massive gravitons on curved spaces}

We now study some new features that emerge when the Fierz-Pauli action is put onto a curved space.  One new feature is the existence of partially massless theories.  These are theories with a scalar gauge symmetry that propagate 4 degrees of freedom in $D=4$.  Another is the absence of the vDVZ discontinuity in curved space. 

\subsection{Fierz-Pauli gravitons on curved space and partially massless theories}

We now study the linear action for a massive graviton propagating on a fixed curved background with metric $g_{\mu\nu}$.  As in the flat space case, the massless part of the action will be the Einstein-Hilbert action with a cosmological constant, ${1\over 2\kappa^2}\sqrt{-g}(R-2\bar\Lambda)$, expanded to second order in the metric perturbation $\delta g_{\mu\nu}=2\kappa h_{\mu\nu}$, about a solution $g_{\mu\nu}$.  The solution must be an Einstein space, satisfying
\be R_{\mu\nu}={R\over D}g_{\mu\nu}, \ \ \  \bar\Lambda=\left(D-2\over 2D\right)R.\ee
Appending the Fierz-Pauli mass term, we have the action
\bea \nn S&=&\int d^Dx\ \sqrt{-g}\left[ -{1\over 2}\nabla_\alpha h_{\mu\nu} \nabla^\alpha h^{\mu\nu}+\nabla_\alpha h_{\mu\nu} \nabla^\nu h^{\mu\alpha}-\nabla_\mu h\nabla_\nu h^{\mu\nu}+\half \nabla_\mu h\nabla^\mu h\right. \\ &&\left. +{R\over D}\left( h^{\mu\nu}h_{\mu\nu}-\half h^2\right)-\frac{1}{2}m^2(h_{\mu\nu}h^{\mu\nu}-h^2)+\kappa h_{\mu\nu}T^{\mu\nu}\right]. \label{curvedmassivelin}\eea
Here the metric, covariant derivatives and constant curvature $R$ are those of the background.  Notice the term, proportional to $R$, that kind of looks like a mass term, but does not have the Fierz-Pauli tuning.  There's some representation theory behind this \cite{1967JMP8170E}, and a long discussion about what it means for a particle to be ``massless'' in a curved space time \cite{Deser:1983mm}, but at the end of the day, (\ref{curvedmassivelin}) is the desired generalization of the flat space Fierz-Pauli action, which, for most choices of $m^2$, propagates 5 degrees of freedom in $D=4$.  See \cite{Fang:1978se,Aragone:1979bm,Bengtsson:1994vn,Cucchieri:1994tx,Naqvi:1999va,Polishchuk:1999nh,Buchbinder:1999ar,Brink:2000ag,Buchbinder:2000fy,Deser:2001wx,Porrati:2003sa,Porrati:2004mz,Deser:2006sq,Izumi:2007pb,Iglesias:2010xz} for some other aspects of massive gravity on curved space.  

For some choices of $m^2$, (\ref{curvedmassivelin}) propagates fewer degrees of freedom.  For $m=0$, the action has the gauge symmetry 
\be\label{curvedgaugesy} \delta h_{\mu\nu}=\nabla_\mu\xi_\nu+\nabla_\nu\xi_\mu,\ee 
and the action propagates 2 degrees of freedom in $D=4$.  As we will see momentarily, for $R={D(D-1)\over D-2}m^2$, $m\not=0$, the action has a scalar gauge symmetry, and propagates $4$ degrees of freedom in $D=4$.  For all other values of $m^2$ and $R$, it has no gauge symmetry and propagates 5 degrees of freedom in $D=4$.  This is summarized in Figure \ref{dsstability}.

We introduce a St\"uckelberg field, $A_{\mu}$, patterned after the $m=0$ gauge symmetry,
\be \label{astukcurve} h_{\mu\nu}\rightarrow h_{\mu\nu}+\nabla_\mu A_\nu+\nabla_\nu A_\mu.\ee
The $\mathcal{L}_{m=0}$ term remains invariant, the source term does not change because we will assume covariant conservation of $T^{\mu\nu}$, so all that changes is the mass term, 
\bea \nn S=\int d^Dx &&\mathcal{L}_{m=0}+\sqrt{-g}\bigg[-\frac{1}{2}m^2(h_{\mu\nu}h^{\mu\nu}-h^2)\\ &-&\half m^2 F_{\mu\nu}F^{\mu\nu}+{2\over D}m^2RA^{\mu}A_{\mu}-2m ^2\left(h_{\mu\nu}\nabla^\mu A^\nu-h\nabla_\mu A^\mu\right)+\kappa h_{\mu\nu}T^{\mu\nu}\bigg],\nn\\ \eea
where $F_{\mu\nu}\equiv\partial_\mu A_\nu-\partial_\nu A_\mu=\nabla_\mu A_\nu-\nabla_\nu A_\mu,$
 and we have used the relation
$\nabla_\mu A_\nu \nabla^\nu A^\mu=(\nabla_\mu A^\mu)^2-R_{\mu\nu}A^\mu A^\nu$
 to see that there is now a term that looks like a mass for the vector, proportional to the background curvature.  There is now a gauge symmetry 
\be  \delta h_{\mu\nu}=\nabla_\mu \xi_\nu+\nabla_\nu \xi_\mu,\ \ \delta A_\mu=-\xi_\mu, \ee
and fixing the gauge $\xi_\mu=0$ recovers the original action (\ref{curvedmassivelin}).  

Introducing the St\"ukelberg scalar and its associated gauge symmetry,
\be \label{phistukcurve} A_{\mu}\rightarrow A_{\mu}+\nabla_\mu\phi, \  \ \ \ \delta A_{\mu}=\nabla_\mu \Lambda,\ \ \ \delta\phi=-\Lambda,\ee
we have
\bea\nn S=\int d^Dx && \mathcal{L}_{m=0}+\sqrt{-g}\bigg[-\frac{1}{2}m^2(h_{\mu\nu}h^{\mu\nu}-h^2)\\ \nn &-&\half m^2 F_{\mu\nu}F^{\mu\nu}+{2\over D}m^2RA^{\mu}A_{\mu}-2m ^2\left(h_{\mu\nu}\nabla^\mu A^\nu-h\nabla_\mu A^\mu\right)\nn\\ 
&&+{4m^2 R\over D} A^\mu \nabla_\mu \phi+{2m^2 R\over D}(\partial\phi)^2-2m^2\(h_{\mu\nu}\nabla^\mu\nabla^\nu\phi-h\square\phi\) +\kappa h_{\mu\nu}T^{\mu\nu}\bigg]. \nn\\
\eea

Under the conformal transformation 
\be \label{curvedconfo} h_{\mu\nu}= h^\prime_{\mu\nu}+\pi g_{\mu\nu},\ee
where $\pi$ is any scalar, the change in the massless part is (no integration by parts here)
\bea \nn \mathcal{L}_{m=0}(h) =&& \mathcal{L}_{m=0}(h^\prime)+\sqrt{-g}\bigg[(D-2)\left(\nabla_\mu\pi\nabla^\mu h^\prime-\nabla_\mu\pi\nabla_\nu h^{\prime\mu\nu}+\half(D-1)\nabla_\mu\pi\nabla^\mu\pi\right) \\ && -R{D-2\over D}\(h'\pi+{D\over 2} \pi^2\)\bigg]. \eea

Applying this in the case $\pi= {2\over D-2}m^2\phi$ yields,
\bea \nn S=\int d^Dx\ \mathcal{L}_{m=0}(h^\prime)&+&\sqrt{-g}\bigg[-\frac{1}{2}m^2(h^\prime_{\mu\nu}h^{\prime\mu\nu}-h^{\prime 2})-\half m^2F_{\mu\nu}F^{\mu\nu}+{2\over D}m^2RA^{\mu}A_{\mu} \\
 \nn &-&2m^2 \left(h^\prime_{\mu\nu}\nabla^\mu A^\nu-h^\prime\nabla_\mu A^\mu\right)+2m^2\({D-1\over D-2}m^2-{R\over D}\)\left(2\phi\nabla_\mu A^\mu+h'\phi\right)\\ \label{massivenolimit}
&-&2m^2 \({D-1\over D-2}m^2-{R\over D}\)\left((\partial\phi)^2-m^2 {2D\over D-2}\phi^2\right) \nn \\
&+&\kappa h^\prime_{\mu\nu}T^{\mu\nu}+{2\over D-2}m^2\kappa \phi T\bigg].\eea
The gauge symmetry reads
\bea \delta h^\prime_{\mu\nu}&=&\partial_\mu \xi_\nu+\partial_\nu \xi_\mu+{2\over D-2}\Lambda g_{\mu\nu},\ \ \delta A_\mu=-\xi_\mu \\ 
\delta A_\mu&=&\partial_\mu\Lambda,\ \ \ \delta\phi=-\Lambda. \eea

Note that for the special value
\be \label{partialmasslessvalue} R={D(D-1)\over D-2}m^2,\ee
the dependence on $\phi$ completely cancels out of (\ref{massivenolimit}).  Setting unitary gauge $A_\mu=0$, and given the replacements (\ref{astukcurve}), (\ref{phistukcurve}) and the conformal transformation (\ref{curvedconfo}), this implies that the original lagrangian (\ref{curvedmassivelin}) with the mass (\ref{partialmasslessvalue}) has the gauge symmetry
\be \label{partialmasslesssym} \delta h_{\mu\nu}=\nabla_\mu\nabla_\nu\lambda+{1\over D-2}m^2\lambda g_{\mu\nu},\ee
where $\lambda(x)$ is a scalar gauge parameter.  The theory at the value (\ref{partialmasslessvalue}) is called \textit{partially massless} \cite{Higuchi:1986py,Deser:2001us,Deser:2001xr,Gabadadze:2008ha,Gabadadze:2008uc}.  Due to the gauge symmetry (\ref{partialmasslesssym}), this theory propagates one fewer degree of freedom than usual so for $D=4$ it carries four degrees of freedom rather than five.  Consistency demands that the trace of the stress tensor vanish for this theory (if it is conserved).  In addition, it marks a boundary in the $R,m^2$ plane between stable and unstable theories, see Figure \ref{dsstability}.

\begin{figure}[h!]
\begin{center}
\epsfig{file=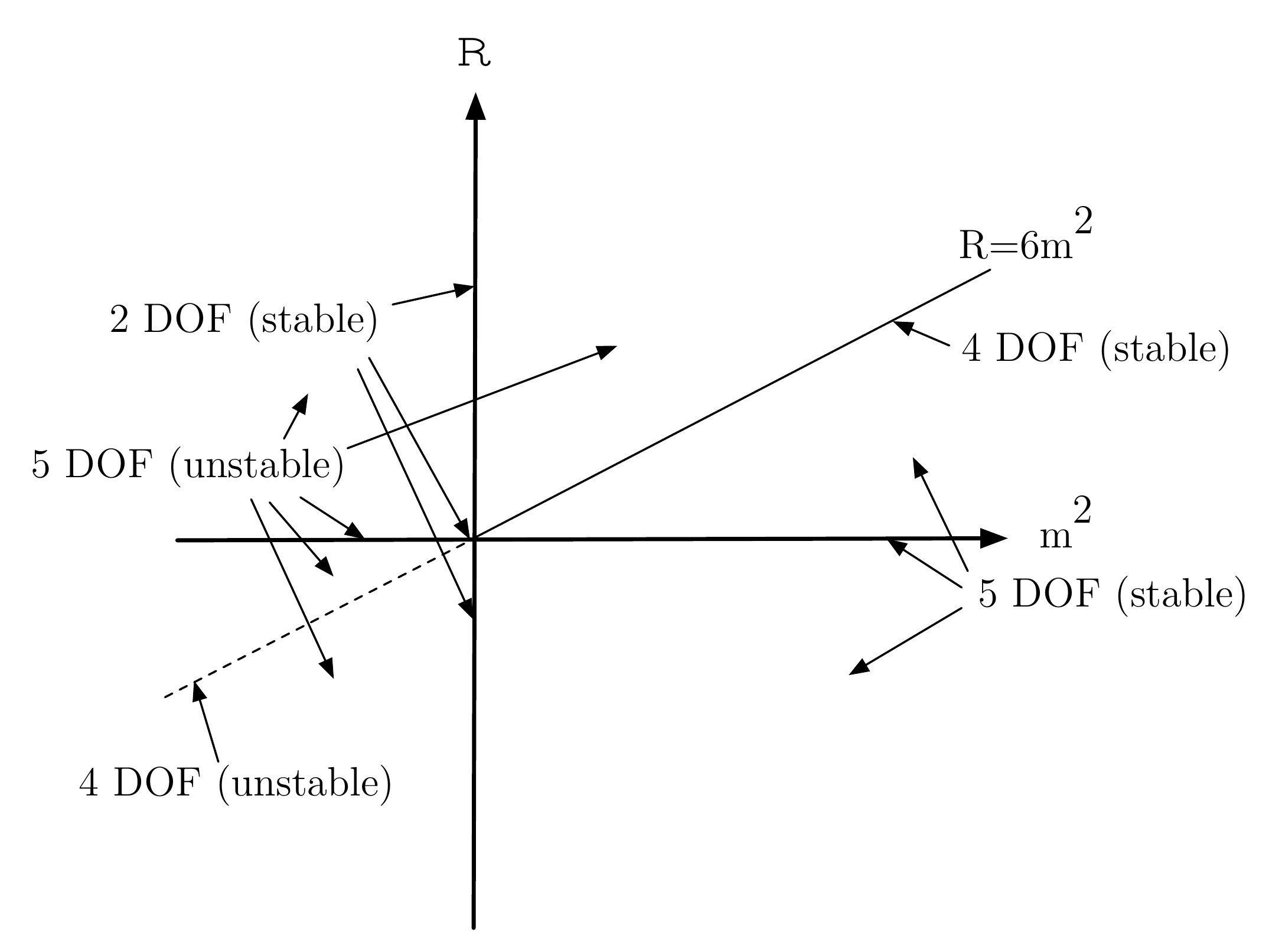,height=3in,width=3.7in}
\caption{\small Degrees of freedom and their stability for values in the $R,m^2$ plane for massive gravity on an Einstein space (shown for $D=4$, the other dimensions follow similarly).  The line $R=6m^2, \ m^2\not=0$ is where a scalar gauge symmetry appears, reducing the number of degrees of freedom by one.  The line $m^2=0$ is where the vector gauge symmetries appear, reducing the number of degrees of freedom by three.}
\label{dsstability}
\end{center}
\end{figure}

\subsection{Absence of the vDVZ discontinuity on curved space}
 
To study the fate of the vDVZ discontinuity, we now take a massless limit, while preserving the number of degrees of freedom.  However,  there are many paths in the $R, m^2$ plane and correspondingly, different ways to take the massless limit. 

For example, let's take the $m\rightarrow 0$ limit for fixed but non-zero $R$.  Here we will see that the vDVZ discontinuity is absent \cite{Higuchi:1986py,Porrati:2000cp,Kogan:2000uy,Karch:2001jb}.  First we go to canonical normalization for the vector by taking $A_{\mu}\rightarrow {1\over m}A_\mu$.  Then we notice that we can immediately take the $m\rightarrow 0$ limit, without the need to introduce the second St\"uckelberg field $\phi$.  This is because a mass term for the vector is present in this limit, so no degrees of freedom are lost.  Thus our limiting action is 
\be S=\int d^Dx\ \mathcal{L}_{m=0}+\sqrt{-g}\bigg[-{1\over 2}  F_{\mu\nu}F^{\mu\nu}+{2R\over D}A^{\mu}A_{\mu}+\kappa h_{\mu\nu}T^{\mu\nu}\bigg].\ee
The massive vector completely decouples from the stress tensor, so there is no vDVZ discontinuity.  Notice that the vector is a tachyon in dS space but healthy in AdS, consistent with the stability regions shown in Figure \ref{dsstability}.  These regions can all be investigated in similar fashion.  Finally, note also that the $R\rightarrow 0$ and $m\rightarrow 0$ limits do not commute.

\section{Non-linear interactions}

Up to this point, we have only studied the linear theory of massive gravity, which is determined by the requirement that it propagate only one massive spin 2 degree of freedom.  We now turn to the study of the possible interactions and non-linearities for massive gravity.

\subsection{\label{GRsection}General relativity}

We start by reviewing the story of non-linearities in GR.  We will then repeat it for massive gravity, to see exactly where things differ. 
General relativity is the theory of a dynamical metric $g_{\mu\nu}$, with the action
\be \label{GRaction1} S={1\over 2\kappa^2}\int d^D x\ \sqrt{-g}R.\ee
The action is invariant under (pullback) diffeomorphism gauge symmetries $f^\mu(x)$,
\be\label{GRfullgauge} g_{\mu\nu}(x)\rightarrow {\partial f^\alpha \over \partial  x^\mu}{\partial f^\beta \over \partial  x^\nu}g_{\alpha\beta}\left(f(x)\right).\ee
Infinitesimally, for $f^\mu(x)=x^\mu+\xi^\mu(x),$ this reads
\be \delta g_{\mu\nu}={\cal L}_\xi g_{\mu\nu}=\nabla_\mu\xi_\nu+\nabla_\nu\xi_\mu,
\ee
where $\xi^\mu$ is the gauge parameter, ${\cal L}_\xi$ is the Lie derivative, and indices are lowered by the metric. 

The field equation for the metric is
\be\label{eisteinequations} R_{\mu\nu}-\half Rg_{\mu\nu}=0,\ee
and the most symmetric solution is flat space $g_{\mu\nu}=\eta_{\mu\nu}$.

To see that this is a theory of an interacting massless spin 2 field, we expand the action around the flat space solution $\eta_{\mu\nu}$,
\[g_{\mu\nu}=\eta_{\mu\nu}+h_{\mu\nu}.\]
To second order in $h_{\mu\nu}$ the action is 
\be S_2={1\over 2\kappa^2}\int d^Dx\ \half\delta^2(\sqrt{-g}R)={1\over 4\kappa^2}\int d^Dx\ -\frac{1}{2}\partial_\lambda h_{\mu\nu}\partial^\lambda h^{\mu\nu}+\partial_\mu h_{\nu\lambda}\partial^\nu h^{\mu\lambda}-\partial_\mu h^{\mu\nu}\partial_\nu h+\frac{1}{2}\partial_\lambda h\partial^\lambda h ,
\ee
where indices on $h_{\mu\nu}$ are raised and traced with the flat background metric $\eta_{\mu\nu}$ and we have ignored total derivatives.
After canonical normalization, $h_{\mu\nu}= 2\kappa \hat h_{\mu\nu}$,
this linear action for GR is exactly that of the $m=0$ spin two particle in Minkowski space (\ref{massivefreeaction}).

If we continue the expansion around flat space to higher non-linear order in $\hat h_{\mu\nu}$, we get a slew of interaction terms, all with two derivatives and increasing powers of $\hat h$, and coefficients all precisely fixed so that the result is diffeomorphism invariant and sums up to (\ref{GRaction1}).  Schematically,
\be \label{GRexpand1} S=\int d^Dx\ \partial^2\hat h^2+{\kappa}\partial^2\hat h^3+\cdots+{\kappa^n}\partial^2\hat h^{n+2}+\cdots
\ee
The higher and higher powers of $\hat h_{\mu\nu}$ are suppressed by appropriate powers of $\kappa$.  The action is expanded in powers of $\kappa \hat h$ and the linearized expansion is valid when $\kappa \hat  h\ll 1.$

When expanding around the background, we must put all of the gauge symmetry into $h_{\mu\nu}$, so that the transformation rule is
\be h_{\mu\nu}(x)\rightarrow  \partial_\mu f^\alpha\partial_\nu f^\beta \eta_{\alpha\beta}-\eta_{\mu\nu}+ \partial_\mu f^\alpha\partial_\nu f^\beta h_{\alpha\beta}\left(f(x)\right).\ee
For infinitesimal transformations $f^\mu=x^\mu+\xi^\mu$, this is
\be \label{hgrtransflat}\delta h_{\mu\nu}=\partial_\mu\xi_\nu+\partial_\nu\xi_\mu+{\cal L}_\xi h_{\mu\nu}.\ee
This is the all orders expression in $h_{\mu\nu}$ for the infinitesimal gauge symmetry, which shows that it gets modified at higher order from its linear form (\ref{gaugesymorig}).

This argument can be turned around.  We can start with the massless linear graviton action and ask what higher power interaction terms can be added.  The possible terms can be arranged in powers of the derivatives, and lower derivatives will be more important at lower energies.  Starting with two derivatives, we ask what terms of the form (\ref{GRexpand1}) can be added.  We must add higher order terms in such a way that the linear gauge invariance is preserved, though the form of the gauge transformations may be altered at higher order in $\hat h$.  These requirements are strong enough to force the interactions to be those obtained from expanding the Einstein-Hilbert term \cite{Gupta:1954zz,Kraichnan:1955zz,Weinberg:1965rz,Deser:1969wk,Boulware:1974sr,Fang:1978rc,Wald:1986bj}.  Looking at it from this direction, an amazing thing happens.  The linear action with which we started has a non-dynamical background metric $\eta_{\mu\nu}$, but after adding all the interactions of the Einstein-Hilbert term, the change of variables $h_{\mu\nu}\rightarrow g_{\mu\nu}-\eta_{\mu\nu}$ completely eliminates the background metric from the action.  The fully interacting Einstein-Hilbert action turns out to be background independent.  This will not be the case once we add a mass term, though we will still be able to introduce gauge invariance through the St\"ukelberg trick.

\subsubsection*{Zero derivative interactions mean curved space}

If we ask for interactions terms with fewer than two derivatives, the only option is zero derivatives, and diffeomorphism invariance forces them to sum up to a cosmological constant $\sqrt{-g}=\half h+{\cal O}\(h^2\)$.  This contains a term linear in $h$, which means there is a tadpole and $h=0$ is not a solution to the equations of motion, so we are not expanded around a solution \cite{Gabadadze:2003jq}.  Instead we may consider GR with a cosmological constant $\bar\Lambda$,
\be \label{GRaction2} S={1\over 2\kappa^2}\int d^D x\ \sqrt{-g}\(R-2\bar\Lambda\).\ee
The equations of motion $G_{\mu\nu}+\bar\Lambda g_{\mu\nu}=0$, implies that the background solution $g^{(0)}_{\mu\nu}$ is an Einstein space,
\be R_{\mu\nu}={R\over D}g_{\mu\nu}, \ \ \  \bar\Lambda=\left(D-2\over 2D\right)R.\ee
Expanding around the background $g_{\mu\nu}=g^{(0)}_{\mu\nu}+h_{\mu\nu}$ to quadratic order, we have the linearized action
\bea \nn S_2&=&{1\over 2\kappa^2}\int d^Dx\ \half\delta^2(\sqrt{-g}R) \\ \nn
&=&{1\over 4\kappa^2}\int d^Dx\ \sqrt{-g}\left[ -{1\over 2}\nabla_\alpha h_{\mu\nu} \nabla^\alpha h^{\mu\nu}+\nabla_\alpha h_{\mu\nu} \nabla^\nu h^{\mu\alpha}-\nabla_\mu h\nabla_\nu h^{\mu\nu}+\half \nabla_\mu h\nabla^\mu h\right. \\ \nn &&\left. +{R\over D}\left( h^{\mu\nu}h_{\mu\nu}-\half h^2\right)\right]+(\text{total }d), \eea
where the covariant derivatives, curvature and metric determinant out front are those of the background.  After canonical normalization, $h_{\mu\nu}= 2\kappa \hat h_{\mu\nu}$,
this linear action is exactly that of the $m=0$ spin 2 particle in an Einstein space (\ref{curvedmassivelin}) we used in Section \ref{massivecurvedspacesection}.

Upon expanding around the background, we must put all of the gauge transformations into $h_{\mu\nu}$, so that the transformation rule is
\be h_{\mu\nu}(x)\rightarrow  \partial_\mu f^\alpha\partial_\nu f^\beta g^{(0)}_{\alpha\beta}\left(f(x)\right)-g^{(0)}_{\mu\nu}+ \partial_\mu f^\alpha\partial_\nu f^\beta h_{\alpha\beta}\left(f(x)\right).\ee
For infinitesimal transformations $f^\mu=x^\mu+\xi^\mu$, this is
\be \delta h_{\mu\nu}=\nabla_\mu\xi_\nu+\nabla_\nu\xi_\mu+{\cal L}_\xi h_{\mu\nu},\ee
where the covariant derivatives are of the background.
This is an all orders expression in $h_{\mu\nu}$ for the infinitesimal gauge symmetry.   To linear order, this reproduces the massless curved space gauge symmetry (\ref{curvedgaugesy}).
As in the flat space case, this argument may be reversed.  The only possible interactions for a massless graviton propagating on an Einstein space (an Einstein space is the only space on which a free graviton can consistently propagate \cite{Deser:2006sq}) should be those of Einstein gravity with a cosmological constant. 

\subsubsection*{Spherical solutions}

Returning now to a flat background $\Lambda=0$, and setting $D=4$, we attempt to find spherically symmetric static solutions to the equations of motion $R_{\mu\nu}-\half R g_{\mu\nu}=0$, using an expansion in powers of non-linearity, a method we will repeat for the non-linear massive graviton.
The most general spherically symmetric static metric can be written 
\be g_{\mu\nu}dx^\mu dx^\nu= -B(r)dt^2+C(r)dr^2+A(r)r^2d\Omega^2.\ee
The most general gauge transformation which preserves this ansatz is a reparametrization of the radial coordinate $r$.  We can use this to set the gauge $A(r)=C(r)$, bringing the metric into the form 
\be g_{\mu\nu}dx^\mu dx^\nu= -B(r)dt^2+C(r)\left[dr^2+r^2d\Omega^2\right].\ee
The linear expansion of this around flat space will be seen to correspond to the Lorenz gauge choice.  
Plugging this ansatz into the equations of motion, we get the following from the $tt$ equation and $rr$ equation respectively,

\bea  3 r \left(C'\right)^2-4 C \left(2 C'+r C''\right)&=&0, \\ 
 4 B' C^2+2 \left(2 B+r B'\right) C' C+B r \left(C'\right)^2&=&0. 
\eea
The $\theta\theta$ equation, (which is the same as the $\phi\phi$ equation by spherical symmetry) turns out to be redundant.  It is implied by the $tt$ and $rr$ equations (this happens because of a Noether identity resulting from the radial re-parametrization gauge invariance). 

 We start by doing a linear expansion of these equations around the flat space solution
 \be B_0(r)=1,\ \ \ C_0(r)=1. \ee
We do this by the method of linearizing a non-linear differential equation about a solution.  We introduce the expansion
\bea B(r)&=&B_0(r)+\epsilon B_1(r)+\epsilon^2 B_2(r)+\cdots ,\\ \nn
C(r)&=&C_0(r)+\epsilon C_1(r)+\epsilon^2 C_2(r)+\cdots ,
\eea
where $\epsilon$ will be a parameter that counts the order of non-linearity.  We proceed by plugging into the equations of motion and collecting like powers of $\epsilon$.   The ${\cal O}(0)$ part gives $0=0$ because $B_0,C_0,A_0$ are solutions to the full non-linear equations.  At each higher order in $\epsilon $ we will obtain a linear equation that lets us solve for the next term in terms of the solutions to previous terms.  

At ${\cal O}(\epsilon)$ we obtain
\be C_1''+{2C_1'\over r}=0,\ \ \ B_1'+C_1'=0.\ee
There are three arbitrary constants in the general solution.  Demanding that $B_1$ and $C_1$ go to zero as $r\rightarrow\infty$, so that the solution is asymptotically flat, fixes two.  The other constant remains unfixed, and represents the mass of the solution\footnote{If we had used the gauge $A=1$, these would be 1st order equations and there would be only two constants to fix, i.e. the order of the equation seems to depend on the gauge.  The reason for this is that the Lorenz gauge does not completely fix the gauge.  Under an active diffeomorphism $f(x)$, the metric transforms as $g_{\mu\nu}(x)\rightarrow \partial_\mu f^\lambda \partial_\nu f^\sigma g_{\lambda\sigma}(f(x))$.  Under a radial reparametrization $\bar r(r)$ this becomes \be B(r)\rightarrow B(\bar r(r)),\ \ \ C(r)\rightarrow {\partial \bar r\over \partial r}C(\bar r(r)),\ \ \ A(r)\rightarrow A(\bar r(r)){\bar r(r)\over r}.\ee  
Choosing the gauge $A=1$ amounts to solving $A(\bar r(r)){\bar r(r)\over r}=1$ for $\bar r$, and this is an algebraic equation so the solution is unique and this choice completely fixes the gauge.    Choosing the gauge $A=C$ amounts to solving ${\partial \bar r\over \partial r}C(\bar r(r))= A(\bar r(r)){\bar r(r)\over r}$, which is a differential equation for $\bar r$.  Thus the solution is not unique because there is an integration constant.  The transformations that preserve the gauge choice solve ${\partial \bar r\over \partial r}= {\bar r(r)\over r}$, with solution $\bar r=kr$, i.e. constant scalings of $r$.  This appears as an extra boundary condition that must be specified, because we must fix the scaling.

In addition, our ansatz is also invariant under time scaling, $\bar t=kt$, under which $B(r)\rightarrow kB(r)$.  This represents another unfixed gauge symmetry.  We generally fix this and the radial scaling by demanding that $A,B\rightarrow 1$ as $r\rightarrow \infty$.  Then the only boundary condition is the mass.
}.  We choose it to reproduce the solution (\ref{masslesspoinspher}) we got from the propagator.  We have then,
\be B_1=-{2GM\over r},\ \ \ C_1={2GM\over r}. \ee

At ${\cal O}(\epsilon^2)$ we obtain another set of differential equations
\bea \frac{3 G^2 M^2}{r^4}-\frac{2 C_2'}{r}-C_2''=0 \\  
\frac{7 G^2 M^2}{r^3}+B_2'+C_2'=0. 
\eea
Again there are three arbitrary constants in the general solution.  Demanding that $B_2$ and $C_2$ go to zero as $r\rightarrow\infty$ again fixes two.  The third appears as the coefficient of  a ${1\over r}$ term, and we set it to zero so that the second order term does not compete with the first order term as $r\rightarrow\infty$.  We can continue in this way to any order, and we obtain the expansion 
\bea B(r)-1&=&-{2GM\over r}\left(1-\frac{GM}{r}+\cdots\right),\\
C(r)-1&=&{2GM\over r}\left(1+\frac{3GM}{4r}+\cdots\right).
\eea
The dots represent higher powers in the non-linearity parameter $\epsilon$.  We see that the non-linearity expansion is an expansion in the parameter $r_S/r$, where 
\be r_S=2GM,\ee
is the Schwarzschild radius.  Thus the Schwarzschild radius $r_S\sim M/M_P^2$ represents the radius at which non-linearities become important.  This scale can also be estimated straight from the lagrangian (\ref{GRexpand1}).  The non-linear terms are suppressed relative to the linear terms by powers of the factor $\hat h/M_P$.  The linear solution is $\hat h\sim {M\over M_Pr}$, so $\hat h/ M_P$ becomes order one when $r\sim M/M_P^2\sim r_S$.

In GR, the linearity expansion can be easily summed to all orders by solving the original equations exactly, 
\[ B(r)=\frac{\left(1-\frac{2 r}{G M}\right)^2}{\left(1+\frac{2 r}{G M}\right)^2},\ \ \ \ C(r)={(1+{G M\over 2r})^4}.\]
This is the Schwarzschild solution, in Lorenz gauge.   

\subsubsection*{GR as a quantum effective field theory}

We can understand the previous results from an effective field theory viewpoint, and in the process check that the black hole solution is still valid despite quantum corrections.  Pure Einstein gravity in $D=4$ is not renormalizable.  It contains couplings with negative mass dimension carrying the scale $M_P$.  Thus it must be treated an effective field theory with cutoff at most $M_P$ \cite{Donoghue:1995cz}.  This can also be seen from scattering amplitudes; by dimensional analysis the $2\rightarrow 2$ graviton scattering amplitude at energy $E$ goes like $E^2\over M_P^2$, which becomes order one and violates unitarity at an energy $E\sim M_P$.  

Since we have an effective theory, we expect quantum mechanically the presence of a plethora of other operators in the effective action, suppressed by appropriate powers of $M_P$ and order one coefficients.  Higher derivatives term, those beyond two derivatives, will be associated with higher order effects in powers of some energy scale over the cutoff.  By gauge invariance, all operators with two derivatives sum up to $\sqrt{-g}R$ and correct the Planck mass, naively by order one.  However, we can generate operators with higher numbers of derivatives\footnote{Note that quantum corrections will also generate the terms with no derivatives, the cosmological constant.  We can consistently declare that these are zero at the expense of a fine tuning.  This is the usual cosmological constant problem \cite{Weinberg:1988cp}.}, suppressed by appropriate powers of the Planck scale, for example,
\be {1\over M_P^2}\partial^4 \hat h^2,\ \ \ {1\over M_P^3}\partial^4 \hat h^3,\ \ \ {1\over M_P^5}\partial^6 \hat h^3,\ \,\cdots\ee
By gauge invariance, they must sum up to higher order curvature scalars, multiplied by appropriate powers of $M_P$, for instance, schematically,
\bea 
\sqrt{-g}R^2&\sim& {1\over M_P^2}\partial^4 \hat h^2+{1\over M_P^3}\partial^4 \hat h^3+{1\over M_P^4}\partial^4 \hat h^4+\cdots,\\
{1\over M_P^2}\sqrt{-g}R\nabla^2 R&\sim& {1\over M_P^4}\partial^6 \hat h^2+{1\over M_P^5}\partial^6 \hat h^3+\cdots,\\
{1\over M_P^2}\sqrt{-g}R^3&\sim& {1\over M_P^5}\partial^6 \hat h^3+{1\over M_P^6}\partial^6 \hat h^4+\cdots
\eea
 
These corrections include terms which are second order in the fields, but higher order in the derivatives, which naively lead to new degrees of freedom, some of which may be ghosts or tachyons.  One might worry why these terms are generated here, however the masses of these ghosts and tachyons is always near or above the cutoff $M_P$, so they need not be considered part of the effective theory, since the unknown UV completion may cure them.  In line with this logic, they must not be re-summed into the propagator (this would be stepping outside the $M_P$ expansion), but rather treated as vertices in the effective theory.  

The important observation is that all these higher terms are suppressed relative to any term in the Einstein-Hilbert part by powers of the derivatives
\be {\partial\over M_P}\sim {1\over M_P r}.\ee
Thus, at distances $r\gg {1\over M_P}$, more than a Planck length from the central singularity of our spherical solution, quantum effects are negligible.  Only when approaching within a Planck length of the center does quantum gravity become important.  The regimes of GR are shown in Figure \ref{GRregimes}.

An important fact about GR is that there exists this parametrically large middle regime in which the theory becomes non-linear and yet quantum effects are still small.  This is the region inside the horizon $r=r_S$ but farther than a Planck length from the singularity.   In this region, we can re-sum the linear expansion by solving the full classical Einstein equations, ignoring the higher derivative quantum corrections, and trust the results.  This is the reason why we know what will happen inside a black hole, but we do not know what will happen near the singularity.  As we will see, this crucial separation of scales, in which the scale of non-linearity is well separated from the quantum scale, does not always occur in massive gravity.  It only occurs if the parameters of the interactions are tuned in a certain way.  

\begin{figure}[h!]
\begin{center}
\epsfig{file=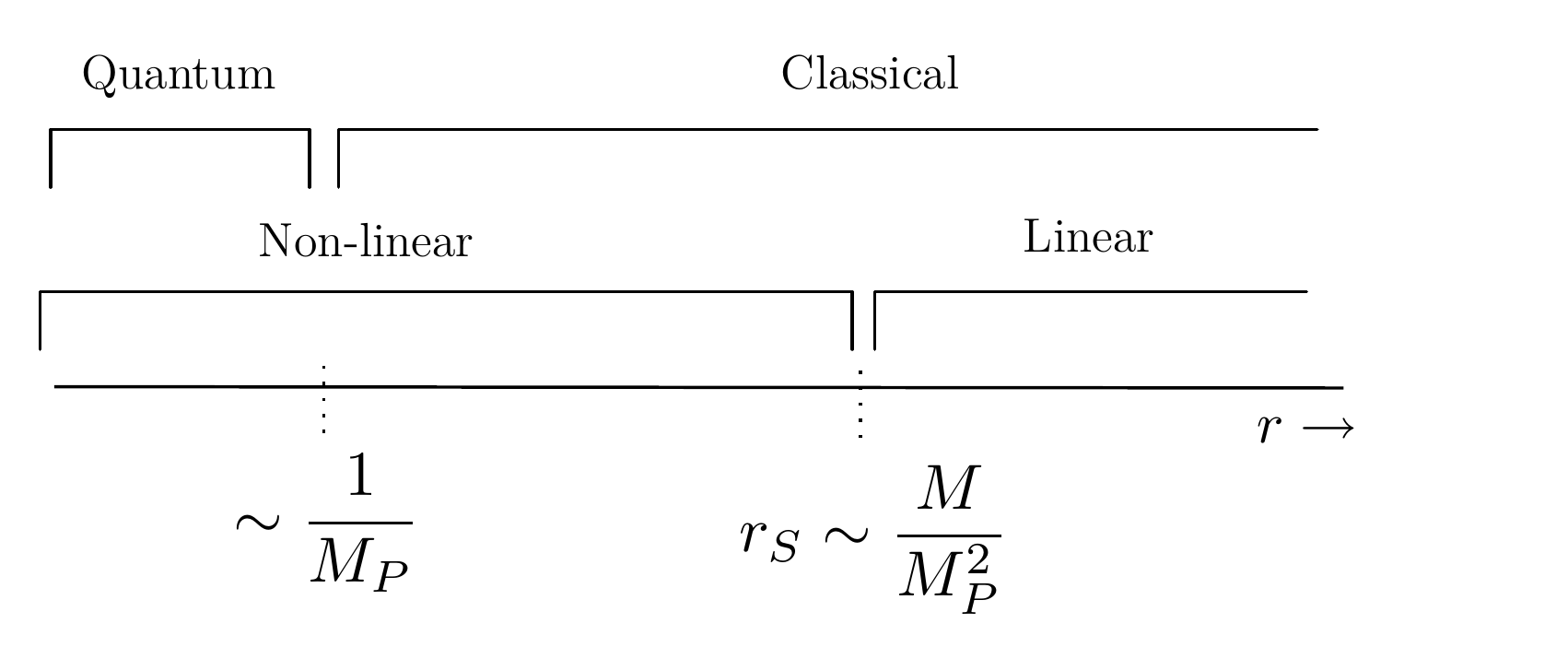,height=1.7in,width=4.0in}
\caption{\small  Regimes for GR.}
\label{GRregimes}
\end{center}
\end{figure}

\subsection{\label{massivegrsec}Massive general relativity}

We now turn to non-linearities in massive gravity.
What we want in a full theory of massive gravity is some non-linear theory whose linear expansion around some background is the massive Fierz-Pauli theory \ref{massivefreeaction}.  Unlike in GR, where the gauge invariance constrains the full theory to be Einstein gravity, the extension for massive gravity is not unique.  In fact, there is no obvious symmetry to preserve, so any interaction terms whatsoever are allowed.

The first extension we might consider would be to deform GR by simply adding the Fierz-Pauli term to the full non-linear GR action, that is, choosing the only non-linear interactions to be those of GR,
\be \label{massiveintfirst} S={1\over 2\kappa^2}\int d^D x\ \left[(\sqrt{-g}R)-\sqrt{-g^0}\frac{1}{4}m^2g^{(0)\mu\alpha}g^{(0)\nu\beta}\left(h_{\mu\nu}h_{\alpha\beta}-h_{\mu\alpha}h_{\nu\beta}\right)\right] .\ee
Here there are several subtleties.  Unlike GR, the lagrangian now explicitly depends on a fixed metric $g^{(0)}_{\mu\nu}$, which we will call the absolute metric, on which the linear massive graviton propagates.  We have $h_{\mu\nu}=g_{\mu\nu}-g^{(0)}_{\mu\nu}$ as before.  The mass term is unchanged from its linear version, so the indices on $h_{\mu\nu}$ are raised and traced with the absolute metric.  The presence of this absolute metric in the mass term breaks the diffeomorphism invariance of the Einstein-Hilbert term.  Note that there is no way to introduce a mass term using only the full metric $g_{\mu\nu}$, since tracing it with itself just gives a constant, so the non-dynamical absolute metric is required to create the traces and contractions.  

Varying with respect to $g_{\mu\nu}$ we obtain the equations of motion 
\be \label{massiveGReom} \sqrt{-g}(R^{\mu\nu}-\half Rg^{\mu\nu})+ \sqrt{-g^{(0)}}{m^2\over 2}\left(g^{(0)\mu\alpha}g^{(0)\nu\beta}h_{\alpha\beta}-g^{(0)\alpha\beta}h_{\alpha\beta}g^{(0)\mu\nu}\right)=0.
\ee
Indices on $R_{\mu\nu}$ are raised with the full metric, and those on $h_{\mu\nu}$ with the absolute metric.  We see that if the absolute metric $g^{(0)}_{\mu\nu}$ satisfies the Einstein equations (\ref{eisteinequations}), then $g_{\mu\nu}=g^{(0)}_{\mu\nu}$,  i.e. $h_{\mu\nu}=0$,  is a solution.  When dealing with massive gravity and more complicated non-linear solutions thereof, there can be at times two different background structures.  On the one hand, there is the absolute metric, the structure which breaks explicitly the diffeomorphism invariance.  On the other hand, there is the background metric, which is a solution to the full non-linear equations, about which we may expand the action.  Often, the solution metric we are expanding around will be the same as the absolute metric, but if we were expanding around a different solution, say a black hole, there would be two distinct structures, the black hole solution metric and the absolute metric.  

If we add matter to the theory and agree to use only minimal coupling to the metric $g_{\mu\nu}$, then the absolute metric does not directly influence the matter.  It is the geodesics and lengths as measured by the solution metric that we care about.  Unlike in GR, if we have a solution metric, we cannot perform a diffeomorphism on it to obtain a second solution to the same theory.  What we obtain instead is a solution to a different massive gravity theory, one whose absolute metric is related to the original absolute metric by the same diffeomorphism.  

Going to more general interactions beyond (\ref{massiveintfirst}), our main interest will be in adding interactions terms with no derivatives, since these are most important at low energies.  The most general such potential which reduces to Fierz-Pauli at quadratic order involves adding terms cubic and higher in $h_{\mu\nu}$ in all possible ways
\begin{equation}\label{potentialflat}
S={1\over 2\kappa^2}\int d^D x\ \left[(\sqrt{-g}R)-\sqrt{-g^0}\frac{1}{4}m^2U(g^{(0)},h)\right],
\end{equation}
where the interaction potential $U$ is the most general one that reduces to Fierz-Pauli at linear order,
\be U(g^{(0)},h)=U_2(g^{(0)},h)+U_3(g^{(0)},h)+U_4(g^{(0)},h)+U_5(g^{(0)},h)+\cdots,\ee
\bea
U_2(g^{(0)},h)&=& \lb h^2\rb-\lb h\rb^2, \\ 
U_3(g^{(0)},h)&=&+C_1\lb h^3\rb+C_2 \lb h^2\rb \lb h\rb +C_3\lb h \rb ^3, \\ 
U_4(g^{(0)},h)&=&+ D_1\lb h^4\rb+D_2 \lb h^3\rb \lb h\rb +D_3\lb h^2 \rb^2 +D_4\lb h^2\rb\lb h\rb^2+D_5\lb h\rb^4,\\ 
U_5(g^{(0)},h)&=&+ F_1\lb h^5\rb+F_2 \lb h^4\rb \lb h\rb +F_3\lb h^3 \rb \lb h\rb^2 +F_4\lb h^3\rb\lb h^2\rb+F_5\lb h^2\rb^2\lb h\rb\nn \\ &&+F_6\lb h^2\rb\lb h\rb^3+F_7\lb h\rb^5,\label{U5equation}\\
&\vdots& \nn
\eea
The square bracket indicates a trace, with indices raised with $g^{(0),\mu\nu}$,  i.e. $[h] = g^{(0)\mu\nu} h_{\mu\nu}$, $[h^2] = g^{(0)\mu\alpha} h_{\alpha\beta}g^{(0)\beta\nu}h_{\nu\mu}$, etc.  The coefficients $C_1,C_2,$ etc. are generic coefficients.  Note that the coefficients in $U_n(g^{(0)},h)$ for $n>D$ are redundant by one, because there is a combination of the various contractions, the characteristic polynomial ${\cal L}_n^{\rm TD}(h)$ (see Appendix \ref{totalDappendix}), which vanishes identically.  Thus one of the coefficients  in $U_n(g^{(0)},h)$ for $n>D$ (or any one linear combination) can be set to zero.

If we like, we can re-organize the terms in the potential by raising and lowering with the full metric $g^{\mu\nu}$ rather than the absolute metric $g^{(0)\mu\nu}$,
\begin{equation}\label{potentialfull}
S={1\over 2\kappa^2}\int d^D x\ \left[(\sqrt{-g}R)-\sqrt{-g}\frac{1}{4}m^2V(g,h)\right],
\end{equation}
where 
\be V(g,h)=V_2(g,h)+V_3(g,h)+V_4(g,h)+V_5(g,h)+\cdots,\ee
\bea
V_2(g,h)&=& \la h^2\ra-\la h\ra^2, \\ 
V_3(g,h)&=&+c_1\la h^3\ra+c_2 \la h^2\ra \la h\ra +c_3\la h \ra ^3, \\ 
V_4(g,h)&=&+ d_1\la h^4\ra+d_2 \la h^3\ra \la h\ra +d_3\la h^2 \ra^2  +d_4\la h^2\ra\la h\ra^2+d_5\la h\ra^4,\\ 
V_5(g,h)&=&+ f_1\la h^5\ra+f_2 \la h^4\ra \la h\ra +f_3\la h^3 \ra \la h\ra^2 +f_4\la h^3\ra\la h^2\ra+f_5\la h^2\ra^2\la h\ra\nn \\ &&+f_6\la h^2\ra\la h\ra^3+f_7\la h\ra^5, \\
&\vdots& \nn
\eea
where the angled brackets are traces with the indices raised with respect to $g^{\mu\nu}$.
It does not matter whether we use potential (\ref{potentialflat}) with indices raised by $g^{(0)\mu\nu}$, or the potential (\ref{potentialfull}) with indices raised by $g^{\mu\nu}$.  The two carry the same information and we can easily relate the coefficients of the two by expanding the inverse full metric and the full determinant in powers of $h_{\mu\nu}$ raised with the absolute metric,
\bea g^{\mu\nu}&=&g^{(0)\mu\nu}-h^{\mu\nu}+h^{\mu\lambda}h_{\lambda}^{\ \nu}-h^{\mu\lambda}h_{\lambda}^{\ \sigma}h_\sigma^{\ \nu}+\cdots, \\
\sqrt{-g}&=&\sqrt{-g^{(0)}}\left[1+\half h-{1\over 4}\left( h^{\mu\nu}h_{\mu \nu}-\half h^2\right)+\cdots\right].
\eea
The following is useful for this purpose,
\be \la h^n\ra=\sum_{l=0}^\infty (-1)^l \left(\begin{array}{c}l+n-1 \\ l\end{array}\right) [h^{l+n}].\ee

While the zero derivative interaction terms we have written in (\ref{potentialflat}) are general, the two derivative terms are not, since we have demanded they sum up to Einstein-Hilbert.  The potential has broken the diffeomorphism invariance, so there is no symmetry reason for the two derivative interaction terms to take the Einstein-Hilbert form.  We could deviate from it if we like, but we will see later that there are good reasons why it is better not to.  We may also conceivably add general interactions with more than two derivatives, but we will omit these for the same reasons we omit them in GR, because they are associated with higher order effective field theory effects which we hope will be small in suitable regimes. 

\subsection{\label{sphersolmassection}Spherical solutions and the Vainshtein radius}

We will now look at static spherical solutions, doing for massive gravity what we did for GR in Section \ref{GRsection}.
We specialize to four dimensions, and for definiteness we pick the action (\ref{massiveintfirst}) with the minimal mass term.   We attempt to find spherically symmetric solutions to the equations of motion \ref{massiveGReom}, in the case where the absolute metric is flat Minkowski in spherical coordinates,
\[ g^{(0)}_{\mu\nu}dx^\mu dx^\nu=-dt^2+dr^2+r^2 d\Omega^2.\] 

We consider a spherically symmetric static ansatz for the dynamical metric\footnote{In general, when there are two metrics staticity and spherical symmetry are not enough to put both in diagonal form.  An $r$ dependent off diagonal $drdt$ term can remain in one of them.  We will not seek such off-diagonal metrics, and will limit ourselves to the diagonal ansatz.} 
\be g_{\mu\nu}dx^\mu dx^\nu= -B(r)dt^2+C(r)dr^2+A(r)r^2d\Omega^2.\ee
Plugging this ansatz into the equations of motion, we get the following from the $tt$ equation,  $rr$ equation and $\theta\theta$ equation (which is the same as the $\phi\phi$ equation by spherical symmetry) respectively,
\bea &&4 B C^2 m^2 r^2 A^3+\left(2 B (C-3) C^2 m^2 r^2-4 \sqrt{A^2 B C} \left(C-r C'\right)\right)
   A^2\nn \\ 
  &&+2 \sqrt{A^2 B C} \left(2 C^2-2 r \left(3 A'+r A''\right) C+r^2 A' C'\right) A+C \sqrt{A^2 B
   C} r^2 \left(A'\right)^2=0,\nn\\ \\ 
   && \frac{4 \left(B+r B'\right) A^2+\left(2 r^2 A' B'-4 B \left(C-r A'\right)\right)
   A+B r^2 \left(A'\right)^2}{A^2 B C^2 r^2}-\frac{2 (2 A+B-3) m^2}{\sqrt{A^2 B C}}=0,\nn \\ \\
   &&-2 B^2 C^2 m^2 r A^4-2 B^2 C^2 (B+C-3) m^2 r A^3\nn \\ 
   &&-\sqrt{A^2 B C} \left(2 C' B^2+\left(r B' C'-2 C
   \left(B'+r B''\right)\right) B+C r \left(B'\right)^2\right) A^2 \nn \\
   &&+B \sqrt{A^2 B C} \left(C r A'
   B'+B \left(4 C A'-r C' A'+2 C r A''\right)\right) A-B^2 C \sqrt{A^2 B C} r \left(A'\right)^2=0.\nn \\
   \eea 
 In the massless case, $A(r)$ could be removed by a coordinate gauge transformation, and the last equation was redundant -- it was a consequence of the first two.  With non-zero $m$, there is no diffeomorphism invariance, so no such coordinate change can be made, and the last equation is independent.  
 
 As we did in Section \ref{GRsection} for GR, we expand these equations around the flat space solution
 \be B_0(r)=1,\ \ \ C_0(r)=1,\ \ \ A_0(r)=1. \ee
We introduce the expansion
\bea B(r)&=&B_0(r)+\epsilon B_1(r)+\epsilon^2 B_2(r)+\cdots ,\\ \nn
C(r)&=&C_0(r)+\epsilon C_1(r)+\epsilon^2 C_2(r)+\cdots ,\\ \nn
A(r)&=&A_0(r)+\epsilon A_1(r)+\epsilon^2 A_2(r)+\cdots.
\eea
Plugging into the equations of motion and collecting like powers of $\epsilon$,  the ${\cal O}(0)$ part gives $0=0$ because $B_0,C_0,A_0$ are solutions to the full non-linear equations.  At each higher order in epsilon we will obtain a linear equation that lets us solve for the next term.  At ${\cal O}(\epsilon)$ we obtain
\bea &&2 \left(m^2 r^2-1\right) A_1+\left(m^2 r^2+2\right) C_1+2 r \left(-3 A_1'+C_1'-r A_1''\right)=0,\\
&& -\frac{1}{2} B_1 m^2+\left(\frac{1}{r^2}-m^2\right) A_1+\frac{r \left(A_1'+B_1'\right)-C_1}{r^2}=0,\\
&&  r A_1 m^2+r B_1 m^2+r C_1 m^2-2 A_1'-B_1'+C_1'-r A_1''-r B_1''=0.
\eea
One way to solve these equations is as follows.  Algebraically solve them simultaneously for $A_1,A_1',A_1''$ in terms of $B_1$'s and $C_1$'s and their derivatives.  Then write the equations ${d\over dr}A_1(B,C)=A_1'(B,C)$ and ${d\over dr}A'(B,C)=A''(B,C)$.  
Solve these two equations for $C_1$ and $C_1'$ in terms of $B_1$'s its derivatives.  Then write ${d\over dr}C_1(B)=C_1'(B)$. 
What is left is 
\be\label{Bequ} -3 r B_1 m^2+6 B_1'+3 r B_1''=0.\ee
There are two integration constants in the solution to (\ref{Bequ}), one is left arbitrary and the other must be sent to zero to prevent the solutions from blowing up at infinity.  We then recursively determine $C_1$ and $A_1$.  Thus the whole solution is determined by two pieces of initial data\footnote{Naively, it is a second order equation in $A_1$ and $B_1$, first order in $C_1$ and we might think this requires $5$ initial conditions, but in fact it is a degenerate system, and there are second class constraints bringing the required boundary data to 2.}.

The solution is 
\bea B_1(r)&=&-{8GM\over 3}{e^{-mr}\over r},\\
C_1(r)&=&-{8GM\over 3}{e^{-mr}\over r}{1+mr\over m^2r^2},\\
A_1(r)&=&{4GM\over 3}{e^{-mr}\over r}{1+mr+m^2r^2\over m^2r^2},
\eea
where we have chosen the integration constant so that we agree with the solution (\ref{masssphersol}) obtained from the Green's function.  

We can now proceed to ${\cal O}(\epsilon^2)$.  Going through the same procedure, we find for the solution, when $mr\ll1$, 
\bea B(r)-1&=&-{8\over 3}{GM\over r}\left(1-{1\over 6}\frac{GM}{ m^4  r^5}+\cdots\right),\\
C(r)-1&=&-{8\over 3}{GM\over m^2 r^3}\left(1-{14}\frac{GM}{ m^4  r^5}+\cdots\right),\\
A(r)-1&=&{4\over 3}{GM\over4\pi m^2 r^3}\left(1-{4}\frac{GM}{ m^4  r^5}+\cdots\right).
\eea
The dots represent higher powers in the non-linearity parameter $\epsilon$.  We see that the non-linearity expansion is an expansion in the parameter $r_V/r$, where 
\be r_V\equiv \left(\frac{GM}{ m^4 }\right)^{1/5},\ee
is known as the \textit{Vainshtein radius}.  As the mass $m$ approaches 0, $r_V$ grows, and hence the radius beyond which the solution can be trusted gets pushed out to infinity.  As argued by Vainshtein \cite{Vainshtein:1972sx}, this perturbation expansion breaks down, and says nothing about the true non-linear behavior of massive gravity in the massless limit.  Thus there is reason to hope that the vDVZ discontinuity is merely an artifact of linear perturbation theory, and that the true non-linear solutions show a smooth limit \cite{Vainshtein:1972sx,Gruzinov:2001hp,Porrati:2002cp,Deffayet:2001uk}.  

One might hope that a smooth limit could be seen by setting up an alternative expansion in the mass $m^2$.  We take a solution to the massless equations, the ordinary Schwarzschild solution, with metric coefficients $B_0,C_0,A_0$, and then plug an expansion
\bea B(r)&=&B_0(r)+m^2 B_1(r)+m^4 B_2(r)+\cdots ,\nn \\ \nn
C(r)&=&C_0(r)+m^2 C_1(r)+m^4 C_2(r)+\cdots ,\\
A(r)&=&A_0(r)+m^2 A_1(r)+m^4 A_2(r)+\cdots,
\eea
into the equations of motion.  Collecting powers of $m$ yields a new perturbation equation at each order, but in this case the equations are generally non-linear.  Even the equation we obtain at ${\cal O}(m^2)$ for the first correction to Schwarzschild is non-linear, so working with this expansion is much more difficult than working with the linearized expansion.  

The linearity expansion is valid is the region $r\gg r_V$.  If general relativity is restored at distances near the source, the mass expansion should be valid in the opposite regime $r\ll r_V$, and the full solutions should interpolate between the two expansions.  There have been several extensive numerical studies of the full non-linear solutions.  At first, troubles were encountered trying to find a complete and satisfactory solution that interpolates between the two regimes \cite{Damour:2002gp}.  Later, the problem was revisited with more sophisticated methods, both in the decoupling limit \cite{Babichev:2009us}, and more extensively in the full theory \cite{Deffayet:2008zz,Babichev:2009jt,Babichev:2010jd}, with the final result being that the non-linearities can in fact work to restore continuity with GR.  We will see later the mechanism by which this occurs.  Other solutions, including some analytic solutions in various cases, are claimed in \cite{Berezhiani:2008nr,Koyama:2011xz,Nieuwenhuizen:2011sq,Koyama:2011yg,Comelli:2011yd,Gruzinov:2011mm}.

\subsection{\label{hamsection}Non-linear hamiltonian and the Boulware-Deser mode}

We now go on to study the hamiltonian of the non-linear massive gravity action (\ref{massiveintfirst}) with flat absolute metric $\eta^{\mu\nu}$,
\be \label{massiveintfirst3} S={1\over 2\kappa^2}\int d^D x\ \left[(\sqrt{-g}R)-\frac{1}{4}m^2\eta^{\mu\alpha}\eta^{\nu\beta}\left(h_{\mu\nu}h_{\alpha\beta}-h_{\mu\alpha}h_{\nu\beta}\right)\right] .\ee
We saw in Section \ref{canonicalanalysis} that the free theory carries five degrees of freedom in $D=4$, due to the fact that the time components $h_{00}$ appeared as a Lagrange multiplier in the action.  We will see that this no longer remains true once the non-linearities of (\ref{massiveintfirst3}) are taken into account, so there is now an extra degree of freedom. 

A particularly nice way to study gravity hamiltonians is through the ADM formalism \cite{Arnowitt:1960es,Arnowitt:1962hi}.  A spacelike slicing of spacetime by hypersurfaces $\Sigma_t$ is chosen, and we change variables from components of the metric $g_{\mu\nu}$ to the spatial metric $g_{ij}$, the lapse $N_i$ and the shift $N$, according to
\bea g_{00}&=&- N^2+g^{ij}N_i N_{j} , \\
g_{0i}&=&N_{i},\\ 
g_{ij}&=&g_{ij}.
\eea
Here $i,j,\ldots$ are spatial indices, and $g^{ij}$ is the inverse of the spatial metric $g_{ij}$ (not the $ij$ components of inverse metric $g^{\mu\nu}$).  

The Einstein-Hilbert part of the action in these variables reads (see \cite{Poisson,Dyer:2008hb} for detailed derivations and formulae)
\be {1\over 2\kappa^2}\int d^D x \sqrt{g} N\left[ ^{(d)}R-K^{2}+K^{ij}K_{ij}\right],\ee
where $^{(d)}R$ is the curvature of the spatial metric $g_{ij}$.  The quantity $K_{ij}$ is the extrinsic curvature of the spatial hypersurfaces, defined as
\be \label{kdefinition} K_{ij}={1\over 2N}\left(\dot g_{ij}-\nabla_i N_j-\nabla_j N_i\right),\ee  
where dot means time derivative, and the covariant derivatives are with respect to the spatial metric $g_{ij}$.  We then Legendre transform the spatial variables $g_{ij}$, defining the canonical momenta
\be \label{momentak} p^{ij}=\frac{\delta{L}}{\delta \dot{g}_{ij}} ={1\over 2\kappa^2}\sqrt{g}\left(K^{ij}-Kg^{ij}\right),
\ee
and writing the action in hamiltonian form
\be 2\kappa^2L=\left(\int_{\Sigma_{t}}d^{d}x \ p^{ij}\dot{g}_{ij}\right) -{H} ,\ee
where the hamiltonian $H$ is defined by
\be H=\left(\int_{\Sigma_{t}}d^{d}x \ p^{ab}\dot{g}_{ab}\right) -{L}= \int_{\Sigma_{t}}d^{d}x\ N{\cal C}+N_i{\cal C}^i,\ee
and the quantities ${\cal C}$ and ${\cal C}_i$ are
\be {\cal C}=\sqrt{g}\left[^{(d)}R+K^{2}-K^{ij}K_{ij}\right],\ \ \ {\cal C}^i=2\sqrt{g}\nabla_j\left(K^{ij}-Kh^{ij}\right),\ee
and here $K_{ij}$ should be thought of as a function of $p^{ij}$ and $g_{ij}$, obtained by inverting (\ref{momentak}) for $\dot g_{ij}$ and plugging into (\ref{kdefinition}),
\be \label{kprelation} K_{ij}={2\kappa^2\over \sqrt{g}}\left(p_{ij}-{1\over D-2}p h_{ij}\right).\ee
All traces and index manipulations are performed with $g_{ij}$ and its inverse.  

For $m=0$, the action is pure constraint, and the hamiltonian vanishes, a characteristic of diffeomorphism invariance.  The shift $N$ and lapse $N_i$ appear as Lagrange multipliers, enforcing the hamiltonian constraint ${\cal C}=0$ and momentum constraints ${\cal C}_i=0$.  It can be checked that these are first class constraints, generating the $D$ diffeomorphism symmetries of the action.  In $D=4$, we have 12 phase space metric components, minus 4 constraints, minus 4 gauge symmetries, leaves 4 phase space degrees of freedom, the same counting as in the linear theory.  The non-linear theory contains the same number of degrees of freedom as the linearized theory.

Now looking at the mass term, in ADM variables we have
\bea \eta^{\mu\alpha}\eta^{\mu\beta}\left(h_{\mu\nu}h_{\alpha\beta}-h_{\mu\alpha}h_{\mu\beta}\right)=&&\delta^{ik}\delta^{jl}\left(h_{ij}h_{kl}-h_{ik}h_{jl}\right)+2\delta^{ij}h_{ij} \nn \\
&&-2N^2\delta^{ij}h_{ij}+2N_i\left(g^{ij}-\delta^{ij}\right)N_i,
\eea
where $h_{ij}\equiv g_{ij}-\delta_{ij}$.
The action becomes
\bea \label{massiveADMaction} S={1\over 2\kappa^2}\int d^D x&& p^{ab}\dot{g}_{ab}-N{\cal C}-N_i{\cal C}^i \\ &&-{m^2\over 4}\left[\delta^{ik}\delta^{jl}\left(h_{ij}h_{kl}-h_{ik}h_{jl}\right)+2\delta^{ij}h_{ij}-2N^2\delta^{ij}h_{ij}+2N_i\left(g^{ij}-\delta^{ij}\right)N_i\right] . \nn
\eea
  
In the $m\not=0$ case, the Fierz-Pauli term brings in contributions to the action that are quadratic in the lapse and shift (but still free of time derivatives).  Thus the lapse and shift no longer serve as Lagrange multipliers, but rather as auxiliary fields, because their equations of motion can be algebraically solved to determine their values,
\be N={{\cal C}\over m^2 \delta^{ij}h_{ij}},\ \ \ \ N_i={1\over m^2}\left(g^{ij}-\delta^{ij}\right)^{-1}{\cal C}^j.\ee

When these values are plugged back into (\ref{massiveADMaction}), we have an action with no constraints or gauge symmetries at all, so all the phase space degrees of freedom are active.  The resulting hamiltonian is
\be \label{massivehamiltoniannonlin} H={1\over 2\kappa^2}\int d^d x\ {1\over 2m^2}{{\cal C}^2\over \delta^{ij}h_{ij}}+{1\over 2m^2}{\cal C}^i\left(g^{ij}-\delta^{ij}\right)^{-1}{\cal C}^j+{m^2\over 4}\left[\delta^{ik}\delta^{jl}\left(h_{ij}h_{kl}-h_{ik}h_{jl}\right)+2\delta^{ij}h_{ij}\right],\ee
which is non-vanishing, unlike in GR.
In 4 dimensions, we thus have 12 phase space degrees of freedom, or 6 real degrees of freedom.  The linearized theory had only five degrees of freedom, and we have here a case where the non-linear theory contains more degrees of freedom than the linear theory.  It should not necessarily be surprising that this can happen, because there is no reason non-linearities cannot change the constraint structure of a theory, or that kinetic terms cannot appear at higher order.

As was argued in \cite{Boulware:1973my}, the hamiltonian (\ref{massivehamiltoniannonlin}) is not bounded, and since the system is non-linear, it is not surprising that it has instabilities \cite{Gabadadze:2003jq}.  The nature of the instability, i.e. whether it is a ghost of a tachyon, what backgrounds it appears around, and its severity, is hard to see in the hamiltonian formalism.  But in Section \ref{ghostsection} we will see that this instability is a ghost, a scalar with a negative kinetic term, and that its mass around a given background can be determined.  It turns out that around flat space, the ghost degree of freedom is not excited because its mass is infinite, but around non-trivial backgrounds its mass becomes finite.  This ghostly extra degree of freedom is referred to as the \textit{Boulware-Deser ghost} \cite{Boulware:1973my}.

There is still the possibility that adding higher order interaction terms such as $h^3$ terms and higher, can remove the ghostly sixth degree of freedom.  Boulware and Deser analyzed a large class of various mass terms, showing that the sixth degree of freedom remained  \cite{Boulware:1973my}, but they did not consider the most general possible potential.  This was addressed in \cite{Creminelli:2005qk}, where the analysis was done perturbatively in powers of $h$.  The lapse is expanded around its flat space values, $N=1+\delta N$.  In this case, $\delta N$ plays the role of the Lagrange multiplier, and it is shown that at fourth order, interaction terms involving higher powers of $\delta N$ cannot be removed.  It is concluded in \cite{Creminelli:2005qk} that the Boulware-Deser ghost is unavoidable, but this conclusion is too quick.  It may be possible that there are field redefinitions under which the lapse is made to appear linearly.  Alternatively, it may be possible that after one solves for the shift using its equation of motion, then replaces into the action, the resulting action is linear in the lapse, even though it contained higher powers of the lapse before integrating out the shift.  It is also possible that the lapse appears linearly in the full non-linear action, even though at any finite order the action contains higher powers of the lapse. (For discussions and examples of these points, see \cite{deRham:2010ik,deRham:2010kj}.)  

As it turns out, it is in fact possible to add appropriate interactions that eliminate the ghost \cite{Hassan:2011hr}.  In $D$ dimensions, there is a $D-2$ parameter family of such interactions.  We will study these in Section \ref{lambda3section}, where we will see that they also have the effect of raising the maximum energy cutoff at which massive gravity is valid as an effective field theory.\footnote{Note that merely finding a ghost free interacting Lorentz invariant massive gravity theory is not hard -- take for instance  $U(\eta,h)=-2\left[\det\left(\delta_\mu^{\ \nu}+ h_\mu^{\ \nu}\right)-h\right]$ in (\ref{potentialflat}), while letting the kinetic interactions be those of the linear graviton only.  A hamiltonian analysis just like that of Section (\ref{canonicalanalysis}) shows that $h_{00}$ and $h_{0i}$ both remain Lagrange multipliers.  The problem is that this theory does not go to GR in the $m\rightarrow 0$ limit, it goes to massless gravity.  The real challenge is to construct a ghost free Lorentz invariant massive gravity that reduces to GR.}

\section{\label{nonlinearstukelbergtrick}The non-linear St\"ukelberg formalism}

In this section we will extend the St\"ukelberg trick to full non-linear order.   This will be a powerful tool with which to elucidate the non-linear dynamics of massive gravity.  It will allow us to trace the breakdown in the linear expansion to strong coupling of the longitudinal mode.  It will also tell us about quantum corrections, the scale of the effective field theory and where it breaks down, as well as the nature of the Boulware-Deser ghost and whether it lies within the effective theory or can be consistently ignored.

\subsection{\label{spinonesection}Yang-Mills example}

We will first warm up with the spin 1 case, and we will set $D=4$.   The unique theory of interacting massless spin 1 particles is Yang-Mills theory \cite{Henneaux:1997bm}.  Analogously to what we've done with gravity in Section \ref{massivegrsec}, consider a non-abelian $SU(N)$ gauge theory with gauge coupling $g$, and add a non-gauge invariant mass term with mass $m$ for the gauge bosons, while leaving the kinetic structure unchanged from the massless case,
\be\label{massiveSUNunitary} S=\int d^4x\ {1\over 2g^2} \Tr \left(F_{\mu\nu}F^{\mu\nu}\right)+{m^2\over g^2}\Tr \left(A_{\mu}A^\mu\right).\ee
The gauge fields are $A_{\mu}=-igA_\mu^a T_a$, taking values in a Lie algebra with generators $T_a$, with adjoint index $a=1,\ldots, N^2-1$.
 The generators satisfy the usual Lie algebra commutation and orthogonality relations $\left[T_a,T_b\right]=if_{ab}^{\ \ c}T_c,\ \ \  Tr(T_{a}T_{b})=\half \delta_{ab}.$
 The field strength is $F_{\mu\nu}\equiv\partial_\mu A_\nu-\partial_\nu
A_\mu+\left[A_\mu,A_\nu\right]=-igF_{\mu\nu}^a T_a.$
The theory (\ref{massiveSUNunitary}) naively appears renormalizable, since there are no interaction terms with mass dimension greater than 4.  But the propagators are those of a massive vector which do not go like $\sim 1/ p^2$, so naive power counting does not apply.
 
  In the absence of the mass term, the action is invariant under the gauge transformations
   \be \label{sungauge} A_{\mu}\rightarrow RA_\mu R^\dag+R \partial_\mu R^\dag,\ee
where $R=e^{-i\alpha^{a}T_{a}}\in SU(N),$ and $\alpha^{a}(x)$ are gauge parameters.  
This reads infinitesimally $\delta A_\mu^a=-\frac{1}{g}\partial_\mu\alpha^a-f_{bc}^{\ \ a}A_\mu^b\alpha^c.$
The field strength transforms covariantly $F_{\mu\nu}\rightarrow RF_{\mu\nu} R^\dag$,
which reads infinitesimally $\delta F_{\mu\nu}^a=f_{bc}^{\ \ a}\alpha^b F_{\mu\nu}^c$.

The mass term breaks the gauge symmetry (\ref{sungauge}) (though it remains invariant under the global version), so we will restore it by introducing St\"ukelberg fields.
We pattern the introduction of the fields after the gauge symmetry we wish to restore, so we make the replacement
\be\label{nonlinvectstuk} A_\mu\rightarrow UA_\mu U^\dag+U\partial_\mu U^\dag,\ee
where 
\be U=e^{-i\pi^{a}T_{a}}\in SU(N),\ee
and the $\pi^a(x)$ are scalar Goldstone fields.  The action now becomes gauge invariant under right gauge transformations\footnote{Making the replacement $A_\mu\rightarrow U^\dag A_\mu U-U^\dag\partial_\mu U$ would have led to left gauge transformations.},
\be A_\mu \rightarrow RA_\mu R^\dag+R \partial_\mu R^\dag,\ \ \ U\rightarrow UR^\dag.\ee
The gauge kinetic term is invariant under this replacement, since it is gauge invariant, so the Goldstones appear only through the mass term
\be {m^2\over g^2}\Tr \left(A_{\mu}A^\mu\right)\rightarrow -{m^2\over g^2}\Tr \left(D_{\mu}U^\dag D^\mu U\right),\label{sigmamodmass} \ee
where  $D_\mu U\equiv \partial_\mu U-UA_\mu$ is a covariant derivative, which transforms covariantly under right gauge transformations\footnote{The sigma model mass term (\ref{sigmamodmass}) is invariant under $SU(N)_L\times SU(N)_R$ global symmetry, $U\rightarrow LUR^\dag$, of which the $SU(N)_R$ part is gauged.  The $SU(N)$ subgroup $L=R$ is realized linearly, and the rest is realized non-linearly.},
$D_\mu U\rightarrow (D_\mu U)R^\dag.$
We can go to the unitary gauge $U=1$, and recover the massive vector action we started with, so the new action is equivalent.

Expanding the terms in (\ref{sigmamodmass}), we find kinetic terms for the vectors and scalars that require them to be canonically normalized as follows,
\be\label{YMcannorm} A\sim g\hat A,\ \ \ \pi\sim {g\over m}\hat \pi.\ee
Note that to lowest order in the fields, the non-linear St\"ukelberg expansion (\ref{nonlinvectstuk}) is the same as the linear one of Section \ref{vectorstukelberg}.  Thus the propagators all go like $\sim 1/p^2$, ordinary power counting applies, and we can read off strong coupling scales from any non-renormalizable terms. 

For interactions, we have the usual normalizable Yang-Mills interaction terms with three and four fields, coming from the gauge kinetic term,
\be \sim g \ \partial \hat A^3,\ \ \ \sim g^2 \hat A^4.\ee
From the mass term we find the non-renormalizable terms
\be \sim \left(g\over m\right)^{n-2}\partial^2 \hat \pi^n ,\ \ \ \sim g\left(g\over m\right)^{n-2}\partial \hat A\ \hat \pi^n,\ \ \  \sim g^2\left(g\over m\right)^{n-2}\hat A^2 \hat \pi^n.\ee

 For $g<1$, the lowest energy scale suppressing the non-renormalizable terms is $\sim {m\over g}$, which comes from the terms with only $\pi$ fields.  The tree level amplitude for $\pi\pi\rightarrow\pi\pi$ scattering at energy $E$ calculated from these terms goes like ${\cal A}\sim {g^2 E^2\over m^2}$.   This amplitude becomes order one and unitarity is violated when $E$ exceeds ${m\over g}$, thus the Goldstones become strongly coupled at this energy, and this scale is the maximal cutoff for the theory,
\be \Lambda={m\over g}.\ee
Note that when $g$ is small this scale is parametrically larger than the vector masses $m$.

We can take the \textit{decoupling limit} which keeps this lowest scale fixed, while sending all the higher scales to infinity,
\be g,m\rightarrow 0,\ \ \Lambda\ {\rm fixed}.\ee
The only terms that survives this limit are the scalar self-interactions (along with the free vector fields),
\be S_{\rm decoupling}=\int d^4x\ -{\Lambda^2}\Tr \left(\partial_{\mu}U^\dag \partial^\mu U\right).\ee
 This is a limit which focuses in on the cutoff of the theory, ignoring all other scales\footnote{Note that the lowest scale is the only scale for which it is possible to take a decoupling limit.  If we try to zoom in on a higher scale in a similar fashion, the terms with lower scales will diverge.}.  For this to be a valid limit, we should be looking at energies higher than the vector masses, and the coupling should be small.  In this limit, the $\pi$'s becomes gauge invariant, but due to the way the Goldstones were introduced through traces of the combination $U\partial_\mu U^\dag$,  they retain a spontaneously broken $SU(N)_L\times SU(N)_R$ global symmetry, $U\rightarrow LUR^\dag$, of which the $SU(N)$ subgroup $L=R$ is realized linearly.

Since we have an effective theory with cutoff $\Lambda$, there will be quantum corrections of all types compatible with the spontaneously broken $SU(N)_L\times SU(N)_R$ global symmetry, suppressed by appropriate powers of the cutoff.  For example we should find the operators 
\be \sim \Tr \left(\left(\partial_\mu U^\dag \partial^\mu U\right)^2\right),\ \ \ \sim \Tr \left(\partial^2 U^\dag \partial^2 U\right),\ldots\ee
which in unitary gauge look like the non-gauge invariant terms
\be\label{aefops} \sim\Tr \(A^4\), \ \ \ \sim\Tr\((\partial A)^2\),\ldots\ee
Notice that the second operator in (\ref{aefops}) modifies the gauge kinetic term in a non-gauge invariant way, and naively leads to ghosts.  However, the mass of the ghost is $m_g^2\sim m^2/g^2=\Lambda^2$, so it is safely at the cutoff.

We might worry about the hierarchy between the small mass $m$ and the high cutoff $\Lambda\sim m/g$.  If quantum corrections to the mass were to go like $\delta m^2\sim \Lambda^2$, then the mass is pushed to the cutoff and there is a hierarchy problem which generally requires a solution in the form of fine-tuning of new physics at the cutoff.  However, this does not happen here.  There are only order one quantum corrections to mass $\delta m^2\sim m^2$, coming from the generated operator $-{\Lambda^2}\Tr \(\partial_{\mu}U^\dag \partial^\mu U\)$.  Thus the small mass is technically natural, and can be consistently incorporated in the effective theory.  

This nice state of affairs is a consequence of the fact that gauge symmetry is restored in the limit as $m\rightarrow 0$, so that mass corrections must be proportional to the mass itself.  For this to be true, it was important that there were no modifications to the kinetic structure of (\ref{massiveSUNunitary}) not proportional to $m$, even though symmetry considerations would suggest that we are free to make such modifications.  For example, suppose we try to calculate the mass correction to $A_\mu$ directly in unitary gauge by constructing Feynman diagrams with vertices read straight from (\ref{massiveSUNunitary}).  There are two interaction vertices $\sim g\partial \hat A^3$ and $\sim g^2 \hat A^4$ coming from the kinetic term.  The mass term contributes no vertices but alters the propagator so that its high energy behavior is $\sim {1\over m^2}$.  At one loop, there are two 1PI diagrams correcting the mass; one containing two cubic vertices and two propagators and one containing a single quartic vertex and a single propagator.  Cutting off the loop at the momenta $k_{\rm max}\sim \Lambda$, the former diagram gives the largest naive correction $\delta m^2\sim {g^2\over m^4}\Lambda^6\sim {\Lambda^2\over g^2}$. (The latter diagram gives the smaller correction $\delta m^2\sim {g^2\over m^2}\Lambda^4\sim {\Lambda^2}$.)  This is above the cutoff, dangerously higher than the order one correction $\delta m^2\sim m^2$ we found in the Goldstone formalism.  

What this means is that there must be a non-trivial cancellation of these leading divergences in unitary gauge, so that we recover the Goldstone result.  This cancellation happens because the kinetic interactions of (\ref{massiveSUNunitary}) are gauge invariant, implying that the dangerous $k^\mu k^\nu/m^2$ terms in the vector propagator do not contribute.  Without these terms, the propagator goes like $1/k^2$ and the estimate for the first diagram is $\delta m^2\sim g^2\Lambda^2\sim m^2$, in agreement with the Goldstone prediction (the second diagram gives the smaller correction $\delta m^2\sim g^2m^2 \log g $).  Non-parametrically altering the coefficients in the kinetic structure would spoil this cancellation and the resulting technical naturalness of the small mass (though such alterations could be done without spoiling technical naturalness if the alterations to the kinetic terms are suppressed by appropriate powers of $m$).  
For these reasons, it is desirable not to mess with the kinetic structure, and to introduce gauge symmetry breaking only through masses and potentials.  The same will be true of massive gravity.

This whole story, more than merely being a toy, can be thought of as a microcosm for the standard model.  The fundamental particles seen so far in the electroweak sector are (the Higgs hasn't been seen as of this writing) spin 1/2 fermions and massive SU(2) spin 1 gauge bosons (never mind the massless U(1) and complications of mixing).  The gauge bosons masses are of order $m\sim 10^2$ GeV, and the couplings $g\sim 10^{-1}$.  Their interactions at energies above $m$ are well described by the above sigma model, up to an energy cutoff $\Lambda\sim m/g\sim 1$ TeV.  The reason for building the Large Hadron Collider is that something must happen at the scale $\Lambda$ to UV complete the theory. 

If one demands that the UV completion be weakly coupled (as is suspected to be the case for the electroweak sector), one is led to introduce a new physical scalar, the Higgs, which unitarizes the amplitudes at energies above $\Lambda$.  This UV completion is the standard model, a spontaneously broken gauge theory, where the Higgs has a mass $\mu$, and a perturbative quartic coupling $\lambda<1$, and gets a VEV $v\sim \mu/\sqrt{\lambda}\sim \Lambda$.  The Higgs mass $\mu\sim \sqrt{\lambda}\Lambda$ sits somewhere between $m\sim g^2 v$ and the cutoff $\Lambda$.  At the scale $\mu$, the four point amplitude reaches the value ${\cal A}\sim \lambda$, the Higgs theory takes over, and the amplitudes cease growing with the energy, so that unitarity is not violated.  From this perspective, studying the addition of a mass term to a gauge theory is not just an idle theoretical exercise.  It leads one to uncover the Higgs mechanism and a new weakly coupled UV completion which is likely realized in nature.  

We will find an analogous story in the case of massive gravity.  There is an effective field theory with a cutoff parametrically higher than the graviton mass, and the hierarchy is technically natural.  The only missing part is the UV completion, which remains an unsolved problem.

\subsection{St\"ukelberg for gravity and the restoration of diffeomorphism invariance}

We will now construct the gravitational analogue of the above.  This method was brought to attention by \cite{ArkaniHamed:2002sp,Schwartz:2003vj}, but was in fact known previously from work in string theory \cite{Green:1991pa,Siegel:1993sk}.  

The full finite gauge transformation for gravity is (\ref{GRfullgauge}), 
\be \label{Grefullgauge2} g_{\mu\nu}(x)\rightarrow {\partial f^\alpha \over \partial  x^\mu}{\partial f^\beta \over \partial  x^\nu}g_{\alpha\beta}\left(f(x)\right),\ee
where $f(x)$ is the arbitrary gauge function, which must be a diffeomorphism.  In massive gravity this gauge invariance is broken only by the mass term.  To restore it, we introduce a St\"uckelberg field $Y^\mu(x)$, patterned after the gauge symmetry (\ref{Grefullgauge2}), and we apply it to the metric $g_{\mu\nu}$, 
\be g_{\mu\nu}(x)\rightarrow G_{\mu\nu}={\partial Y^\alpha \over \partial  x^\mu}{\partial Y^\beta \over \partial  x^\nu}g_{\alpha\beta}\left(Y(x)\right).\ee
The Einstein-Hilbert term $\sqrt{-g}R$ will not change under this substitution, because it is gauge invariant, and the substitution looks like a gauge transformation with gauge parameter $Y^\mu(x)$, so no $Y$ fields are introduced into the Einstein-Hilbert part of the action.    

The graviton mass term, however, will pick up dependence on $Y$'s, in such a way that it will now be invariant under the following gauge transformation
\be g_{\mu\nu}(x)\rightarrow {\partial f^\alpha \over \partial  x^\mu}{\partial f^\beta \over \partial  x^\nu}g_{\alpha\beta}\left(f(x)\right),\ \ \ Y^\mu(x)\rightarrow f^{-1}\left(Y(x)\right)^\mu.\ee
with $f(x)$ the gauge function.  
This is because the combination $G_{\mu\nu}$ is gauge \textit{in}variant (not \textit{co}variant).  To see this, first transform\footnote{The transformation of fields that depend on other fields is potentially tricky.  To get it right, it is sometimes convenient to tease out the dependencies using delta functions.  For example, suppose we have a scalar field $\phi(x)$, which we know transforms according to $\phi(x)\rightarrow\phi(f(x))$.  How should $\phi(Y(x))$ transform?  To make it clear, write 
\be \phi(Y(x))=\int dy \phi(y)\delta(y-Y(x)).\ee
Now the field $\phi$ appears with coordinate dependence, which we know how to deal with,
\be \rightarrow \int dy \phi(f(y))\delta(y-Y(x))=\phi\left(f(Y(x))\right).\ee
Going through an identical trick for the metric, which we know transforms as $g_{\mu\nu}(x)\rightarrow {\partial f^\alpha \over \partial  x^\mu}{\partial f^\beta \over \partial  x^\nu}g_{\alpha\beta}\left(f(x)\right)$, we find
\be  g_{\alpha\beta}\left(Y(x)\right)\rightarrow \partial_\alpha f^\lambda|_{Y}\partial_\beta f^\sigma|_{Y}g_{\lambda\sigma}\left(f(Y(x))\right).\ee} $g_{\mu\nu}$,
\be\partial_\mu Y^\alpha \partial_\nu Y^\beta g_{\alpha\beta}\left(Y(x)\right)\rightarrow \partial_\mu Y^\alpha \partial_\nu Y^\beta \partial_\alpha f^\lambda|_{Y}\partial_\beta f^\sigma|_{Y}g_{\lambda\sigma}\left(f(Y(x))\right),\ee
and then transform $Y$,
\bea 
&& \rightarrow \partial_\mu \left[f^{-1}(Y)\right]^\alpha \partial_\nu  \left[f^{-1}(Y)\right]^\beta \partial_\alpha f^\lambda|_{f^{-1}(Y)}\partial_\beta f^\sigma|_{f^{-1}(Y)}g_{\lambda\sigma}\left(Y(x)\right)\nn\\ 
&&=\partial_\rho \left[f^{-1}\right]^\alpha|_{Y}\partial_\mu Y^\rho \partial_\tau \left[f^{-1}\right]^\beta|_{Y}\partial_\nu Y^\tau\partial_\alpha f^\lambda|_{f^{-1}(Y)}\partial_\beta f^\sigma|_{f^{-1}(Y)}g_{\lambda\sigma}\left(Y(x)\right)\nn\\
&&= \delta_\rho^\lambda\delta^\sigma_\tau\ \partial_\mu Y^\rho\partial_\nu Y^\tau g_{\lambda\sigma}\left(Y(x)\right)=\partial_\mu Y^\lambda \partial_\nu Y^\sigma g_{\lambda\sigma}\left(Y(x)\right).
\eea

We now expand $Y$ about the identity,
\begin{equation}\label{Yexp}
Y^\alpha(x) = x^\alpha + A^\alpha(x).
\end{equation}

The quantity $G_{\mu\nu}$ is expanded as
\bea
G_{\mu \nu} &=& \frac{\partial Y^{\alpha} ( x )}{\partial x^{\mu}}
\frac{\partial Y^{\beta} ( x )}{\partial x^{\nu}} g_{\alpha \beta} ( Y ( x
) )
= \frac{\partial ( x^{\alpha} + A^{\alpha} )}{\partial x^{\mu}}
\frac{\partial ( x^{\beta} + A^{\beta} )}{\partial x^{\nu}} g_{\alpha
 \beta} ( x + A ) \nn\\
&=& ( \delta_{\mu}^{\alpha} + \partial_\mu A^{\alpha} ) ( \delta_{\nu}^{\beta
} + \partial_\nu A^{\beta} ) ( g_{\alpha \beta} + A^{\lambda}
\partial_ \lambda g_{\alpha \beta } + \frac{1}{2} A^{\lambda} A^{\sigma} \partial_ \lambda \partial_ \sigma g_{\alpha
\beta } + \cdots ) \nn\\
&=& g_{\mu \nu} + A^{\lambda} \partial_ \lambda g_{\mu \nu} + \partial_\mu A^{\alpha} g_{\alpha \nu} + \partial_\nu A^{\alpha} g_{\alpha \mu} +
\frac{1}{2} A^{\alpha} A^{\beta} \partial_ \alpha \partial_ \beta g_{\mu \nu } \nn\\
&&\quad\quad+
\partial_\mu A^{\alpha} \partial_\nu A^{\beta} g_{\alpha \beta} + \partial_\mu A^{\alpha} A^{\beta} \partial_ \beta g_{\alpha \nu} + \partial_\nu A^{\alpha} A^{\beta} \partial_ \beta g_{\mu \alpha }+ \cdots 
\label{Gexp}
\eea

We now look at the infinitesimal transformation properties of $g$, $Y$, $G$, and
$Y$, under infinitesimal general coordinate transformations generated by
$f(x) = x +  \xi(x)$.  The metric transforms in the usual way,
\be
\delta g_{\mu \nu} = \xi^{\lambda} \partial_ \lambda g_{\mu \nu } +
\partial_\mu \xi^{\lambda} g_{\lambda \nu} + \partial_\nu \xi^{\lambda} g_{\mu \lambda} \label{dgi}.\ee

The transformation law for the $A$'s comes from the transformation
of $Y$,
\bea &&Y ^\mu( x )\nn
\rightarrow f^{-1}(Y(x))^\mu \approx Y^\mu(x)-\xi^\mu(Y(x)), \\ \nn
&& \delta  Y^\mu =-\xi^\mu(Y),\\
&& \delta  A^\mu =-\xi^\mu(x+A)=-\xi^\mu-A^\alpha\partial_\alpha\xi^\mu-{1\over 2}A^\alpha A^\beta\partial_\alpha\partial_\beta\xi^\mu-\cdots. \label{stuk1Atrans}
\eea
The $A^\mu$ are the Goldstone bosons that non-linearly carry the broken diffeomorphism invariance in massive gravity.  
The combination $G_{\mu\nu}$, as we noted before, is gauge invariant 
 \be
\delta G_{\mu \nu} =0.\ee

We now have a recipe for St\"ukelberg-ing the general massive gravity action of the form (\ref{potentialflat}).  We leave the Einstein-Hilbert term alone.  In the mass term, we write all the $h_{\mu\nu}$'s with lowered indices to get rid of the dependence on the absolute metric, and then we replace all occurrences of $h_{\mu\nu}$ with
\bea
H_{\mu \nu} (x)& =& G_{\mu \nu}(x)-g^{(0)}_{\mu \nu}(x).
\eea
We then expand $G_{\mu\nu}$ as in (\ref{Gexp}), and $Y^\mu$ as in (\ref{Yexp}). To linear order in $h_{\mu\nu}=g_{\mu\nu}-g^{(0)}_{\mu\nu}$ and $A_\mu$, the expansion reads 
\be H_{\mu \nu} = h_{\mu \nu} +\nabla^{(0)}_\mu A_\nu+\nabla^{(0)}_\nu A_\mu,\ee
where indices on $A$ are lowered with the background metric.  This is exactly the Goldstone substitution we made in Section \ref{massivecurvedspacesection} in the linear case.

In the case where the absolute metric is flat, $g^{(0)}_{\mu\nu}=\eta_{\mu\nu}$, we have from (\ref{Gexp}),
\be \label{goldstoneplus} H_{\mu \nu} = h_{\mu \nu} +\partial_\mu A_\nu+\partial_\nu A_\mu+\partial_\mu A^\alpha\partial_\nu A_\alpha+\cdots\ee
Here indices on $A^\mu$ are lowered with $\eta_{\mu\nu}$ and the ellipsis are terms quadratic and higher in the fields and containing at least one power of $h$. This takes into account the full non-linear gauge transformation.  

As in the linear case, we will usually want to do another scalar St\"ukelberg replacement to introduce a $U(1)$ gauge symmetry,
\be   A_\mu\rightarrow A_\mu+\partial_\mu\phi.\ee
Then the expansion for flat absolute metric takes the form
\be \label{stukelbergfirstreplace} H_{\mu \nu} = h_{\mu \nu} +\partial_\mu A_\nu+\partial_\nu A_\mu+2\partial_\mu\partial_\nu\phi+\partial_\mu A^\alpha\partial_\nu A_\alpha+\partial_\mu A^\alpha\partial_\nu \partial_\alpha\phi+\partial_\mu \partial^\alpha\phi\partial_\nu A_\alpha+\partial_\mu\partial^\alpha\phi\partial_\nu\partial_\alpha\phi+ \cdots,\ee
where again the ellipsis are terms quadratic and higher in the fields and containing at least one power of $h$.  The gauge transformation laws are (\ref{stuk1Atrans}), (\ref{hgrtransflat}),
\bea\delta h_{\mu\nu}&=&\partial_\mu \xi_\nu+\partial_\nu \xi_\mu+{\cal L}_\xi h_{\mu\nu}, \nn\\
 \delta A_\mu&=&\partial_\mu\Lambda-\xi_\mu-A^\alpha\partial_\alpha\xi_\mu-{1\over 2}A^\alpha A^\beta\partial_\alpha\partial_\beta\xi_\mu-\cdots, \nn\\ 
 \delta \phi&=&- \Lambda. \label{gaugesyms1}
 \eea  

This method of St\"ukelberg-ing can be extended to any number of gravitons and general coordinate invariances, as was done in \cite{ArkaniHamed:2002sp,ArkaniHamed:2003vb}, in analogy with the gauge theory little Higgs models and dimensional de-construction \cite{ArkaniHamed:2001ca,ArkaniHamed:2001nc}.  When multiple gravitons are present, all but one must become massive, since there are no non-trivial interactions between multiple massless gravitons \cite{Boulanger:2000rq}, and these gravitons mimic the Kaluza-Klein spectrum of a discrete extra dimension.   Other work in this area, including applications to bi-gravity and multi-gravity models, can be found in \cite{Jejjala:2002we,Kan:2002rp,Deffayet:2003zk,GrootNibbelink:2004hg,Nibbelink:2006sz,Deffayet:2011uk}.

\subsection{\label{stukel2section}Another way to St\"ukelberg}

In the last section, we introduced gauge invariance and the St\"ukelberg fields by replacing the metric $g_{\mu\nu}$ with the gauge \textit{invariant} object $G_{\mu\nu}$.  This is well suited to the case where we have a potential arranged in the form (\ref{potentialflat}), because all the background $g^{(0)\mu\nu}$'s appearing in the contractions and determinant of the mass term do not need replacing.  The drawback is that the Stu\"kelberg expansion involves an infinite number of terms higher order in $h_{\mu\nu}$.  If we wish to keep track of the $h_{\mu\nu}$'s, this is not very convenient.  

Instead, we develop another method, which is to introduce the St\"ukelberg fields through the background metric $g^{(0)}_{\mu\nu}$, and then allow $g_{\mu\nu}$ to transform \textit{co}variantly.  This method will be better suited to a potential arranged in the form (\ref{potentialfull}), and will have the advantage that the St\"ukelberg expansion contains no higher powers of $h_{\mu\nu}$.  

We make the replacement
\be g^{(0)}_{\mu\nu}(x)\rightarrow g^{(0)}_{\alpha\beta}\(Y(x)\)\partial_\mu Y^{\alpha}\partial_\nu Y^{\beta}.\ee
The $Y^{\alpha}(x)$ that are introduced are four fields, which despite the index $\alpha$, are to transform as \textit{scalars} under diffeomorphisms
\be Y^{\alpha}(x)\rightarrow Y^\alpha(f(x)),\ee
or infinitesimally,
\be \delta Y^{\alpha}=\xi^\nu\partial_\nu Y^{\alpha}.\ee
This is to be contrasted with the transformation rule $\delta Y^{\alpha}=\xi^\nu\partial_\nu Y^{\alpha}-\left(\partial_\nu \xi^{\alpha}\right)Y^{\nu}$ which would hold if $Y^{\mu}$ were a vector.  Given this scalar transformation rule for $Y^\alpha$, the replaced $g^{(0)}_{\mu\nu}$ now transforms like a metric tensor.  If we now assign the usual diffeomorpshim transformation law to the metric $g_{\mu\nu}$ (so that it is now \textit{co}variant), quantities like $g^{(0)}_{\mu\nu}g^{\mu\nu}$ and other contractions will transform as diffeomorphism scalars.  We can take any action which is a scalar function of $g^{(0)}_{\mu\nu}$ and $g_{\mu\nu}$, and introduce gauge invariance in this way\footnote{This is essentially the technique of spurion analysis, where a coupling constant is made to transform as a field.  A quantity which is normally a background quantity, a coupling constant in the case of spurions, or the background $g^{(0)}_{\mu\nu}$ in this case, is made to transform in some way that gives the action more symmetries.  Note that this method of introducing gauge invariance can be carried out on \textit{any} Lorentz invariant action, even one that does not contain a dynamical metric $g_{\mu\nu}$.  For example, a plain old scalar field in flat space can be made diffeomorphism invariant in this way.  This highlights the fact that general coordinate invariance is not the critical ingredient that leads one to a theory of gravity, since it can be made to hold in \textit{any} theory.}.

This is convenient when we have a potential of the form (\ref{potentialfull}).  First we lower all indices on the $h_{\mu\nu}$'s in the potential.
Now the background metric $g^{(0)}_{\mu\nu}$ appears only through $h_{\mu\nu}=g_{\mu\nu}- g^{(0)}_{\mu\nu}$, so we replace all occurrences of $h_{\mu\nu}$ with
\be H_{\mu\nu}(x)= g_{\mu\nu}(x)- g^{(0)}_{\alpha\beta}\(Y(x)\)\partial_\mu Y^{\alpha}\partial_\nu Y^{\beta}.\ee
Note that we need make no replacement on the $g^{\mu\nu}$'s used to contract the indices, or on the $\sqrt{-g}$ out front of the potential in (\ref{potentialfull}).

Expanding,
\be Y^{\alpha}=x^\alpha-A^{\alpha},\ee
and using $g_{\mu\nu}=g^{(0)}_{\mu\nu}+h_{\mu\nu}$, we have
\be H_{\mu\nu}= h_{\mu\nu}+g^{(0)}_{\nu\alpha}\partial_\mu A^\alpha+g^{(0)}_{\mu\alpha}\partial_\nu A^\alpha-g^{(0)}_{\alpha\beta}\partial_\mu A^\alpha \partial_\nu A^\beta+\cdots,\ee
where the ellipses denote terms that contain derivatives of $g_{\mu\nu}^{(0)}$ (and so vanish in the usual case of interest where $g^{(0)}_{\mu\nu}=\eta_{\mu\nu}$).  Note the difference in sign for the term quadratic in $A^\mu$ compared with (\ref{goldstoneplus}).  

Under infinitesimal gauge transformations we have
\bea \delta A^\alpha&=&-\xi^\alpha+\xi^\nu\partial_\nu A^\alpha, \\
 \delta h_{\mu\nu}&=&\nabla_\mu^{(0)}\xi_\nu+\nabla_\nu^{(0)}\xi_\mu+{\cal L}_\xi h_{\mu\nu},
 \eea
where the covariant derivatives are with respect to $g^{(0)}_{\mu\nu}$ and the indices on $\xi^\mu$ are lowered using $g^{(0)}_{\mu\nu}$.  To linear order, the transformations are
\bea \delta A^\alpha&=&-\xi^\alpha, \\
 \delta h_{\mu\nu}&=&\nabla_\mu^{(0)}\xi_\nu+\nabla_\nu^{(0)}\xi_\mu,
 \eea
which reproduces the linear St\"ukelberg expansion used in Section \ref{massivecurvedspacesection}.

In the case of a flat background, $g^{(0)}_{\mu\nu}=\eta_{\mu\nu}$, the replacement is
\be H_{\mu\nu}= h_{\mu\nu}+\partial_\mu A_\nu+\partial_\nu A_\mu-\partial_\mu A^\alpha \partial_\nu A_\alpha,\ee
with indices on $A^\alpha$ lowered by $\eta_{\mu\nu}$.
Notice that this is the complete expression, there are no higher powers of $h_{\mu\nu}$, unlike (\ref{goldstoneplus}).

We will often follow this with the replacement $A_\mu\rightarrow A_\mu+\partial_\mu\phi$ to extract the helicity 0 mode.  The full expansion thus reads
\be \label{stukelbergsecondreplace} H_{\mu\nu}= h_{\mu \nu} +\partial_\mu A_\nu+\partial_\nu A_\mu+2\partial_\mu\partial_\nu\phi-\partial_\mu A^\alpha\partial_\nu A_\alpha-\partial_\mu A^\alpha\partial_\nu \partial_\alpha\phi-\partial_\mu \partial^\alpha\phi\partial_\nu A_\alpha-\partial_\mu\partial^\alpha\phi\partial_\nu\partial_\alpha\phi .\ee
Under infinitesimal gauge transformations,
\bea\delta h_{\mu\nu}&=&\partial_\mu \xi_\nu+\partial_\nu \xi_\mu+{\cal L}_\xi h_{\mu\nu}, \\
 \delta A_\mu&=&\partial_\mu\Lambda-\xi_\mu+\xi^\nu\partial_\nu A_\mu, \\ 
 \delta \phi&=&- \Lambda. \label{gaugesyms2}
 \eea
 
 Yet another way to introduce St\"ukelberg fields is advocated in \cite{Chamseddine:2010ub,Alberte:2010it,Alberte:2010qb}, in which they make the inverse metric $g^{\mu\nu}$ covariant through the introduction of scalars $g^{\mu\nu}(x)\rightarrow g^{\alpha\beta}\(Y^{-1}(x)\)\partial_\alpha Y^\mu \partial_\beta Y^\nu$.  There have also been many studies, initiated by 't Hooft, of the so-called gravitational Higgs mechanism, which is also essentially a St\"ukelberg-ing of different forms of massive gravity \cite{tHooft:2007bf,Kirsch:2005st,Leclerc:2005qc,Kakushadze:2007dj,Kakushadze:2007zn,Demir:2009ig,Kluson:2010qf,Oda:2010gn,Oda:2010wn}.  All of these are equivalent to the theories we study, as can be seen simply by going to unitary gauge \cite{Berezhiani:2010xy}.  At the end of the day, (\ref{potentialflat}) is the most general Lorentz invariant graviton potential, and any Lorentz invariant massive gravity theory will have a unitary gauge with a potential which is equivalent to it for some choice of the coefficients $C_1,C_2,$ etc.

\section{St\"ukelberg analysis of interacting massive gravity}

In this section, we will set $D=4$ and apply the St\"ukelberg analysis to the massive GR action (\ref{massiveintfirst}) in the case of a flat absolute metric.  The mass term reads
\begin{equation}
S_{mass} = -{M_P^2\over 2}{m^2\over 4}\int d^4 x \eta^{\mu \nu} \eta^{\alpha \beta} 
\left(  h_{\mu \alpha} h_{\nu \beta}- h_{\mu \nu} h_{\alpha \beta} 
\right)  .
\end{equation}
The St\"ukelberg analysis instructs us to make the replacement (\ref{stukelbergfirstreplace}), 
\be \label{stukelbergfirstreplaceagain} h_{\mu \nu} \rightarrow H_{\mu\nu}=h_{\mu \nu} +\partial_\mu A_\nu+\partial_\nu A_\mu+\partial_\mu A^\alpha\partial_\nu A_\alpha+2\partial_\mu\partial_\nu\phi+\partial_\mu\partial^\alpha\phi\partial_\nu\partial_\alpha\phi \cdots.\ee
The extra terms with $h$ in the ellipsis will not be important for this theory, as we will see.  

At the linear level, this replacement is exactly the linear St\"ukelberg expansion of Section \ref{stukelbergtrick}.  We will have to canonically normalize the fields here to match the fields of the linear analysis.  Using a hat to signify the canonically normalized fields with the same coefficients as used in Section \ref{stukelbergtrick} (although there we omitted the hats), we have
\be \label{canonicalfields}\hat{h}=\half M_Ph,\ \ \ \hat A=\half mM_P A,\ \ \ \hat \phi=\half m^2 M_P\phi.\ee

We also get a whole slew of interaction terms, third order and higher in the fields, suppressed by various scales.  We always assume $m<M_P$.  $\phi$ always appears with two derivatives, $A$ always appears with one derivative, and $h$ always appears with none, so a generic term, with $n_h$ powers of $h_{\mu\nu}$, $n_A$ powers of $A_\mu$ and $n_\phi$ powers of $\phi$, reads
\be \sim m^2M_P^2 h^{n_h} (\partial A)^{n_A} (\partial^2\phi)^{n_\phi}\sim \Lambda_\lambda^{4-n_h-2n_A-3n_\phi}  \hat h^{n_h} (\partial\hat A)^{n_A}(\partial^2\hat \phi)^{n_\phi},\ee
where the scale suppressing the term is
\be\label{scalesformula} \Lambda_{\lambda}=\left(M_Pm^{\lambda-1}\right)^{1/\lambda},\ \ \ \lambda={3n_\phi+2n_A+n_h-4\over n_\phi+n_A+n_h-2}.\ee
The larger $\lambda$, the smaller the scale, since $m<M_P$.  We have $n_\phi+n_A+n_h\geq 3$, since we are only considering interaction terms.  
The term suppressed by the smallest scale is the cubic scalar term, $n_\phi=3$, $n_A=n_h=0$, which is suppressed by the scale $\Lambda_5=( M_Pm^4 )^{1/5}$,
\be \label{scalcubschem}\sim {(\partial^2\hat\phi)^3\over \Lambda_5^5},\ \ \ \ \Lambda_5=( M_Pm^4 )^{1/5} .\ee

In terms of the canonically normalized fields (\ref{canonicalfields}), the gauge symmetries (\ref{gaugesyms1}) read
\bea 
 \delta h_{\mu\nu}&=&\partial_\mu \hat\xi_\nu+\partial_\nu \hat \xi_\mu+{2\over M_P}{\cal L}_{\hat \xi}\hat h_{\mu\nu}, \nn\\
 \delta \hat A_\mu&=&\partial_\mu\hat \Lambda-m\hat\xi_\mu+{2\over M_P}\hat\xi^\nu\partial_\nu \hat A_\mu-{2\over mM_P^2}\hat A^\alpha \hat A^\beta\partial_\alpha\partial_\beta \hat\xi_\mu-\cdots,\nn \\
 \delta \phi&=&- m\hat\Lambda, \label{gaugesymscanon1}
 \eea
where we have rescaled $\hat\Lambda={mM_P\over 2}\Lambda$ and $\hat\xi^\mu={M_P\over 2}\xi^\mu$.

Finally, note that since the scalar field $\phi$ always appears with at least two derivatives in the St\"ukelberg replacement (\ref{stukelbergfirstreplaceagain}), the resulting action is automatically invariant under the global \textit{galilean symmetry}
\be \label{galileansym} \phi(x)\rightarrow c+b_\mu x^\mu,\ee
where $c$ and $b_\mu$ are constants.  In addition, the action is automatically invariant under global shifts in $A_\mu\rightarrow A_\mu+c_\mu$ for constant $c_\mu$.   These symmetries are the gravitational analogs of the $SU(N)_L\times SU(N)_R$ global symmetry of the Yang-Mills model in Section \ref{spinonesection}.  It will persist even in limits where the gauge symmetries on $A_\mu$ and $\phi$ no longer act. 

\subsection{Decoupling limit and breakdown of linearity}

As seen in Section \ref{gravitonstukelsec}, the propagators have all been made to go like $\sim 1/p^2$, so normal power counting applies, and the lowest scale, $\Lambda_5$, is the cutoff of the effective field theory.  To focus in on the cutoff scale, we take the decoupling limit
\be m\rightarrow 0,\ \ M_P\rightarrow\infty,\ \ T\rightarrow \infty,\ \ \ \ \ \Lambda_5,\ {T\over M_P}\ \text{fixed}.\ee
All interaction terms go to zero, except for the scalar cubic term (\ref{scalcubschem}) responsible for the strong coupling, which we may calculate using the replacement  $H_{\mu\nu}=2\partial_\mu\partial_\nu\phi+\partial_\mu\partial^\alpha\phi\partial_\nu\partial_\alpha\phi $ since we do not need the vector and tensor terms.  As discussed in Section \ref{gravitonstukelsec}, we must also do the conformal transformation $h_{\mu\nu}= h^\prime_{\mu\nu}+m^2\phi\eta_{\mu\nu}$.  This will diagonalize all the kinetic terms (except for various cross terms proportional to $m$ which are eliminated with appropriate gauge fixing terms, as discussed in Section \ref{gravitonstukelsec}, and which go to zero anyway in the decoupling limit).  

After all this, the lagrangian for the scalar reads, up to a total derivative,
\be\label{phiaction} S_\phi=\int d^4x\ -3(\partial\hat\phi)^2+{2\over \Lambda_5^5}\left[(\square \hat\phi)^3-(\square\hat\phi)(\partial_\mu\partial_\nu \hat\phi)^2\right]+{1\over M_P} \hat\phi T.\ee 
The free graviton coupled to the source via ${1\over M_P}\hat h^\prime_{\mu\nu}T^{\mu\nu}$ also survives the limit, as does the free decoupled vector.  

We can now understand the origin of the Vainshtein radius at which the linear expansion breaks down around heavy point sources.  The scalar couples to the source through the trace, ${1\over M_P}\hat\phi T$.  To linear order around a central source of mass $M$, we have 
\be \hat\phi\sim {M\over M_P}{1\over r}.\ee
The non-linear term is suppressed relative to the linear term by the factor
\be {\partial^4 \hat\phi\over \Lambda_5^5}\sim {M\over M_P}{1\over \Lambda_5^5 r^5}.\ee
Non-linearities become important when this factor becomes of order one, which happens at the radius 
\be r_V\sim \left(M\over M_P\right)^{1/5}{1\over \Lambda_5}\sim \left(GM\over m^4\right)^{1/5}.\ee
When $r\lesssim r_V$, linear perturbation theory breaks down and non-linear effects become important.  This is exactly the Vainshtein radius found in Section \ref{sphersolmassection} by directly calculating the second order correction to spherical solutions.   

In the decoupling limit, the gauge symmetries (\ref{gaugesymscanon1}) reduce to their linear forms,
\bea 
 \delta h_{\mu\nu}&=&\partial_\mu \hat\xi_\nu+\partial_\nu \hat \xi_\mu, \nn\\
 \delta \hat A_\mu&=&\partial_\mu\hat \Lambda, \nn \\
 \delta \phi&=&0.
 \eea
Even though $\phi$ is gauge invariant in the decoupling limit, the fact that it always comes with two derivatives means that the global galileon symmetry (\ref{galileansym}) is still present, as is the shift symmetry on $A_\mu$.  

\subsection{\label{ghostsection}Ghosts}

Note that the lagrangian (\ref{phiaction}) is a higher derivative action, and its equations of motion are fourth order.  This means that this lagrangian actually propagates two lagrangian degrees of freedom rather than one, since we need to specify twice as many initial conditions to uniquely solve the fourth order equations of motion \cite{deUrries:1998bi}, and by Ostrogradski's theorem \cite{Ostrogradski,Woodard:2006nt}, one of these degrees of freedom is a ghost.  The decoupling limit contains six degrees of freedom -- two in the massless tensor, two in the free vector, and two in the scalar.  This matches the number of degrees of freedom in the full theory as determined in Section \ref{hamsection}, so the decoupling limit we have taken is smooth.  The extra ghostly scalar degree of freedom is the Boulware-Deser ghost.  Note that at linear order, the higher derivative scalar terms for the scalar are not visible, so the linear theory has only 5 degrees of freedom.

Following \cite{Creminelli:2005qk}, let's consider the stability of the classical solutions to (\ref{phiaction}) around a massive point source.  We have a classical background $\Phi (r)$, which is a solution of the $\hat\phi$ equation of motion, and we expand the lagrangian of (\ref{phiaction})  to
quadratic order in the fluctuation $\varphi \equiv \hat\phi - \Phi$.  The
result is schematically 
\be \label{L_ghost} 
{\cal
  L}_\varphi \sim -(\partial \varphi)^2 + \frac {(\partial^2 \Phi)} {\Lambda_5^5}
(\partial^2 \varphi)^2.
\ee 
There is a
four-derivative contribution to the $\varphi$ kinetic term, signaling that this theory propagates two linear degrees of freedom.  As shown in Section 2 of \cite{Creminelli:2005qk}, one is stable and massless, and the other is a ghost with a mass of order the scale appearing in front of the higher derivative terms.  So in this case the
ghost has an $r$-dependent mass 
\be \label{m_ghost} 
m^2_{\rm ghost}(r) \sim \frac{\Lambda_5^5}{\partial^2 \Phi(r)} \; .  
\ee 
This shows that around a flat background, or far from the source, the ghost mass goes to infinity and the ghost freezes, explaining why it was not seen in the linear theory.  It is only around non-trivial backgrounds that it becomes active.  Notice, however, that the backgrounds around which the ghost becomes active are perfectly nice, asymptotically flat configurations sourced by compact objects like the Sun, and not disconnected in any way in field space (this is in contrast to the ghost in DGP, which occurs only around asymptotically de Sitter solutions).

We are working in an effective field theory with a UV cutoff
$\Lambda_5$, therefore we should not worry about instabilities until the mass of the ghost
drops below $\Lambda_5$.  This happens at the distance $r_{\rm ghost}$ where
$\partial^2 \Phi^c \sim \Lambda_5^3$.   For a source of mass $M$, at distances $r \gg r_V$ the background field goes like $\Phi
(r) \sim {M\over M_P}{ 1\over r}$, so 
\be  \label{R_ghost}
r_{\rm ghost} \sim
\left(\frac{M}{M_P}\right)^{1/3}\frac1{\Lambda_5}  \; \gg \;
r_V \sim  \left(\frac{M}{M_P}\right)^{1/5}\frac1{\Lambda_5}
.  
\ee 
$r_{\rm ghost}$ is parametrically larger than the Vainshtein
radius $r_V$. 

As we will see in Section \ref{quantumL5}, the distance $r_{\rm ghost}$ is the same distance at which quantum effects become important.  Whatever UV completion takes over should cure the ghost instabilities that become present at this scale, so we will be able to consistently ignore the ghost.  We see already that we cannot trust the classical solution even in regions parametrically farther than the Vainshtein radius.  The best we can do is make predictions outside $r_{\rm ghost}$, and we will have more to say later about this.

\subsection{\label{ghostscreensection}Resolution of the vDVZ discontinuity and the Vainshtein mechanism}

We are now in a position to see the mechanism by which non-linearities can resolve the vDVZ discontinuity.  This is known as the \textit{Vainshtein mechanism}.  It turns out to involve the ghost in a critical role.  

Far outside the Vainshtein radius, where the linear term of (\ref{phiaction}) dominates, the field has the usual Coulombic $1/r$ form.  But inside the Vainshtein radius, where the cubic term dominates, it is easy to see by power counting that the field gets an $r^{3/2}$ profile,
\be \label{phibehavior} \begin{cases}  \hat\phi\sim {M\over M_P}{1\over r},  & r\gg r_V, \\
\hat\phi\sim \left({M\over M_P}\right)^{1/2}\Lambda_5^{5/2}r^{3/2}, &   r\ll r_V.
\end{cases} 
\ee

At distances much below the Vainshtein radius, the ghost mass (\ref{m_ghost}) becomes very small, and the ghost starts to mediate a long range force.  Usually a scalar field mediates an attractive force, but due to the ghost's wrong sign kinetic term, the force mediated by it is repulsive.  In fact, it cancels the attractive force due to the longitudinal mode, the force responsible for the vDVZ discontinuity, and so general relativity is restored inside the Vainshtein radius.

We will now see this more explicitly.  Following \cite{Deffayet:2005ys}, some field re-definitions can be done on the scalar action (\ref{phiaction}), and the result is an action schematically of the form ${\cal L}=-(\partial \tilde \phi)^2+(\partial \psi)^2+\Lambda_5^{5/2}\psi^{3/2}+{1\over M_P}\tilde\phi T+{1\over M_P}\psi T.$  Here $\tilde\phi$ is the healthy longitudinal mode, $\psi$ is the ghost mode, and the original scalar can be found from $\hat\phi=\tilde \phi-\psi$.  Both are coupled gravitationally to the stress tensor.  Note that the self-interactions appear in these variables as a peculiar non-analytic $\psi^{3/2}$ term (we can also see that the ghost mass around a background $\la \psi\ra$ will be $\Lambda_5^{5/2}/\la \psi\ra^{1/2}$).  The $\tilde\phi$ field is free and has the profile $\tilde\phi\sim {M\over M_P}{1\over r}$ everywhere, mediating an attractive force.  

The $\psi$ field however has two competing terms, which becomes comparable at the Vainshtein radius.  The linear term dominates at radii \textit{smaller} than the Vainshtein radius, so $\psi\sim {M\over M_P}{1\over r}$ for $r\ll r_V$.  This profile generates a repulsive Coulomb force that exactly cancels the attractive force mediated by $\tilde \phi$, so in sum there are no extra forces beyond gravity in this region.  (The leading correction to the profile is found by treating the $\psi^{3/2}$ term as a perturbation, $\psi\sim \psi_{0}+\psi_{(1)}+\cdots$, with $\psi_{0}\sim  {M\over M_P}{1\over r}$ , plugging in the equation of motion $\partial^2\psi_{(1)}+\Lambda_5^{5/2}\psi_{(0)}^{1/2}=0$ obtaining $\psi_{(1)}\sim \left({M\over M_P}\right)^{1/2}\Lambda_5^{5/2}r^{3/2}$, in agreement with (\ref{phibehavior}).) The funny non-linear term dominates at radii larger than the Vainshtein radius, so $\psi\sim \left(M\over M_P\right)^2{1\over \Lambda_5^5 r^6}$ for $r\gg r_V$, and so the ghost profile is negligible in this region compared to the $\tilde\phi$ profile.  Thus the ghost ceases to be active beyond the Vainshtein radius, and the longitudinal mode generates a fifth force.

This is known as a \textit{screening mechanism}, a mechanism for rendering a light scalar inactive at short distances through non-linearities (see the introduction and references in \cite{Hinterbichler:2010es,Hinterbichler:2010wu}, and in a different context \cite{Gabadadze:2007as}).

One can think of this as a kind of classical version of a weakly coupled UV completion via a Higgs.  Above the Vainshtein radius (low energies), there is only the long distance scalar, which starts to become non-linear (strongly coupled) around the Vainshtein radius, so one can think of this regime in terms of an effective field theory with cutoff the Vainshtein radius.  Below the Vainshtein radius (high energies), a new degree of freedom, the ghost (analogous to the physical Higgs in the standard model), kicks in.  Much below the Vainshtein radius, everything is again linear and weakly coupled, with the difference that there are now two active degrees of freedom, so one can think of this as a classical UV completion of the effective theory.

Of course, this ghostly mechanism for restoring continuity with GR relies on an instability, which would become apparent were we to investigate small fluctuations beyond the gross-scale features described here.  Furthermore, as we will see in the next section, the ghost issue is moot, since the classical mechanism described in this section occurs outside the regime of validity of the quantum effective theory and is swamped by unknown quantum corrections.

\subsection{\label{quantumL5}Quantum corrections and the effective theory}

 Quantum mechanically, massive gravity is an effective field theory, since there are non-renormalizable operators suppressed by the mass scale $\Lambda_5$.  The amplitude for $\pi\pi\rightarrow \pi\pi$ scattering at energy $E$, coming from the cubic coupling in (\ref{phiaction}), goes like ${\cal A}\sim\left(E\over \Lambda_5\right)^{10}$.  This amplitude should correspond to the scattering of longitudinal gravitons.  The wave function of the longitudinal graviton (\ref{longitudinalmode}) for a large boost is proportional to $m^{-2}$, while the largest term at high momentum in the graviton propagator (\ref{massmprop}) is
proportional to $m^{-4}$, so naive power counting would suggest that the amplitude at energies much larger than $m$ goes like ${\cal A}\sim {E^{14}\over M_P^2 m^{12}}$.  However, as recognized in \cite{ArkaniHamed:2002sp}, and calculated explicitly in \cite{Aubert:2003je}, there is a cancellation in the diagrams so that the result agrees with the result of the Goldstone description.  We will encounter these kinds of cancellations again in loops, and part of the usefulness of the Goldstone description is that they are made manifest.
  
The amplitude becomes order one and hence strongly coupled when $E\sim \Lambda_5$.  Thus $\Lambda_5$ is the maximal cutoff of the theory. 
We expect to generate all operators compatible with the symmetries, suppressed by appropriate powers of the cutoff.  In the unitary gauge, there are no symmetries, so we will generate all operators of the form
 \begin{equation}\label{unitarterms}
c_{p,q} \partial^q h^p.
\end{equation}
We wish to determine the scales in the coefficient $c_{p,q}$.  

After St\"ukelberg-ing, the decoupling limit theory contains only the scalar $\hat\phi$, and the single coupling scale $\Lambda_5$.  In addition, there is the galileon symmetry $\hat\phi\rightarrow\hat\phi+c+c_\mu x^\mu$.  Quantum mechanically, we expect to generate in the quantum effective action all possible operators with this symmetry, suppressed by the appropriate power of the cutoff $\Lambda_5$.  The galileon symmetry forces each $\hat\phi$ to carry at least two derivatives\footnote{Actually, there are a finite number of terms which have fewer than two derivatives per field, the so-called galileon terms \cite{Nicolis:2008in} which change by a total derivative under the galileon symmetry \cite{Nicolis:2008in}.  However, there is a non-renormalization theorem that says these are not generated at any loop by quantum corrections \cite{Hinterbichler:2010xn}, so we need not include them.  We will encounter them later when we raise the cutoff to $\Lambda_3$.}, so the general term we can have is
\begin{equation}\label{pqscalarterm}
\sim \frac{\partial^q (\partial^2 \hat\phi)^p}{\Lambda_5^{3p + q - 4}}.
\end{equation}
To compare to (\ref{unitarterms}), we go back to the original normalization for the fields by replacing $\hat\phi \sim m^2 M_P \phi$ 
and recall that
$\partial_{\mu}\partial_\nu\phi$ comes from an $h_{\mu \nu}$ to find that
in unitary gauge the coefficients $c_{p,q}$ go like
\begin{equation}
c_{p,q} \sim 
\Lambda_5^{-3p-q+4} M_P^p  m^{2p} =
\left(m^{16 - 4q - 2p} M_P^{2p - q + 4}\right)^{1/5} .
\end{equation}
This comparison is possible because the operations of taking the decoupling limit and computing quantum corrections should commute.

Notice that the term with $p=2,\ q=0$ is a mass term, $\sim {M_P^2m^4\over \Lambda_5^2}h^2$, corresponding to a mass correction $\delta m^2=m^2\(m^2\over \Lambda_5^2\)$.  This is down by a factor of $m^2/\Lambda_5^2$ from the tree level mass term.  Thus a small mass graviton mass $m\ll \Lambda_5$ is technically natural, and there is no quantum hierarchy problem associated with a small mass.  This is in line with the general rule of thumb that a small term is technically natural if a symmetry emerges as the term is dialed to zero.  In this case, it is the diffeomorphism symmetry of GR which is restored as the mass term goes to zero.   The quantum mass correction will also generically ruin the Fierz-Pauli tuning, but its coefficient is small enough that ghost/tachyons associated to the tuning violation are postponed to the cutoff -- indeed the resulting ghost mass, using the relations in the second paragraph of Section \ref{Massive}, is $\sim \Lambda_5$.

Similar to the spin 1 case in Section \ref{spinonesection},  it is important that there were no non-parametric modifications to the kinetic structure of the Einstein-Hilbert term, even though the lack of gauge symmetry would suggest that we are free to make such modifications.  Suppose we try to calculate the mass correction directly in unitary gauge.  The graviton mass term contributes no vertices but alters the propagator so that its high energy behavior is $\sim {k^2\over m^4}$ (the next leading terms go like ${1\over m^2}$ and then ${1\over k^2}$).  At one loop, there are two 1PI diagrams correcting the mass; one containing two cubic vertices ${1\over M_P}\partial^2 \hat h^3$ from the Einstein-Hilbert action and two propagators, and another containing a single quartic vertex ${1\over M_P^2}\partial^2 \hat h^4$ from the Einstein-Hilbert action and a single propagator.  Cutting off the loop at the momenta $k_{\rm max}\sim \Lambda_5$, the first diagram gives the largest naive correction $\delta m^2\sim {1\over M_P^2m^8}\Lambda_5^{12}\sim {\Lambda_5^2}$. (The second diagram gives a smaller correction.)  This is at the cutoff, dangerously higher than the small correction $\delta m^2\sim m^2{m^2\over \Lambda_5^2}$ we found in the Goldstone formalism.  

This means that there must be a non-trivial cancellation of this leading divergence in unitary gauge, so that we recover the Goldstone result.  This cancellation happens because the kinetic interactions of Einstein-Hilbert are gauge invariant, implying that the dangerous $k^\mu k^\nu k^\alpha k^\beta/m^4$ terms in the graviton propagator do not contribute.  Without these terms, the propagator goes like $1/m^2$ and the estimate for the first diagram is $\delta m^2\sim {1\over M_P^2m^4}\Lambda_5^8\sim m^2{m^2\over \Lambda_5^2}$, in agreement with the Goldstone prediction (again the second diagram again gives a smaller correction).  Non-parametrically altering the coefficients in the kinetic structure would spoil this cancellation and the resulting technical naturalness of the small mass (though such alterations could be done without spoiling technical naturalness if the alterations to the kinetic terms are parametrically suppressed by appropriate powers of $m$).  These kinds of cancellations can be seen explicitly in the calculations of \cite{Aubert:2003je}.  Some loop calculations for massive gravity have been done in \cite{Park:2010rp,Park:2010zw}.

In summary, in unitary gauge the theory (\ref{massiveintfirst}) in $D=4$ is a natural effective field theory with a cutoff parametrically larger than the graviton mass, with the effective action 
\begin{equation}
S=\int d^4x\ {M_P^2\over 2} \left[\sqrt{-g} R - 
{m^2\over 4}  (h_{\mu \nu}^2 - h^2)\right] +
\sum_{p,q} c_{p,q} \partial^q h^p,
\end{equation}
and a cutoff $\Lambda_5 = (m^4 M_P)^{1/5}$.

We should take into account the effect that the unknown quantum operators have on the solution around a heavy source.  Given that the linear field goes like $\hat \phi\sim {M\over M_P}{1\over r}$, the radius $r_{p,q}$ at which 
the term (\ref{pqscalarterm}) becomes comparable to the kinetic term $(\partial\hat\phi)^2$ is
\begin{equation}
r_{p,q} \sim\left(\frac{M}{M_{Pl}}\right)^{\frac{p-2}{3p + q - 4}}{1\over \Lambda_5}.
\end{equation}
This distance increases with $p$, and asymptotes to its highest value
\begin{equation}
r_Q\sim \left(\frac{M}{M_{Pl}}\right)^{1/3}{1\over \Lambda_5}.
\end{equation}

Thus we cannot trust the classical solution at distances below $r_Q$, since quantum operators become important there. This distance is parametrically larger than the Vainshtein radius, where classical non-linearities become important.  Unlike the case in GR, there is no intermediate regime where the linear approximation breaks down but quantum effects are still small, so there is no sense in which a non-linear solution to massive gravity can be trusted for making real predictions in light of quantum mechanics.  

In particular, the entire ghost screening mechanism of Section \ref{ghostscreensection} is in the non-linear regime, and so it becomes swamped in quantum corrections.  Thus there is no regime for which GR is a good approximation -- the theory transitions directly from the linear classical regime with a long range fifth force scalar, to the full quantum regime.  Note that it is the higher dimension operators that become important first, so there is no hope of finding the leading quantum corrections.  Finally, the radius $r_Q$ is the same as the radius $r_{\rm ghost}$ where the ghost mass drops below the cutoff, so it is consistent to ignore the ghost since it lies beyond the reach of the quantum effective theory.  The various regions are shown in Figure \ref{plot1}.  Note that in the decoupling limit we are working in, the Schwarzschild radius (and the radii associated to all scales larger than $\Lambda_5$) are sent to zero, while the scale $r\sim {1\over m}$ where Yukawa suppression takes hold is sent to infinity.

\begin{figure}[h!]
\begin{center}
\epsfig{file=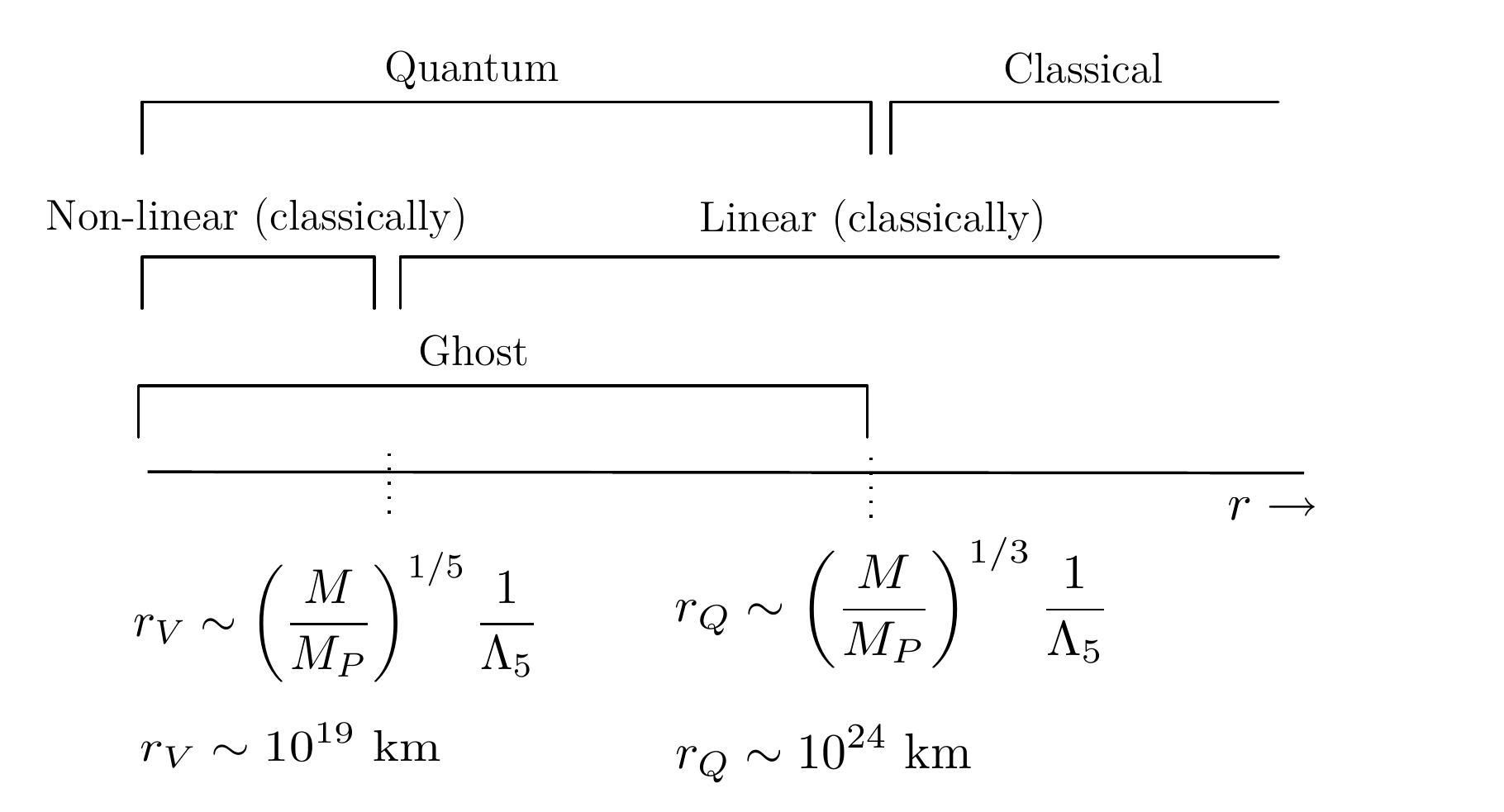,height=2.5in,width=4.0in}
\caption{\small  Regimes for massive gravity with cutoff $\Lambda_5=(M_Pm^4)^{1/5}$, and some values within the solar system, for which $\Lambda_5^{-1}\sim 10^{11}\ {\rm km}$.  Note that $r_Q$ is a bit larger than the observable universe, i.e. this theory makes no observable predictions within its range of validity.}
\label{plot1}
\end{center}
\end{figure}

\section{\label{lambda3section}The $\Lambda_3$ theory}

We have seen that the theory (\ref{massiveintfirst}) containing only the linear graviton mass term has some undesirable features, including a ghost instability and quantum corrections that become important before classical non-linearities can restore continuity with GR.  In this section, we consider the higher order potential terms in (\ref{potentialflat}) and ask whether they can alleviate these problems.  It turns out that there is a special choice of potential that cures all these problems, at least in the decoupling limit. 

This choice also has the advantage of raising the cutoff.
With only the Fierz-Pauli mass term, the strong coupling cutoff was set by the cubic scalar self coupling 
$\sim {(\partial^2\hat\phi)^3\over \Lambda_5^5}$.  The cutoff $\Lambda_5=(M_Pm^4)^{1/5}$ is very low, and as we will see, generically any interaction term will have this cutoff.  But by choosing this special tuning of the higher order interactions, we end up raising the cutoff to the higher scale $\Lambda_3=(M_Pm^2)^{1/3}$.  

It was recognized already in \cite{ArkaniHamed:2002sp}, that if the scalar self-interactions could be eliminated, the cutoff would be raised to $\Lambda_3$.  This was studied more fully in \cite{Creminelli:2005qk}, where the cancelation was worked through and it was (mistakenly) concluded that ghosts would be unavoidable once the cutoff was raised.  Motivated by constructions of massive gravity with auxiliary extra dimensions (see Section \ref{auxdimsection}), this was revisited in \cite{deRham:2010ik,deRham:2010kj}, where the decoupling limit lagrangian was calculated explicitly and was seen to be ghost free.  In \cite{Hassan:2011hr}, the full theory was shown to be ghost free\footnote{The objections of \cite{Alberte:2010qb} and \cite{Folkerts:2011ev} are addressed in \cite{deRham:2011rn} and \cite{deRham:2011qq} respectively.}.

\subsection{Tuning interactions to raise the cutoff}

Looking back at the scales (\ref{scalesformula}), the term suppressed by the smallest scale is the cubic scalar term, which is suppressed by the scale $\Lambda_5=( M_Pm^4 )^{1/5}$,
\be \sim {(\partial^2\hat\phi)^3\over M_P m^4}.\ee
The next highest scale is $\Lambda_4=( M_P m^3 )^{1/4}$, carried by a quartic scalar interaction, and a cubic term with a single vector and two scalars,
\be \sim {(\partial^2\hat\phi)^4\over M_P^2 m^6 },\ \ \ \sim {\partial \hat A(\partial^2\hat \phi)^2\over M_Pm^3}.\ee
The next highest is a quintic scalar, and so on.  The only terms which carry a scale less than $\Lambda_3=( M_P m^2 )^{1/3}$ are terms with only scalars $(\partial^2\hat\phi)^n$, and terms with one vector and the rest scalars $\partial \hat A(\partial^2\hat \phi)^n$.

The scale $\Lambda_3$ is carried by only the following terms
\be\label{lambda3terms} \sim {\hat h(\partial^2\hat\phi)^n\over M_P^{n+1} m^{2n+2}},\ \ \ \sim{(\partial \hat A)^2(\partial^2\hat\phi)^n\over M_P^{n+2} m^{2n+4}}.\ \  \ee
All other terms carry scales higher than $\Lambda_3$.

It turns out that we can arrange to cancel all of the scalar self couplings by appropriately choosing the coefficients of the higher order terms.  We will work with the form of the potential in (\ref{potentialfull}) where indices are raised with the full metric, and the St\"ukelberg formalism of Section \ref{stukel2section}.  We do because we eventually want to keep track of powers of $h$, so the form of the St\"ukelberg replacement in Section \ref{stukel2section} is simpler.
We are interested only in scalar self interactions, so we may make the replacement (\ref{stukelbergsecondreplace}) with the vector field set to zero,
\be \label{hscalarreplace} H_{\mu\nu} \rightarrow 2 \, \partial_{\mu} \partial_{\nu} \phi 
- \partial_{\mu} \partial_{\alpha} \phi \, \partial_{\nu} \partial^{\alpha} \phi.\ee

The interaction terms are a function of the matrix of second derivatives $\Pi_{\mu\nu}\equiv \partial_\mu\partial_\nu\phi$.  As reviewed in Appendix \ref{totalDappendix}, there is at each order in $\phi$ a single polynomial in $\Pi_{\mu\nu}$ which is a total derivative.  By choosing the coefficients (\ref{potentialfull}) correctly, we can arrange for the $\phi$ terms to appear in these total derivative combinations.  The total derivative combinations have at each order in $\phi$ as many terms as there are terms in the potential of (\ref{potentialfull}), so all the coefficients must be fixed, except for one at each order which becomes the overall coefficient of the total derivative combination.

The choice of coefficients in the potential (\ref{potentialfull}) which removes the scalar self interactions is, to fifth order \cite{deRham:2010ik},
\be c_1=2c_3+\frac 1 2,\ \ \ c_2=-3c_3-\frac 12,\ee
\bea
d_1&=&-6d_5+\frac{1}{16}(24c_3+5),\ \ \ d_2=8d_5-\frac{1}{4}(6c_3+1)\,\nn ,\\
d_3&=&3d_5-\frac{1}{16}(12c_3+1),\ \ \ \ d_4=-6d_5+\frac34 c_3,
\eea

\begin{eqnarray}
\label{fs}
\begin{array}{ccc}
f_1=\frac{7}{32}+\frac{9}{8}c_3-6d_5+24f_7\ ,
& \hspace{20pt}& f_2 = -\frac{5}{32} -\frac{15}{16}c_3+6d_5-30f_7\ ,\\
f_3=\frac38 c_3-3d_5+20 f_7\ ,
& \hspace{20pt}& f_4=-\frac{1}{16}-\frac34 c_3+5d_5-20f_7\ ,\\
f_5=\frac{3}{16} c_3-3d_5+15f_7\ , & \hspace{20pt}& f_6=d_5-10f_7\,.
\end{array}
\end{eqnarray}
At each order, there is a one-parameter family of choices that works to create a total derivative.  Here $c_3$, $d_5$ and $f_7$ are chosen to carry that parameter at order 3, 4 and 5 respectively.  Note however that at order 5 and above (or $D+1$ and above if we were doing this in $D$ dimensions), there is one linear combination of all the terms, the characteristic polynomial of $h$ mentioned below (\ref{U5equation}), that vanishes identically.  This means that one of the coefficients is redundant, and we can in fact set $d_5$ and its higher counterparts to any value we like without changing the theory.  Thus there is only a two parameter family ($D-2$ parameter in dimension $D$) of theories with no scalar self-interactions.
This can be carried through at all orders, and at the end there will be no terms $\sim (\partial^2\phi)^n$.

The only terms with interaction scales lower than $\Lambda_3$ were the scalar self-interactions $(\partial^2\hat\phi)^n$, and terms with one vector and the rest scalars $\partial \hat A(\partial^2\hat \phi)^n$.  We have succeeded in eliminating the scalar self-interactions, but since these always came from combinations $(A_\mu+\partial_\mu)$ the terms $\partial \hat A(\partial^2\hat \phi)^n$ are automatically of the form $\partial^\mu A^\nu X^{(n)}_{\mu\nu}$, where the $X^{(n)}_{\mu\nu}$ are the functions of $\partial_\mu\partial_\nu\phi$ described in Appendix \ref{totalDappendix}, which are identically conserved $\partial^\mu X^{(n)}_{\mu\nu}=0$.  Thus, once the scalar self-interactions are eliminated, the $\partial \hat A(\partial^2\hat \phi)^n$ terms are all total derivatives and are also eliminated.

 Now the lowest interaction scale will be due to the terms in (\ref{lambda3terms}),
\be\label{lambda3terms2} \sim {\hat h(\partial^2\hat\phi)^n\over M_P^{n+1} m^{2n+2}},\ \ \ \sim{(\partial \hat A)^2(\partial^2\hat\phi)^n\over M_P^{n+2} m^{2n+4}},\ \  \ee
which are suppressed by the scale $\Lambda_3=( M_P m^2 )^{1/3}$,
 so the cutoff has been raised to $\Lambda_3$, carried by the terms (\ref{lambda3terms2}). 

The decoupling limit is now 
\be\label{decouplambda3} m\rightarrow 0,\ \ M_P\rightarrow\infty,\ \ \ \ \Lambda_3\  \text{fixed},\ee
and the only terms which survive are those in (\ref{lambda3terms}).  To find these terms we must now go back to the full St\"ukelberg replacement (\ref{stukelbergsecondreplace}), and we must also expand the inverse metric and determinant in the potential of (\ref{potentialfull}) in powers of $h$.
The $h(\partial^2\phi)^n$ terms, up to quintic order in the decoupling limit, and up to total derivatives are \cite{deRham:2010ik},
\bea S=\int d^4x\ && \frac{1}{2}\hat h_{\mu\nu}{\cal E}^{\mu\nu,\alpha\beta} \hat h_{\alpha\beta}-{1\over 2} \hat h^{\mu\nu}\left[-4X^{(1)}_{\mu\nu}(\hat \phi)+{4(6c_3-1)\over \Lambda_3^3}X^{(2)}_{\mu\nu}(\hat\phi)+{16(8d_5+c_3)\over \Lambda_3^6}X^{(3)}_{\mu\nu}(\hat \phi)\right] \nn \\
&&+{1\over M_P} \hat h_{\mu\nu}T^{\mu\nu}. \label{lambda3decoupling}\eea
Here the $X^{(n)}_{\mu\nu}$ are the identically conserved combinations of $\partial_\mu\partial_\nu\hat\phi$ described in Appendix \ref{totalDappendix}.  The $(\partial A)^2(\partial^2\phi)$ terms can be found to cubic order in \cite{deRham:2010gu}.  The terms with $A$'s can in any case be consistently set to zero at the classical level, since they never appear linearly in the lagrangian, so we will focus only on the terms involving $h$ and $\phi$.  Properties of this lagrangian, including its cosmological solutions, degravitation effects and phenomenology are studied in \cite{deRham:2010tw}.  Spherical solutions are studied in \cite{Chkareuli:2011te}.  The cosmology of a covariantized version is studied in \cite{deRham:2011by}.

In terms of the canonically normalized fields (\ref{canonicalfields}), the gauge symmetries (\ref{gaugesyms2}) of the full theory are 
\bea \delta \hat A_\mu&=&\partial_\mu\hat \Lambda-m\hat\xi_\mu+{2\over M_P}\hat\xi^\nu\partial_\nu \hat A_\mu, \\
 \delta h_{\mu\nu}&=&\partial_\mu \hat\xi_\nu+\partial_\nu \hat \xi_\mu+{2\over M_P}{\cal L}_{\hat \xi}\hat h_{\mu\nu}, \\
 \delta \phi&=&- m\hat\Lambda, \label{gaugesymscanon}
 \eea
 where we have rescaled $\hat\Lambda={mM_P\over 2}\Lambda$ and $\hat\xi^\mu={M_P\over 2}\xi^\mu$.
In the decoupling limit (\ref{decouplambda3}), this gauge symmetry reduces to its linear form,
\bea \delta \hat A_\mu&=&\partial_\mu\hat \Lambda, \\
 \delta h_{\mu\nu}&=&\partial_\mu \hat\xi_\nu+\partial_\nu \hat \xi_\mu, \\
 \delta \phi&=&0. \label{gaugesymscanondec}
 \eea
The lagrangian (\ref{lambda3decoupling}) should be invariant under the decoupling limit gauge symmetries (\ref{gaugesymscanondec}).  Indeed, the identity $\partial^\mu X^{(n)}_{\mu\nu}=0$ ensures that it is.  The scalar $\phi$ is gauge invariant in the decoupling limit, but the fact that it always comes with two derivatives means that the global galileon symmetry (\ref{galileansym}) is still present, as is the shift symmetry on $A_\mu$.

Note that for the specific choice $c_3=1/6$ and $d_5=-1/48$, all the interaction terms disappear.  This could mean that the theory becomes strongly coupled at some scale larger than $\Lambda_3$, or there could be no lowest scale, since there are scales arbitrarily close to but above $\Lambda_3$.  In the later case, the theory would have no non-linear behavior, and so no mechanism to recover continuity with GR, and it would therefore be ruled out observationally.

\subsection{Exactness of the decoupling limit}

In \cite{deRham:2010kj}, a nice trick was used to show that the decoupling limit lagrangian (\ref{lambda3decoupling}) is exact to all orders in the fields, that is, there are no further terms $h(\partial^2\phi)^n$ for $n\geq 4$.  In fact, the method extends easily to any dimension, showing that in dimension $D$, there are no further terms $h(\partial^2\phi)^n$ for $n\geq D$.  

Define a new tensor, 
\be
{\cal K}^\mu_{\ \nu} (g,H)=\delta^\mu_{\ \nu} -\sqrt{\delta^\mu_{\ \nu} - H^\mu_{\ \nu}}=\sum_{n=1}^{\infty}d_n ( H^n)^\mu_{\ \nu}, \ \ \  d_n=-\frac{(2n)!}{(1-2n)(n!)^2 4^n}\,.
\ee
Here indices are raised using the full metric $g_{\mu\nu}$, $H^\mu_{\ \nu} = g^{\mu\alpha}H_{\alpha\nu}$, and
$(H^n)^\mu_{\ \nu}=H^\mu_{\ \alpha_1}H^{\alpha_1}_{\ \alpha_2}\cdots H^{\alpha_{n-1}}_{\ \nu}$ denotes the matrix product
of $n$ tensors. The
tensor ${\cal K}_{\mu\nu}= g_{\mu\alpha}{\cal K}^\alpha_{\ \nu}$ is defined so that, given the St\"ukelberg replacement with no vectors $ H_{\mu\nu} =h_{\mu\nu}+ 2 \, \partial_{\mu} \partial_{\nu} \phi 
- \partial_{\mu} \partial_{\alpha} \phi \, \partial_{\nu} \partial^{\alpha} \phi$, we have
\be \label{Kproperty}
{\cal K}_{\mu\nu}(g,H)\Big|_{h_{\mu\nu}=0}= \partial_\mu\partial_\nu\phi\,.
\ee

Using ${\cal K}$ as a new variable, the most general potential in (\ref{potentialfull}) can be written as
\be \label{Kpotential} W(g,{\cal K})= \la {\cal K}^2\ra-\la {\cal K} \ra^2 +\tilde c_1\la {\cal K}^3\ra+ \tilde c_2 \la  {\cal K}^2\ra \la  {\cal K}\ra + \tilde c_3\la  {\cal K} \ra ^3+\cdots\ee
where brackets mean traces with respect to the full metric, and the tilde coefficients are arbitrary and can be related to those in (\ref{potentialfull}) by expanding.
Because of the property (\ref{Kproperty}), this reorganization of the potential makes it easy to see what the scalar self-interactions look like -- they are simply $W(g,{\cal K})=W(g,\Pi)$.  Thus the $\Lambda_3$ theory corresponds to choosing coefficients in (\ref{Kpotential}) such that the ${\cal K}$ terms appear in the total derivative combinations of Appendix \ref{totalDappendix}.  Each total derivative combination can have an arbitrary overall coefficient, so the $\Lambda_3$ theory corresponds to
\be  \label{KpotentialTD} W(g,{\cal K})=\sum_{n\geq 2}\alpha_n {\cal L}_n^{\rm TD}({\cal K}),\ee
with arbitrary coefficients $\alpha_n$.  These coefficients correspond to the free coefficients of the $\Lambda_3$ theory, $\alpha_2=-2^2$, $\alpha_3=2^3c_3$, $\alpha_4=2^4d_5,$ etc.  In (\ref{KpotentialTD}), the sum is finite and stops at $n=D$, since the total derivative combinations vanish for $n>D$.

The decoupling limit interactions contain only one power of $h$, so the entire decoupling limit action is given by (note that in this section the fields are not canonically normalized)
\be S=\int d^4x\ \frac{1}{2\kappa^2}\({1\over 4}h_{\mu\nu}{\cal E}^{\mu\nu,\alpha\beta}h_{\alpha\beta}-{m^2\over 4}h^{\mu\nu}\bar X_{\mu\nu}\),\ee
where
\be \bar X^{\mu\nu}={\delta\over \delta h_{\mu\nu}}\left(\sqrt{-g}W(g,{\cal K})\right)\big|_{h_{\mu\nu}=0}.\ee
Using the relation
\be {\delta\over \delta h^{\mu\nu}}\la {\cal K}^n\ra\big|_{h_{\mu\nu}=0}={n\over 2}\left(\Pi^{n-1}_{\mu\nu}-\Pi^n_{\mu\nu}\right),\ee
we calculate
\be {\delta\over \delta h_{\mu\nu}}\left(\sqrt{-g}{\cal L}_n^{\rm TD}({\cal K})\right)\big|_{h_{\mu\nu}=0}=\sum_{m=0}^n{(-1)^mn!\over 2(n-m)!}\left(\Pi^{m}_{\mu\nu}-\Pi^{m-1}_{\mu\nu}\right){\cal L}_{n-m}^{\rm TD}(\Pi)={1\over 2}\left(X_{\mu\nu}^{(n)}+nX_{\mu\nu}^{(n-1)}\right),\ee
where we have used the definitions $\Pi\equiv\partial_\mu\partial_\nu\phi$, as well as $\Pi^0_{\mu\nu}\equiv \eta_{\mu\nu}$ and $\Pi^{-1}_{\mu\nu}\equiv 0$, and the $X^{(n)}_{\mu\nu}$ are the identically conserved combinations of $\partial_\mu\partial_\nu\hat\phi$ described in Appendix \ref{totalDappendix}.  Thus we have 
\be \bar X^{\mu\nu}={1\over 2}\sum_{n\geq2}\alpha_n\left(X_{\mu\nu}^{(n)}+nX_{\mu\nu}^{(n-1)}\right).\ee
For $D=4$ this agrees with (\ref{lambda3decoupling}), showing that (\ref{lambda3decoupling}) contains all the scalar and tensor terms of the decoupling limit.  Some other re-summations are discussed in \cite{Hassan:2011vm,Nieuwenhuizen:2011sq}.

\subsection{The appearance of galileons and the absence of ghosts}

We can partially diagonalize the interaction terms in (\ref{lambda3decoupling}) by using the properties (\ref{Eopergalrel}).  
First, we perform the conformal transformation needed to diagonalize the linear terms, $\hat h_{\mu\nu}\rightarrow \hat h_{\mu\nu}+\hat\phi\eta_{\mu\nu}$, after which the lagrangian takes the form
\bea \label{lambda3decouplingdiag1} S=\int d^4x &&\frac{1}{2} \hat h_{\mu\nu}{\cal E}^{\mu\nu,\alpha\beta} \hat h_{\alpha\beta}-{1\over 2} \hat h^{\mu\nu}\left[{4(6c_3-1)\over \Lambda_3^3} \hat X^{(2)}_{\mu\nu}+{16(8d_5+c_3)\over \Lambda_3^6} \hat X^{(3)}_{\mu\nu}\right]+{1\over M_P} \hat h_{\mu\nu}T^{\mu\nu} \nn\\
&& -3(\partial \hat\phi)^2+{6(6c_3-1)\over \Lambda_3^3}(\partial \hat\phi)^2\square \hat\phi+{16(8d_5+c_3)\over \Lambda_3^6}(\partial \hat\phi)^2\([\hat\Pi]^2-[\hat\Pi^2]\)+{1\over M_P} \hat\phi T.\nn\\
\eea
Here the brackets are traces of $\hat\Pi_{\mu\nu}\equiv\partial_\mu\partial_\nu\hat\pi$ and its powers (the notation is explained at the end of the Introduction).

The cubic $h\phi\phi$ couplings can be eliminated with a field redefinition $\hat h_{\mu\nu}\rightarrow \hat h_{\mu\nu}+{2(6c_3-1)\over \Lambda_3^3}\partial_\mu \hat\phi\partial_\nu \hat\phi$, after which the lagrangian reads,
\bea \label{lambda3decouplingdiag2}S=\int d^4x &&\frac{1}{2} \hat h_{\mu\nu}{\cal E}^{\mu\nu,\alpha\beta} \hat h_{\alpha\beta}-{8(8d_5+c_3)\over \Lambda_3^6} \hat h^{\mu\nu} \hat X^{(3)}_{\mu\nu}+{1\over M_P} \hat h_{\mu\nu}T^{\mu\nu} \nn\\
&& -3(\partial \hat\phi)^2+{6(6c_3-1)\over \Lambda_3^3}(\partial \hat\phi)^2\square \hat\phi-4{(6c_3-1)^2-4(8d_5+c_3)\over \Lambda_3^6}(\partial \hat\phi)^2\([\hat\Pi]^2-[\hat\Pi^2]\) \nn \\
&&-{40(6c_3-1)(8d_5+c_3)\over \Lambda_3^9}(\partial \hat\phi)^2\([\hat\Pi]^3-3[\hat\Pi^2][\hat\Pi]+2[\hat\Pi^3]\) \nn \\
&&+{1\over M_P} \hat\phi T+{2(6c_3-1)\over \Lambda_3^3M_P}\partial_\mu \hat\phi\partial_\nu \hat\phi T^{\mu\nu}.\nn\\
\eea
There is no local field redefinition that can eliminate the $h\phi\phi\phi$ quartic mixing (there is a non-local redefinition that can do it), so this is as unmixed as the lagrangian can get while staying local.  

The scalar self-interactions in (\ref{lambda3decouplingdiag2}) are given by the following four lagrangians,
\begin{align}
\mathcal{L}_{2}&=-{1\over 2}(\partial\phi)^2 \ , \nn \\
\mathcal{L}_{3}&=-{1\over 2}(\partial\phi)^2[\Pi] \ ,\nn  \\
\mathcal{L}_{4}& =-{1\over 2}(\partial\phi)^2\left([\Pi]^2-[\Pi^2]\right) \ ,\nn \\
\mathcal{L}_{5}& =-{1\over 2}(\partial\phi)^2\left([\Pi]^3-3[\Pi][\Pi^2]+2[\Pi^3]\right)\ .
\label{normalGalileons}
\end{align}
 These are known as the \textit{galileon} terms \cite{Nicolis:2008in} (see also Section II of \cite{Hinterbichler:2010xn} for a summary of the galileons).  They share two special properties: their equations of motion are purely second order (despite the appearance of higher derivative terms in the lagrangians), and they are invariant up to a total derivative under the galilean symmetry (\ref{galileansym}), $\phi(x)\rightarrow \phi(x)+c+b_\mu x^\mu$.  As shown in \cite{Nicolis:2008in}, the terms (\ref{normalGalileons}) are the only polynomial terms in four dimensions with these properties.
 
 The galileon was first discovered in studies of the DGP brane world model \cite{Dvali:2000hr} (which we will explore in more detail in Section \ref{DGPsection}), for which the cubic galileon, ${\cal L}_3$, was found to describe the leading interactions of the brane bending mode \cite{Luty:2003vm,Nicolis:2004qq}.  The rest of the galileons were then discovered in \cite{Nicolis:2008in}, by abstracting the properties of the cubic term away from DGP.  They have some other very interesting properties, such as a non-renormalization theorem (see e.g. Section VI of \cite{Hinterbichler:2010xn}), and a connection to the Lovelock invariants through brane embedding \cite{deRham:2010eu}.  Due to these unexpected and interesting properties, they have since taken on a life of their own.  They have been generalized in many directions \cite{Deffayet:2009mn,Deffayet:2009wt,Deffayet:2010zh,Deffayet:2011gz,Goon:2011uw,Khoury:2011da,Padilla:2010ir}, and are the subject of much recent activity (see for instance the $>100$ papers citing \cite{Nicolis:2008in}).

The fact that the equations are second order ensures that, unlike (\ref{phiaction}), no extra degrees of freedom propagate.  In fact, as pointed out in \cite{deRham:2010ik}, the properties (\ref{xtensorprops}) of the tensors $X_{\mu\nu}$ guarantee that there are no ghosts in the lagrangian (\ref{lambda3decoupling}) of the decoupling limit theory.\footnote{
This is contrary to \cite{Creminelli:2005qk}, which claims that a ghost is still present at quartic order.  As remarked however in \cite{deRham:2010ik}, they arrive at the incorrect decoupling limit lagrangian, which can be traced to a minus sign mistake in their Equation 5, which should be as in (\ref{hscalarreplace}).}  By going through a hamiltonian analysis similar to that of Section \ref{canonicalanalysis}, we can see that $h_{00}$ and $h_{0i}$ remain Lagrange multipliers enforcing first class constraints (as they should since the lagrangian (\ref{lambda3decoupling}) is gauge invariant.  In addition, the equations of motion remain second order, so the decoupling limit lagrangian (\ref{lambda3decoupling}) is free of the Boulware-Deser ghost and propagates 3 degrees of freedom around any background. 

Once the two degrees of freedom of the vector $A_\mu$ are included, and if there are no ghosts in the vector part or its interactions, the total number degrees of freedom goes to 5, the same as the linear massive graviton.  The vector interactions were shown to be ghost free at cubic order in \cite{deRham:2010gu}.   It was shown in \cite{deRham:2010kj} that the full theory beyond the decoupling limit, including all the fields, is ghost free, up to quartic order in the fields.  This guarantees that any ghost must carry a mass scale larger than $\Lambda_3$ and hence can be consistently excluded from the quantum theory.  Finally, in \cite{Hassan:2011hr} it was shown using the hamiltonian formalism that the full theory, including all modes and to all orders beyond the decoupling limit, carries 5 degrees of freedom.  The $\Lambda_3$ theory is therefore free of the Boulware-Deser ghost, around any background.  This can also been seen in the St\"ukelberg language \cite{deRham:2011rn}.  

\subsection{The $\Lambda_3$ Vainshtein radius}

We can now derive the scale at which the linear expansion breaks down around heavy point sources in the $\Lambda_3$ theory.  To linear order around a central source of mass $M$, the fields still have their usual Coulomb form, 
\be\hat \phi, \hat h\sim {M\over M_P}{1\over r}.\ee
The non-linear terms in (\ref{lambda3decoupling}) or (\ref{lambda3decouplingdiag2}) are suppressed relative to the linear term by a different factor than in the $\Lambda_5$ theory, 
\be {\partial^2 \hat\phi\over \Lambda_3^3}\sim {M\over M_P}{1\over \Lambda_3^3 r^3}.\ee
Non-linearities become important when this factor becomes of order one, which happens at the radius 
\be \label{vaisnteinr3} r_V^{(3)}\sim \left(M\over M_P\right)^{1/3}{1\over \Lambda_3}\sim \left(GM\over m^2\right)^{1/3}.\ee
This is parametrically larger than the Vainshtein radius found in the $\Lambda_5$ theory.

It is important that the decoupling limit lagrangian was ghost free. To see what would go wrong if there were a ghost, expand around some spherical background $\hat\phi=\Phi(r)+\varphi$, and similarly for $h_{\mu\nu}$.  The cubic coupling and quartic couplings could possibly give fourth order kinetic contributions of the schematic form, respectively,
\be {1\over \Lambda_3^3}\Phi(\partial^2\varphi)^2,\ \ \ {1\over \Lambda_3^6}\Phi\partial^2\Phi(\partial^2\varphi)^2.\ee
These would correspond to ghosts with $r$-dependent masses,
\be m^2_{\rm ghost}(r)\sim {\Lambda_3^3\over \Phi},\ \ \ { \Lambda_3^6\over \Phi\partial^2\Phi}.\ee
or, given that the background fields go like $\Phi\sim {M\over M_P}{1\over r}$,
\be  m^2_{\rm ghost}(r)\sim {M_P\over M}\Lambda_3^3r,\ \ \ \left(M_P\over M\right)^2\Lambda_3^6r^4.\ee
Thus the ghost mass sinks below the cutoff $\Lambda_3$ at the radius
\be r_{\rm ghost}^{(3)}\sim \left(M\over M_P\right){1\over \Lambda_3},\ \  \left(M\over M_P\right)^{1/2}{1\over \Lambda_3}.\ee
As happened in the $\Lambda_5$ theory, these radii are parametrically larger than the Vainshtein radius.  This is a fatal instability which renders the whole non-linear region inaccessible, unless we lower the cutoff of the effective theory so that the ghost stays above it, in which case unknown quantum corrections would also kick in at $\sim r_{\rm ghost}^{(3)}$, swamping the entire non-linear Vainshtein region.

\subsection{The Vainshtein mechanism in the $\Lambda_3$ theory}

In the $\Lambda_5$ theory, the key to the resolution of the vDVZ discontinuity and recovery of GR was the activation of the Boulware-Deser ghost, which cancelled the force due to the longitudinal mode.  In the $\Lambda_3$ theory, there is no ghost (at least in the decoupling limit), so there must be some other method by which the scalar screens itself to restore continuity with general relativity.  This method uses non-linearities to enlarge the kinetic terms of the scalar, rendering its couplings small.

To see how this works, consider the lagrangian in the form (\ref{lambda3decouplingdiag1}).  Set $d_5=-c_3/8$, $c_3=5/36$ to simplify coefficients, and ignore for a second the cubic $h\phi\phi$ coupling, so that we only have a cubic $\phi$ self-interaction governed by the galileon term ${\cal L}_3$,
\be \label{DGPcubic}S=\int d^4x\ -3(\partial \hat\phi)^2-{1\over \Lambda_3^3}(\partial\hat\phi)^2\square \hat\phi+{1\over M_4} \hat\phi T.\ee
This is the same lagrangian studied in \cite{Nicolis:2004qq} in the DGP context.  

Consider the static spherically symmetric solution, $\hat\phi(r)$, around a point source of mass $M$, $T\sim M\delta^3(r)$.  The solution transitions, at the Vainshtein radius $r_V^{(3)}\equiv\left(M\over M_{Pl}\right)^{1/3}{1\over \Lambda_3}$, between a linear and non-linear regime.  For $r\gg r_V^{(3)}$ the kinetic term in (\ref{DGPcubic}) dominates over the cubic term, linearities are unimportant, and we get the usual $1/r$ Coulomb behavior.  For $r\ll r_V^{(3)}$, the cubic term is dominant, and we get a non-linear $\sqrt{r}$ potential,
\be \label{sphericalsol}
\hat\phi(r)\sim \begin{cases} \Lambda_3^3 {r_V^{(3)}}^{2} \left(\frac{r}{r_V^{(3)}}\right)^{1/2} & r\ll r_V^{(3)}, \\ \Lambda_3^3 {r_V^{(3)}}^2 \left(\frac{r_V^{(3)}}{r}\right) & r\gg r_V^{(3)}.\end{cases} \ 
\ee

We can see the Vainshtein mechanism at work already by calculating the ratio of the fifth force due to the scalar to the force from  ordinary newtonian gravity,
\be {F_\phi\over F_{\rm Newton}}={\hat\phi'(r)/M_P\over M/(M_P^2r^2)}= \begin{cases}\sim  \(r\over r_V^{(3)}\)^{3/2} & r\ll r_V^{(3)}, \\ \sim 1 & r\gg r_V^{(3)}. \end{cases}\ee
There is a gravitational strength fifth force at distances much farther than the Vainshtein radius, but the force is suppressed at distances smaller than the Vainshtein radius.  

This suppression extends to all scalar interactions in the presence of the source.  To see how this comes about, we study perturbations around a given background solution $\Phi(x)$.  Expanding 
\be \hat \phi=\Phi+\varphi, \ \ \ T=T_0+\delta T,\ee
we have after using the identity $(\partial^\mu\varphi)\square\varphi=\partial_\nu\left[\partial^\nu\varphi\partial^\mu\varphi-{1\over 2}\eta^{\mu\nu}(\partial\varphi)^2\right]$ on the quadratic parts and integrating by parts
\be \label{expandedpi} S_\varphi=\int d^4x\ -3(\partial\varphi)^2+{2\over \Lambda^3}\left(\partial_\mu\partial_\nu \Phi-\eta_{\mu\nu}\square \Phi\right)\partial^\mu\varphi\partial^\nu\varphi-{1\over \Lambda^3}(\partial\varphi)^2\square\varphi+{1\over M_4}\varphi \delta T.\ee
Note that expanding the cubic term yields new contributions to the kinetic terms, with coefficients that depend on the background.  Unlike the $\Lambda_5$ lagrangian (\ref{phiaction}), no higher derivative kinetic terms are generated, so no extra degrees of freedom are propagated on any background.  This is a property shared by all the galileon lagrangians (\ref{normalGalileons}) \cite{Endlich:2010zj}.

Around the solution (\ref{sphericalsol}), the coefficient of the kinetic term in (\ref{expandedpi}) is ${\cal O}(1)$ at distances $r\gg r_V^{(3)}$, but goes like $\(r_V^{(3)}\over r\)^{3/2}$ for distances $r\ll r_V^{(3)}$.  Thus the kinetic term is enhanced at distances below the Vainshtein radius, which means that after canonical normalization the couplings of the fluctuations to the source are reduced.  The fluctuations $\varphi$ effectively decouple near a large source, so the scalar force between two small test particles in the presence of a large source is reduced, and continuity with GR is restored.  A more careful study of the Vainshtein screening in the $\Lambda_3$ theory, including numerical solutions of the decoupling limit action, can be found in \cite{Chkareuli:2011te}.

\subsection{Quantum corrections in the $\Lambda_3$ theory}

As in Section \ref{quantumL5}, we expect quantum mechanically the presence of all operators with at least two derivatives per $\phi$, now suppressed by the cutoff $\Lambda_3$ (we ignore for simplicity the scalar tensor interactions),
\begin{equation}\label{pqscalarterm3}
\sim \frac{\partial^q (\partial^2 \hat\phi)^p}{\Lambda_3^{3p + q - 4}}.
\end{equation}
These are in addition to the classical galileon terms in (\ref{lambda3decouplingdiag2}), which have fewer derivatives per $\phi$, and are of the form 
\be \label{galterms} \sim \frac{(\partial\hat\phi)^2 (\partial^2 \hat\phi)^p}{\Lambda_3^{3p}}.\ee
An analysis just like that of Section \ref{quantumL5} shows that the terms (\ref{pqscalarterm3}) become important relative to the kinetic term at the radius $r \sim \left(\frac{M}{M_{Pl}}\right)^{1/3}{1\over \Lambda_3}$.  This is the same radius at which classical non-linear effects due to (\ref{galterms}) become important and alter the solution from its Coulomb form.  Thus we must instead compare the terms (\ref{pqscalarterm3}) to the classical non-linear galileon terms (\ref{galterms}).   We see that the terms (\ref{pqscalarterm3}) are all suppressed relative to the galileon terms (\ref{galterms}) by powers of $\partial/\Lambda_3$, which is $\sim {1\over \Lambda_3 r}$ regardless of the non-linear solution.  Thus, quantum effects do not become important until the radius
\be r_Q\sim  {1\over \Lambda_3},\ee
which is parametrically smaller than the Vainshtein radius (\ref{vaisnteinr3}).

This behavior is much improved from that of the $\Lambda_5$ theory, in which the Vainshtein region was swamped by quantum correction.  Here, there is a parametrically large intermediate classical region in which non-linearities are important but quantum effects are not, and in which the Vainshtein mechanism should screen the extra scalar.  In this region, GR should be a good approximation.  See Figure (\ref{massive3regimes}).  

\begin{figure}[h!]
\begin{center}
\epsfig{file=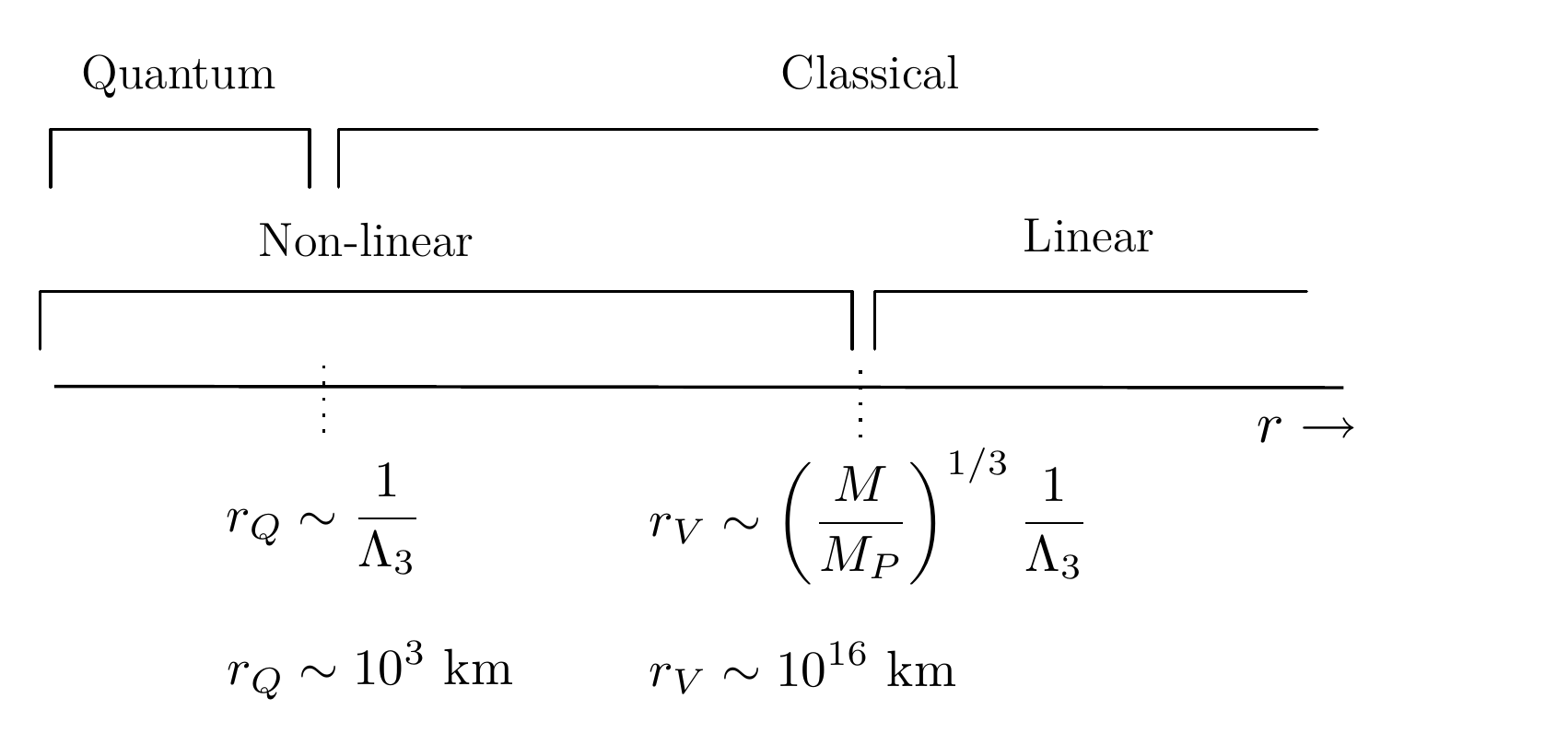,height=1.8in,width=4.0in}
\caption{\small Regimes for massive gravity with cutoff $\Lambda_3=(M_Pm^2)^{1/3}$, and some values within the solar system.  The values are much more reasonable than those of the $\Lambda_5$ theory.}
\label{massive3regimes}
\end{center}
\end{figure}

As in the $\Lambda_5$ theory, quantum corrections are generically expected to ruin the various classical tunings for the coefficients, but the tunings are still technically natural because the corrections are parametrically small.  For example, cutting off loops by $\Lambda_3$, we generate the operator $ \sim {1\over \Lambda_3^2}(\square\hat\phi)^2,$ which corrects the mass term.   The canonically normalized $\hat\phi$ is related to the original dimensionless metric by $h\sim{1\over \Lambda_3^3}\partial\partial\hat\phi,$ so the generated term corresponds in unitary gauge to
 $\Lambda_3 ^4 h^2 = M_p^2 m^2\left(\Lambda_3\over M_p\right) h^2$, representing a mass correction $\delta m^2\sim m^2\left(\Lambda_3\over M_p\right)$.  This mass correction is parametrically smaller than the mass itself and so the hierarchy $m\ll \Lambda_3$ is technically natural.  This correction also ruins the Fierz-Pauli tuning, but the pathology associated with the de-tuning of Fierz-Pauli, the ghost mass, is $ m_g^2\sim {m^2\over \delta m^2/m^2}\sim\Lambda_3^2,$ safely at the cutoff.  

We should mention another potential issue with the $\Lambda_3$ theory.  It was found in \cite{Nicolis:2008in} that lagrangians of the galileon type inevitably have superluminal propagation around spherical background solutions.  No matter what the choice of parameters in the lagrangian, if the solution is stable, then superluminality is always present at distances far enough from the source (see also \cite{Osipov:2008dd}). It has been argued that such superluminality is a sign that the theory cannot be UV completed by a standard local Lorentz invariant theory \cite{Adams:2006sv}, though others have argued that this is not a problem \cite{Babichev:2007dw}.  In addition, the analysis of \cite{Nicolis:2008in} was for pure galileons only, and the scalar-tensor couplings of the massive gravity lagrangian can potentially change the story.  These issues have been studied within massive gravity in \cite{deRham:2011pt}.

\section{Massive gravity from extra dimensions}

So far, we have stuck to the effective field theorist's philosophy.  We have explored the possibility of a massive graviton by simply writing down the most general mass term a graviton may have, remaining agnostic as to its origin.  However, it is important to ask whether such a mass term has a top down construction or embedding into a wider structure, one which would determine the coefficients of all the various interactions.  This goes back to the question of whether it is possible to UV complete (or UV extend) the effective field theory of a massive graviton.

One way in which a massive graviton naturally arises is from higher dimensions.  We will now study several of these higher dimensional scenarios, the Kaluza-Klein reduction, the DGP brane world model, and a model of a non-dynamical auxiliary extra dimension, showing in each case how massive gravitons emerge in a 4d description.

\subsection{Kaluza-Klein theory}

 In the original Kaluza-Klein idea \cite{Kaluza:1921tu,Klein:1926tv} (see \cite{Overduin:1998pn} for a review),  gravity on a 5d space with a single compact direction is dimensionally reduced onto the four non-compact dimensions, where it is found that the lightest modes describe Einstein gravity, electromagnetism, and a massless scalar, all in interaction with each other.  In almost all work on Kaluza-Klein theory (including the rather large subset going by the name of string compactifications), only the lowest energy modes are considered.

Beyond the lowest energy modes, there is an entire tower of massive fields.  In the dimensional reduction of gravity, this tower will consist of massive gravitons.  We will now review the dimensional reduction of pure 5d gravity down to four dimensions.  We will work at the linear level, keeping all the massive modes, and we will see that the massive gravitons which arise are described by the 4d part of the 5d metric obeying precisely the Fierz-Pauli mass term (\ref{massivefreeaction}).  The Fierz-Pauli tuning of coefficients arises automatically from the dimensional reduction.  In addition, the 5d components of the 5d metric become a tower of 4d scalars and a tower of 4d vectors.

There is also the question of gauge symmetry.  The 5d gravity action has 5d diffeomorphism invariance.  The result of the reduction, a tower of massive gravitons in 4d, has no diffeomorphism symmetry, so where does this symmetry go?  We will see that what comes out of the reduction is not the unitary gauge action (\ref{massivefreeaction}), but rather the St\"ukelberg-ed action (\ref{massivenolimit}).  The 5d gauge symmetry becomes the 4d St\"ukelberg gauge symmetry, and the towers of vectors and scalars become the St\"ukelberg fields.

We start with a massless graviton in 5 dimensions, with 5d Planck mass $M_5$,
\be \label{5dmassless} S=M_5^3\int d^5 X -\frac{1}{2}\partial_C H_{AB}\partial^C H^{AB}+\partial_A H_{BC}\partial^B H^{AC}-\partial_A H^{AB}\partial_B H+\frac{1}{2}\partial_A H\partial^A H+ {1\over M_5^3}H_{AB}T^{AB}.
\ee
Here $H_{AB}$ is the dimensionless 5d graviton, with indices $A,B,\ldots$ running over 5d spacetime.  We divide spacetime into 4d coordinates $x^\mu$, and a fifth coordinate $y$,  so that $X^A=\left(x^\mu,y\right)$.  We compactify $y$ so that it runs along a circle of circumference $L$,  $y\in(0,L)$.  $T^{AB}$ is the fixed external 5d stress tensor, which is conserved in 5d, $\partial_BT^{AB}=0$.

Now we change variables by expanding in a Fourier series over the circle,
\bea H_{\mu\nu}(x,y)&=&\sum_{n=-\infty}^\infty h_{\mu\nu,n}(x)e^{i\omega_n y},\nn \\
H_{\mu y}(x,y)&=&\sum_{n=-\infty}^\infty A_{\mu,n}(x)e^{i\omega_n y}, \nn\\
H_{yy}(x,y)&=&\sum_{n=-\infty}^\infty\phi_n(x)e^{i\omega_n y}. \label{varchangekk}
\eea
Here $n$ is an integer, $\omega_n\equiv {2\pi n\over L}$, and we have the usual orthogonality relation $\int_0^Ldy \left(e^{i\omega_m y}\right)^\ast e^{i\omega_n y}=L\delta_{mn}$.

The coefficients in the Fourier expansion, $\phi_n$, $h_{\mu\nu,n}$, and $A_{\mu,n}$, are the new variables and will become the 4d fields.  Reality of the 5d fields imposes the conditions
\be \label{4drealityconditions} h_{\mu\nu,n}^\ast=h_{\mu\nu,-n},\ \ \ A_{\mu,n}^\ast=A_{\mu,-n},\ \ \ \phi_n^\ast=\phi_{-n}.\ee

In addition, we decompose the 5d stress tensor in similar fashion,
\bea T_{\mu\nu}(x,y)&=&\sum_{n=-\infty}^\infty t_{\mu\nu,n}(x)e^{i\omega_n y}, \\
T_{\mu y}(x,y)&=&\sum_{n=-\infty}^\infty j_{\mu,n}(x)e^{i\omega_n y}, \\
T_{yy}(x,y)&=&\sum_{n=-\infty}^\infty j_n(x)e^{i\omega_n y}.
\eea 
The fields $t_{\mu\nu,n}$, $j_{\mu,n}$ and $j_n$, which satisfy reality conditions just like (\ref{4drealityconditions}), 
\be \label{4drealityconditionst} t_{\mu\nu,n}^\ast=t_{\mu\nu,-n},\ \ \ j_{\mu,n}^\ast=j_{\mu,-n},\ \ \ j_n^\ast=j_{-n},\ee
will become the 4d sources.  The equation for 5d stress tensor conservation, $\partial^BT_{AB}=0$, when expanded out in components and in the Fourier series, implies 
\bea &&\partial^\mu j_{\mu,0}=0,\ \ \ \partial^\nu t_{\mu\nu,0}=0, \\
&& j_{\mu,n}={i\over \omega_n}\partial^\nu t_{\mu\nu,n},\ \ \ j_n=-{1\over \omega_n^2}\partial^\mu\partial^\nu t_{\mu\nu,n},\ \ \ \ n\not=0.
\eea

Plugging the Fourier expansions into (\ref{5dmassless}) and doing the $y$ integral, we get the following equivalent 4d action,
\bea S=LM_5^3\int d^4x &&\half h_{\mu\nu,0}{\cal E}^{\mu\nu,\alpha\beta}h_{\alpha\beta,0}-\half F_{\mu\nu,0}^2+h^{\mu\nu}_{0}\left(\partial_\mu\partial_\nu \phi_0-\eta_{\mu\nu}\phi_0\right)\nn \\ 
&&+{1\over M_5^3}h_{\mu\nu,0}t^{\mu\nu}_0+{2\over M_5^3}A_{\mu,0}j^{\mu}_0+{1\over M_5^3}\phi_0j_0\nn \\
&&+\sum_{n=1}^\infty h_{\mu\nu,n}^\ast{\cal E}^{\mu\nu,\alpha\beta}h_{\alpha\beta,n}-\omega_n^2\left(\left|h_{\mu\nu,n}\right|^2-\left| h_n \right|^2\right)-\left|F_{\mu\nu,n}\right|^2 \nn \\
&& +\left[2i\omega_nA_{\mu,n}\left(\partial_\nu h^{\mu\nu\ast}_n-\partial^\mu h_n^\ast\right)+h^{\mu\nu}_n\left(\partial_\mu\partial_\nu\phi_n^\ast-\eta_{\mu\nu}\square\phi_n^\ast\right)+c.c\right] \nn \\
&&+\sum_{n=1}^\infty\left[{1\over M_5^3} h_{\mu\nu,n} t^{\mu\nu\ast}_n+{2\over M_5^3} A_{\mu,n} j^{\mu\ast}_n+{1\over M_5^3}\phi_n j_n^\ast+c.c.\right] .\nn \\ \label{KKreducedaction}
\eea
We have used the reality conditions (\ref{4drealityconditions}) and (\ref{4drealityconditionst}) to change the range of the sum.  This action is exactly equivalent to (\ref{5dmassless}), and describes all the 5d dynamics.  We have not truncated anything or restricted the fields in any way, we have merely changed variables to ones that are more easily recognizable in 4d. 

 From the prefactor we can read off the effective 4d Planck mass
\be M_4^2=LM_5^3.\ee

We now study the fate of the gauge symmetry.  The 5d action has the gauge symmetry $\delta H_{AB}=\partial_A\Xi_B+\partial_B\Xi_A$, for a gauge vector $\Xi_A(X)$.  Fourier decomposing the gauge parameter,
\bea \Xi_\mu(x,y)&=&\sum_{n=-\infty}^\infty\xi_{\mu,n}e^{i\omega y},\\
 \Xi_y(x,y)&=&\sum_{n=-\infty}^\infty\xi_{n}e^{i\omega y},
\eea
where the coefficients have reality properties like those of (\ref{4drealityconditions}). 
The gauge transformations can be decomposed component by component to yield
\bea \delta h_{\mu\nu,n}&=&\partial_\mu \xi_{\nu,n}+\partial_\nu \xi_{\mu,n},\\
 \delta A_{\mu,n}&=&\partial_\mu\xi_n+i\omega_n\xi_{\mu,n}, \\
  \delta \phi_{\mu\nu,n}&=&2i\omega_n\xi_n.
  \eea
The zero mode of the 5d gauge parameter $\Xi_A$ breaks up into a vector and a scalar, which become the linear diffeomorphism invariance $\xi_{\mu,0}$ of the zero mode graviton $h_{\mu\nu,0}$, and the Maxwell gauge invariance $\xi_0$ of the zero mode vector $A_{\mu,0}$ (the zero mode scalar $\phi_0$ is gauge invariant).  The $n\not=0$ modes, on the other hand, get transformations of exactly the St\"ukelberg form (\ref{massivenolimitstuk}).  The 5d gauge symmetry has become the 4d St\"ukelberg symmetry.
  
In fact, the action (\ref{KKreducedaction}) can be written solely in terms of the following gauge invariant combination for $n\not=0$,
\be h_{\mu\nu,n}+{i\over \omega_n}\left(\partial_\mu A_{\nu,n}+\partial_\nu A_{\mu,n}\right)-{1\over \omega_n^2}\partial_\mu\partial_\nu\phi_n,\ee
which is just the linear St\"ukelberg replacement rule.
The action (\ref{KKreducedaction}) for the $n\not=0$ modes is precisely the complex version of our St\"ukelberg action (\ref{massivenolimit}) for a massive graviton.  The higher modes of the $\mu 5$ and $55$ components of the 5d metric $H_{AB}$ have become the non-physical St\"ukelberg fields, and are pure gauge.

Fixing the unitary gauge $A_{\mu,n}=\phi_n=0$ for $n\not=0$, and canonically normalizing, we have
\bea S= \int d^4x &&\half h_{\mu\nu,0}{\cal E}^{\mu\nu,\alpha\beta}h_{\alpha\beta,0}-\half F_{\mu\nu,0}^2+h^{\mu\nu}_{0}\left(\partial_\mu\partial_\nu \phi_0-\eta_{\mu\nu}\phi_0\right)\nn \\ 
&&+{1\over M_4}h_{\mu\nu,0}t^{\mu\nu}_0+{2\over M_4}A_{\mu,0}j^{\mu}_0+{1\over M_4}\phi_0j_0\nn \\
&&+\sum_{n=1}^\infty h_{\mu\nu,n}^{'\ast}{\cal E}^{\mu\nu,\alpha\beta}h'_{\alpha\beta,n}-\omega_n^2\left(\left|h'_{\mu\nu,n}\right|^2-\left| h'_n \right|^2\right)+\left[{1\over M_4} h'_{\mu\nu,n} t^{\mu\nu\ast}_n+c.c.\right]  ,\nn \\
\eea
which shows that the theory (after the conformal transformation (\ref{conformalt}) for the zero modes), consists of a single real massless graviton, a single real massless vector, a single real massless scalar, and a tower of complex massive Fierz-Pauli gravitons with masses
\be m_n=\omega_n={2\pi n\over L},\ \ \ n=1,2,3,\cdots.\ee
These fields are all coupled to the various modes and components of the 5d stress tensor.  In addition, it can be easily shown that the translation symmetry along $y$ in the original theory becomes an internal $U(1)$ rotating the phase of the massive gravitons.  There are interesting issues that arise when one wishes to couple this $U(1)$ to electromagnetism in the case of a single massive graviton \cite{Porrati:2008gv}.

To go beyond linear order, we would put the higher order in $H$ interactions coming from the 5d Einstein-Hilbert term into the 5d action \ref{5dmassless}, make the same change of variables into Fourier components (\ref{varchangekk}), then plug in and do the $y$ integral.  This will give a slew of interaction terms in 4d, involving all the modes interacting with each other.  This should be a consistent, stable, ghost free theory of an infinite number of fully interacting massive gravitons, since it is equivalent to 5d Einstein gravity which we know to be consistent.  There should be no strong coupling problems or low scale cutoffs, and the effective theory should be valid all the way up to the 5d Planck mass.  All the 4d graviton modes should miraculously interact in such a way as to cancel out all the strong coupling effects we have uncovered for a single massive graviton \cite{Schwartz:2003vj}.  It is possible to write these interactions for all the fields to all orders in closed form \cite{Aulakh:1985un,Maheshwari:1985ux,Kol:2010si}.

It turns out to be consistent to truncate the theory to only the zero modes (consistent in the sense that the processes of truncating and deriving the equations of motion commute).  This leaves 4d Einstein-Hilbert self-interactions for the zero mode graviton, in addition to other interactions between the various zero mode fields.  The ansatz that describes this truncation, which was the original Kaluza-Klein ansatz, is to take the 5d metric to be independent of $y$.  It would be desirable to find a consistent truncation that involves only a single massive graviton (or a finite number) so that we could study the resulting consistent interactions.  This does not appear to be possible in this simple model, but may be possible in compactifications involving more complicated manifolds or sets of fields.

\subsection{\label{DGPsection}DGP and the resonance graviton}

The Dvali-Gabadadze-Porrati (DGP) model \cite{Dvali:2000hr} is an extra-dimensional model which has spawned a great deal of interest (see the $>1300$ papers citing \cite{Dvali:2000hr}).  It provides another, more novel realization of a graviton mass.  Unlike the Kaluza-Klein scenario, in DGP the extra dimensions can be infinite in extent, though there must be a brane on which to confine standard model matter (see \cite{Gabadadze:2003ii} for lectures on large extra dimensions).  By integrating out the extra dimensions, we can write an effective 4d action for this scenario which contains a momentum dependent mass term for the graviton.  This provides an example of a graviton resonance, i.e. a continuum of massive gravitons.  

Another model which has revived a great deal of attention ($>4800$ citations) is the Randall-Sundrum brane world \cite{Randall:1999ee}, in which there is a brane floating in large warped extra dimensions.  This model is not as interesting from the point of view of massive gravity at low energies, since the 4d spectrum is similar to ordinary Kaluza-Klein theory, containing ordinary Einstein gravity as a zero mode, and then massive gravitons as higher Kaluza-Klein modes.   See \cite{Maartens:2010ar} for a review on brane world gravity.

\subsubsection*{The DGP action}

 DGP is the model of a $3+1$ dimensional brane (the 3-brane) floating in a $4+1$ dimensional bulk spacetime.  Gravity is dynamical in the bulk and the brane position is dynamical as well, and the action contains both 4d and 5d parts,
 \be \label{DGPaction} S={M_5^3\over 2}\int d^5X\sqrt{-G}R(G)+{M_4^2\over 2}\int d^4x\sqrt{-g}R(g)+S_M.\ee
 Here $X^A$, $A,B,\cdots=0,1,2,3,5$ are the 5d bulk coordinates, $G_{AB}(X)$ is the 5d metric, and $M_5$ is the 5d Planck mass.  $x^\mu$, $\mu,\nu,\ldots=0,1,2,3$ are the 4d brane coordinates, $g_{\mu\nu}(x)$ is the 4d metric which is given by inducing the 5d metric $G_{AB}$ onto the brane, and $M_4$ is the 4d Planck mass.  $S_M$ is the matter action, which we imagine to be localized to the brane,
 \be S_M=\int d^4x \ {\cal L}_M(g,\psi),\ee
 where $\psi(x)$ are the 4d matter fields.  Due to the presence of a brane Einstein-Hilbert term, this scenario is also called \textit{brane induced gravity} \cite{Gabadadze:2007dv}.
  
The dynamical variables are the 5d metric depending on the 5d coordinates, the embedding $X^A(x)$ of the brane depending on the 4d coordinates, and the 4d matter fields depending on the 4d coordinates,
\be G_{AB}(X),\ \ \ X^A(x),\ \ \ \psi(x).\ee
The 4d metric is not independent, but is fixed to be the pullback of the 5d metric,
\be g_{\mu\nu}(x)=\partial_\mu X^A\partial_\nu X^B G_{AB}\left(X(x)\right).\ee
 Note that the dependence of the action on the $X^A$ enters only through the induced metric $g_{\mu\nu}$.  
 
 The action (\ref{DGPaction}) has a lot of gauge symmetry.  First, there are the reparametrizations of the brane given by infinitesimal vector fields $\xi^\mu(x)$, under which the $X^A$ transform as scalars and the matter fields transform as tensors (i.e. with a Lie derivative), 
 \be \delta_\xi X^A=\xi^\mu\partial_\mu X^A,\ \ \ \delta_\xi \psi={\cal L}_\xi \psi.\ee
Second, there are reparametrizations of the bulk given by infinitesimal vector fields $\Xi^A(X)$, under which $G_{AB}$ transforms as a tensor and the $X^A$ shift, 
\be \delta_\Xi G_{AB}=\nabla_A\Xi_B+\nabla_B\Xi_A,\ \ \ \delta_\Xi X^A=-\Xi^A(X).\ee
The induced metric $g_{\mu\nu}$ transforms as a tensor under $\delta_\xi$, and is invariant under $\delta_\Xi$\footnote{To see invariance under $\delta_\Xi$, transform 
\bea \delta_\Xi G_{AB}&=&\delta_\Xi\left(\partial_\mu X^A\partial_\nu X^B G_{AB}\left(X(x)\right)\right)\nn\\
&=&-\partial_\mu \Xi^A\partial_\nu X^B G_{AB}\left(X(x)\right)-\partial_\mu X^A\partial_\nu \Xi^B G_{AB}\left(X(x)\right)+\partial_\mu X^A\partial_\nu X^B \delta_\Xi G_{AB}\left(X(x)\right),\eea
then in transforming $G_{AB}$, remember that both the function and the argument are changing, 
\be  \delta_\Xi G_{AB}\left(X(x)\right)={\cal L}_\Xi  G_{AB}\left(X(x)\right)-\Xi^C\partial_CG_{AB}.\ee
Putting all this together, we find $\delta_\Xi G_{AB}=0$.}.

We first proceed to fix some of this gauge symmetry.  In particular, we will freeze the position of the brane.  Note that the brane coordinate functions, $X^A(x)$, are essentially Goldstone bosons since they shift under the bulk gauge symmetry, $X^A(x)\rightarrow X^A(x)-\Xi^A\left(X(x)\right)$.  We can thus reach a sort of unitary gauge where the $X^A$ are fixed to some specified values.  We will set values so that the brane is the surface $X^5=0$, and the brane coordinates $x^\mu$ coincide with the coordinates $X^\mu$, thus we set
\bea X^\mu(x)&=&x^\mu,\ \ \ \mu=0,1,2,3, \\
X^5(x)&=&0. \label{DGPgaugex}
\eea

There are still residual gauge symmetries which leave this gauge choice invariant.  Acting with the two gauge transformations $\delta_\xi$ and $\delta_\Xi$ on the gauge conditions and demanding that the change be zero, we find
\bea \nn &&\delta_\Xi X^5(x)+\delta_\xi X^5(x)=-\Xi^5\left(X(x)\right)+\xi^\mu\partial_\mu X^5 \underset{X^5(x)=0}{\rightarrow}-\Xi^5\left(X(x)\right)\\
&&\Rightarrow \Xi^5\left(X(x)\right)=0.\\
 \nn &&\delta_\Xi X^\mu(x)+\delta_\xi X^\mu(x)=-\Xi^\mu\left(X(x)\right)+\xi^\nu\partial_\nu X^\mu(x) \underset{X^\mu(x)=x^\mu}{\rightarrow}-\Xi^\mu\left(X(x)\right)+\xi^\mu(x)\\
&& \Rightarrow \Xi^\mu\left(X(x)\right)=\xi^\mu(x).
\eea
The residual gauge transformations are bulk gauge transformations that do not move points onto or off of the brane, but only move brane points to other brane points.  Furthermore, the brane diffeomorphism invariance is no longer an independent invariance but is fixed to be the diffeomorphisms induced from the bulk.  

We now fix this gauge  in the action (\ref{DGPgaugex}), which is permissible since no equations of motion are lost.  This means that the induced metric is now 
\be g_{\mu\nu}(x)=G_{\mu\nu}(x,X^5=0).\ee
We split the action into two regions, region $L$ to the left of the brane, and region $R$ to the right of the brane, with outward pointing normals, as in Figure \ref{branevariation}.  We call the fifth coordinate $X^5\equiv y$.  The brane is at $y=0$.  

\begin{figure}[h!]
\begin{center}
\epsfig{file= 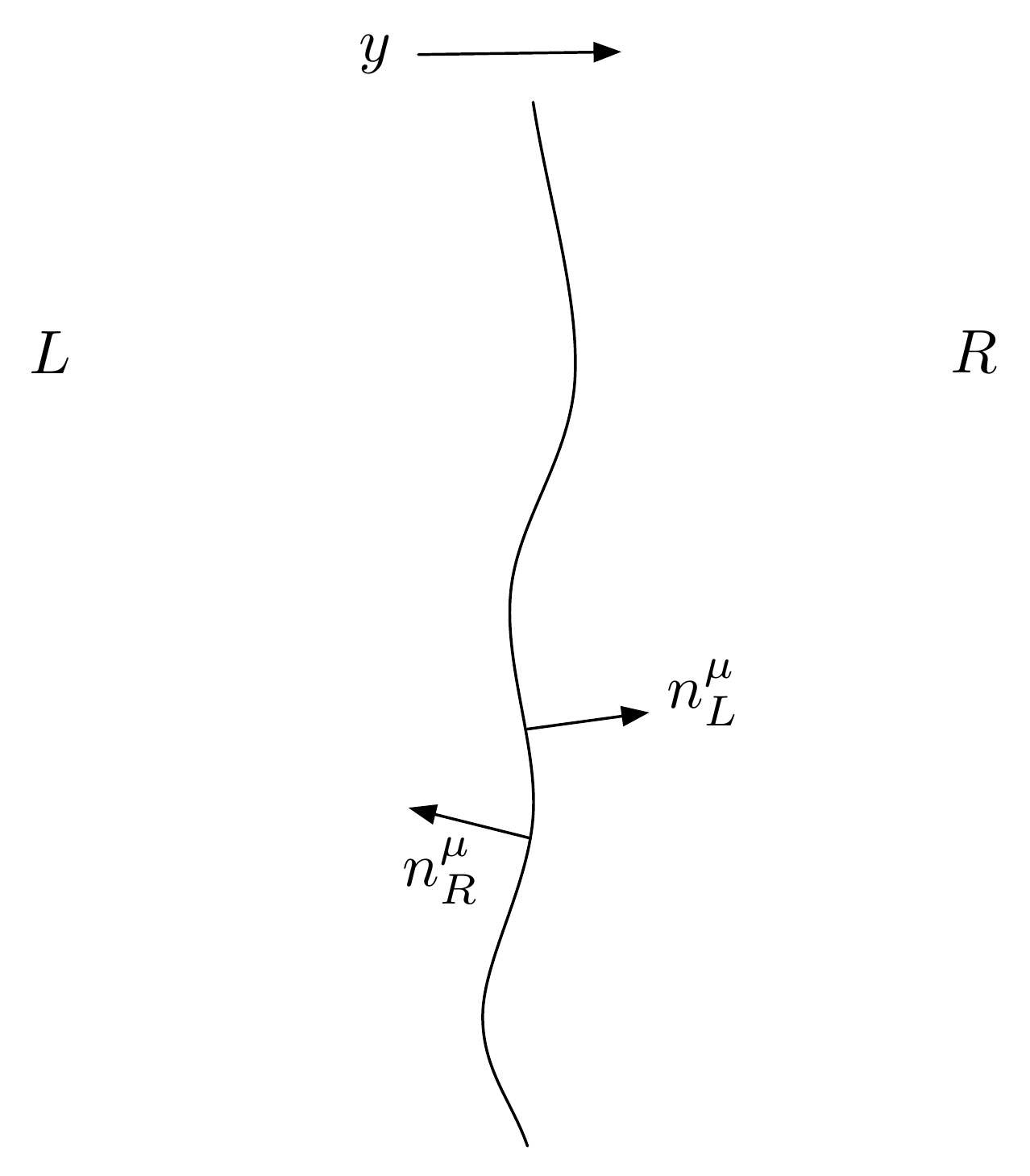,height=3in,width=3.0in}
\caption{\small Splitting the DGP action}
\label{branevariation}
\end{center}
\end{figure}

\be S={M_5^3\over 2}\left(\int_L+\int_R\right) d^4xdy\sqrt{-G}R(G)+\int d^4x\ {\cal L}_4,\ee
where ${\cal L}_4\equiv {M_4^2\over 2}\sqrt{-g}R(g)+{\cal L}_M(g,\psi)$ is the 4d part of the lagrangian.  
To have a well defined variational principle, we must have Gibbons-Hawking terms on both sides \cite{Dyer:2008hb}, corresponding to the outward pointing normals.  Adding these, the resulting action is
\bea \nn S&=&{M_5^3\over 2}\left[\left(\int_L+\int_R\right) d^4xdy\sqrt{-G}R(G)+2\oint_Ld^4x\sqrt{-g}K_L+2\oint_Rd^4x\sqrt{-g}K_R\right] \\
&+&\int d^4x\ {\cal L}_4\ ,\eea
where $K_R$, $K_L$ are the extrinsic curvatures relative to the normals $n_R$ and $n_L$ respectively.  

We now go to spacelike ADM variables \cite{Arnowitt:1960es,Arnowitt:1962hi} adapted to the brane (see \cite{Poisson,Dyer:2008hb} for detailed derivations and formulae).  The lapse and shift relative to $y$ are $N^\mu(x,y)$ and $N(x,y)$, and the 4d metric is $g_{\mu\nu}(x,y)$.  The 5d metric is
\be G_{AB}= \left(\begin{array}{c|c} N^2+N^\mu N_{\mu}  & N_{\mu} \\ \hline N_{\mu} & g_{\mu\nu}  \end{array}\right).\ee
The 4d extrinsic curvature is taken with respect to the positive pointing normal $n_L$, and is given by
\be K_{\mu\nu}={1\over 2N}\left(g'_{\mu\nu}-\nabla_\mu N_\nu-\nabla_\nu N_\mu\right),\ee
where a prime means derivative with respect to $y$.  The action is now\footnote{The Ricci scalar and metric determinant are
\be\label{Radmdec} \ ^{(5)}R=\ ^{(4)}R+\left(K^2-K_{\mu\nu}K^{\mu\nu}\right)+2\nabla_A\left(n^B\nabla_B n^A-n^A K\right),\ee
\be \sqrt{-G}=N\sqrt{-g}.\ee
The total derivatives coming from $2\nabla_A\left(n^B\nabla_B n^A-n^A K\right)$ in the Einstein-Hilbert part of the action exactly cancel the Gibbons-Hawking terms.} 
\bea\label{DGPadm} \nn S&=&{M_5^3\over 2}\left(\int_L+\int_R\right) d^4xdy\ N\sqrt{-g}\left[R(g)+K^2-K_{\mu\nu}K^{\mu\nu}\right] \\
&+&\int d^4x\ {\cal L}_4.\eea

It can be checked that a flat brane living in flat space is a solution to the equations of motion of this action.  This is called the \textit{normal branch}.  There is another maximally symmetric solution with a flat 5d bulk, which contains a de Sitter brane with a 4d Hubble scale $H\sim M_5^3/M_4^2$.  This is called the \textit{self-accelerating branch}, and has caused much interest because the solution exists even though the brane and bulk cosmological constants vanish.

\subsubsection*{Linear expansion}

To see the particle content of DGP, we will expand the action (\ref{DGPadm}) to linear order around the flat space solution, and then integrate out the bulk to obtain an effective 4d action. 
We start by expanding the 5d graviton about flat space
\be G_{AB}=\eta_{AB}+H_{AB}.\ee
We use the lapse, shift and 4d metric variables, with their expansions around flat space,
\be g_{\mu\nu}=\eta_{\mu\nu}+h_{\mu\nu},\ \ \ N_{\mu}=n_{\mu},\ \ \ N=1+n.\ee
We have the relations, to linear order in $h_{\mu\nu},n_\mu,n$, 
\be H_{\mu\nu}=h_{\mu\nu},\ \ \ H_{\mu 5}=n_\mu,\ \ \ H_{55}=2n.\ee
We will first expand the DGP action (\ref{DGPadm}) to quadratic order in $h_{\mu\nu},n_\mu,n$.  We will then solve the 5d equations of motion, subject to arbitrary boundary values on the brane and going to zero at infinity.  We then plug this solution back into the action to obtain an effective 4d theory for the arbitrary brane boundary values.  

The 5d equations of motion away from the brane are simply the vacuum Einstein equations, which read, to linear order,
\be \label{bulkequations} -2 R_{AB}(G)_{\rm linear}=\square^{(5)}H_{AB}+\partial_A\partial _B H-\partial^C\partial_AH_{BC}-\partial^C\partial_B H_{AC}=0.\ee
We will solve (\ref{bulkequations}) in the de Donder gauge,
\be \label{dedondercondition} \partial^BH_{AB}-\half \partial_AH=0.\ee
With this, (\ref{bulkequations}) is equivalent to 
\be \label{laplaceequation} \square^{(5)}H_{AB}=0,\ee
along with the de Donder gauge condition (\ref{dedondercondition}).  
In terms of the ADM variables, (\ref{laplaceequation}) becomes 
\bea &&\square h_{\mu\nu}+\partial_y^2 h_{\mu\nu}=0,\\  && \square n_{\mu}+\partial_y^2 n_{\mu}=0,  \\ && \square n+\partial_y^2 n=0,
\eea
where $\square$ is the 4d laplacian.  These have the following solutions in terms of boundary values $h_{\mu\nu}(x),n_\mu(x),n(x)$,
\bea h_{\mu\nu}(x,y)&=&e^{-y\Delta}h_{\mu\nu}(x), \\  n_{\mu}(x,y)&=&e^{-y\Delta}n_{\mu}(x), \\  n(x,y)&=&e^{-y\Delta}n(x).\eea
Here, the operator $\Delta$ is the formal square root of the 4d laplacian, 
\be \Delta\equiv \sqrt{-\square}.\ee
The $A=\mu$ and $A=5$ components of the gauge condition (\ref{dedondercondition}) are, respectively,
\bea &&\partial^\nu h_{\mu\nu}-\half \partial_\mu h+\partial_y n_\mu -\partial_\mu n=0,\\  && \partial^\mu n_\mu-\half \partial_y h+\partial _y n=0. \eea
For these to be satisfied everywhere, it is necessary and sufficient that the boundary fields satisfy the following at $y=0$,
\bea \nn &&\partial^\nu h_{\mu\nu}-\half \partial_\mu h-\Delta n_\mu -\partial_\mu n=0,\\  && \label{gaugeconstraints} \partial^\mu n_\mu+\half \Delta h-\Delta n=0. \eea
These should be thought of as constraints determining some of the boundary variables in terms of the others\footnote{Note that we cannot think of them as determining $n^\mu$, $n$ in terms of $h_{\mu\nu}$.  Acting with $\partial_\mu$ on the first equation, $\Delta$ on the second, and then adding, we find the equation \be\partial_\mu\partial_\nu h^{\mu\nu}- \square h=0,\ee which is precisely the statement that the 4d linearized curvature vanishes (which is, in turn, the linearized hamiltonian constraint in general relativity).  Thus, we must think of these constraints as determining some of the components of the metric.}.  
We will at this point imagine that we have solved these constraints, and that the action is really a function of the independent variables.  

The deDonder gauge is preserved by any 5d gauge transformation $\Xi^A$ satisfying
\be \label{laplacegauge} \square^{(5)}\Xi^A=0.\ee
The component $\Xi^5$ must vanish at $y=0$ because the position of the brane is fixed.  (\ref{laplacegauge}) then implies that $\Xi^5$ vanishes everywhere.  The other components can have arbitrary values $\Xi^\mu(x,0)=\xi^\mu(x)$ on the brane, which are then extended into bulk in order to satisfy (\ref{laplacegauge}),
\be \Xi^\mu(x,y)=e^{-y\Delta}\xi(x).\ee
The residual gauge transformations acting on the boundary fields are then
\bea \nn \delta h_{\mu\nu}&=&\partial_\mu\xi_\nu+\partial_\nu\xi_\nu,\\ \nn
\delta n_\mu&=&-\Delta \xi _\mu,\\ \label{gaugetransforms}
\delta n&=&0.
\eea
The constraints (\ref{gaugeconstraints}) are invariant under these gauge transformations.  The 4d effective action must and will be invariant under (\ref{gaugetransforms}).

The 5d part of the action reads 
\be S_5={M_5^3\over 2}\int d^4x dy\ N\sqrt{-g}\left[R(g)+K^2-K_{\mu\nu}K^{\mu\nu}\right].\ee
We want to expand this to quadratic order in $h_{\mu\nu},n_{\mu},n$ and then plug in our solution.  We will need the expansion of $K_{\mu\nu}$ to first order,
\be K_{\mu\nu}=\half\left(\partial_y h_{\mu\nu}-\partial_\mu n_\nu-\partial_\nu n_\mu\right).\ee
Expanding, we have (after much integration by parts in 4d)
\bea \nn {2\over M_5^3}S_5=\int d^4xdy && n\partial_\mu\partial_\nu h^{\mu\nu}-n \square h+\half \partial_\lambda h_{\mu\nu}\partial^\nu h^{\mu\lambda}-\half \partial_\mu h\partial_\nu h^{\mu\nu} \\ \nn && -\partial_y h \partial_\mu n^\mu +\half (\partial_\mu n^\mu)^2+\partial_y h_{\mu\nu}\partial^\mu n^\nu+\half n_\mu\square n^\mu \\ \nn && {1\over 4}h_{\mu\nu}\square h^{\mu\nu}-{1\over 4}\partial_y h_{\mu\nu}\partial_y h^{\mu\nu}-{1\over 4}h\square h+{1\over 4}(\partial_y h)^2 .
\eea
Now, in the last line, integrate by parts in $y$, picking up a boundary term at $y=0$, and use (\ref{laplaceequation}) to kill the bulk part,
\bea \nn {2\over M_5^3}S_5=\int d^4xdy && n\partial_\mu\partial_\nu h^{\mu\nu}-n \square h+\half \partial_\lambda h_{\mu\nu}\partial^\nu h^{\mu\lambda}-\half \partial_\mu h\partial_\nu h^{\mu\nu} \\ \nn && -\partial_y h \partial_\mu n^\mu +\half (\partial_\mu n^\mu)^2+\partial_y h_{\mu\nu}\partial^\mu n^\nu+\half n_\mu\square n^\mu \\ \nn +\int d^4x && -{1\over 4} h\partial h+{1\over 4} h_{\mu\nu}\partial_y h^{\mu\nu} .
\eea

We now insert the following term into the action
\be S_{\rm GF}=-{M_5^3\over 4}\int d^5X\ \left(\partial^BH_{AB}-\half \partial_AH\right)^2.\ee
The 5d equations of motion solve the de Donder gauge condition, so this term contributes 0 to the action (thought of as a function of the unconstrained variables) and we are free to add it.  However, we write it in terms of the unconstrained 4d variables for now,
\bea {2\over M_5^3}S_{\rm GF}=\int d^4x dy\ && -\half \left(\partial^\nu h_{\mu\nu}-\half \partial_\mu h+\partial_y n_\mu-\partial_\mu n\right)^2 \\ && -\half\left(\partial_\mu n^\mu -\half \partial_y h+\partial_y n\right)^2.\eea
Adding this to the previous 5d term, we find that after using the 5d Laplace equations, the entire action can be reduced to a boundary term at $y=0$,
\bea \label{intermediateaction} \nn {2\over M_5^3}\left(S_5+S_{GF}\right)=\int d^4x && -{1\over 4} h_{\mu\nu}\Delta h^{\mu\nu}+{1\over 8}h\Delta h-\half n\Delta n-\half n_\mu\Delta n^\mu \\ &&+\half h\Delta n+n^\mu\left(-\partial_\mu n-\half \partial_\mu h+\partial^\nu h_{\mu\nu}\right).\eea

Now a crucial point.  We have been imagining solving the constraints (\ref{gaugeconstraints}) for the independent variables.  But now, consider the action (\ref{intermediateaction}) as a function of the original variables $h_{\mu\nu},n^\mu,n$.   Varying with respect to $n^\mu$ and $n$, we recover precisely the constraints (\ref{gaugeconstraints}).  Thus, we can re-introduce the solved variables as auxiliary fields, since the constraints are then implied.  The action now becomes a function of $h_{\mu\nu},n^\mu,n$.  

Now add in the 4d part of the action,
\be S={M_4^2\over 2}\int d^4x\ \sqrt{-g}R(g)+S_M+2\left(S_5+S_{\rm GF}\right).\ee
where $S_M$ is the 4d matter action and the factor of 2 in front of the 5d parts results from taking into account both sides of the bulk (through boundary conditions at infinity we have thus implicitly imposed a ${\mathbb Z}_2$ symmetry).   

Expanded to quadratic order,
\bea \nn S=\int d^4x\ && {M_4^2\over 4}\frac{1}{2}h_{\mu\nu}{\cal E}^{\mu\nu,\alpha\beta}h_{\alpha\beta}+{M_4^2m\over 4}\bigg[ -{1\over 2} h_{\mu\nu}\Delta h^{\mu\nu}+{1\over 4}h\Delta h-n\Delta n-n_\mu\Delta n^\mu \\ \label{finalquadaction} &&+ h\Delta n+n^\mu\left(-2\partial_\mu n-\partial_\mu h+2\partial^\nu h_{\mu\nu}\right)\bigg]+\half h_{\mu\nu}T^{\mu\nu},
\eea
where ${\cal E}^{\mu\nu,\alpha\beta}$ is the massless graviton kinetic operator (\ref{Eoper}),
and 
\be m\equiv{2 M_5^3\over M_4^2}\ee
is known as the DGP scale. 

The action (\ref{finalquadaction}) should now be compared to the massive gravity action in the form (\ref{auxiliaryaction}).   It is invariant under the gauge transformations (\ref{gaugetransforms}), under which $n^\mu$ plays the role of the vector St\"ukelberg field.  $n$ plays the role of the gauge invariant  auxiliary field.  To get this into Fierz-Pauli form, first eliminate $n$ as an auxiliary field by using its equation of motion.   Then use (\ref{gaugetransforms}) to fix the gauge $n^\mu=0$.  The resulting action is
\be \label{resmassaction} S=\int d^4x\  {M_4^2\over 4}\left[\frac{1}{2}h_{\mu\nu}{\cal E}^{\mu\nu,\alpha\beta}h_{\alpha\beta}-{1\over 2}m\left( h_{\mu\nu}\Delta h^{\mu\nu}-h\Delta h\right)\right]+\half h_{\mu\nu}T^{\mu\nu},
\ee
which is of the Fierz-Pauli form, with an operator dependent mass term $m\Delta$.

One can go on to study interactions terms for DGP, and the longitudinal mode turns out to be governed by interactions which include the cubic galileon term $\sim (\partial\phi)^2\square\phi$ \cite{Luty:2003vm,Nicolis:2004qq,Gabadadze:2006tf}, and are suppressed by the scale $\Lambda_3=(M_4m^2)^{1/3}$ (in fact this was where the galileons were first uncovered).  In this sense, DGP is analogous to the nicer $\Lambda_3$ theories of Section (\ref{lambda3section}).  The theory is free of ghosts and instabilities around solutions connected to flat space \cite{Nicolis:2004qq}, but changing the asymptotics to the self-accelerating de Sitter brane solutions flips the sign of the kinetic term of the longitudinal mode, so there is a massless ghost around the self-accelerating branch \cite{Koyama:2007za}.  This branch is completely unstable, which is bad news for doing cosmology on this branch.  In addition to ghosts, there are other issues with other non-trivial branches, such as superluminal fluctuations \cite{Hinterbichler:2009kq}, and uncontrolled singularities and tunneling \cite{Gregory:2008bf}.

\subsubsection*{Resonance gravitons} 

The operator dependent mass term in (\ref{resmassaction}) is known as a \textit{resonance mass}, or \textit{soft mass} \cite{Dvali:2000rv,Dvali:2001gm,Dvali:2001gx,Gabadadze:2003ck,Gabadadze:2004dq,Dvali:2006su,LopezNacir:2006tn,Gabadadze:2007as,Dvali:2007kt,Patil:2010mq}.
To see the particle content of this theory, we will decompose the propagator into a sum of massive gravity propagators.  The linear St\"ukelberg analysis, leading to the propagators (\ref{gravstukelprops}), goes though identically, with the replacement $m^2\rightarrow m\Delta$.  
The momentum part of the propagators now reads 
\be {-i\over p^2+m\sqrt{p^2}},\ee
Setting $z=-p^2$, the propagator has a branch cut in the $z$ plane from $(0,\infty)$, with discontinuity
\be -{2m\over \sqrt{z}(z+m^2)}.\ee 

A branch cut can be thought of as a string of simple poles, in the limit where the spacing between the poles and their residues both go to zero. The function $f(z)=\int_{-\infty}^\infty d\lambda \rho(\lambda){1\over z-\lambda}$ has a cut along the real axis everywhere that $\rho$ is non-zero, with discontinuity $-2i\pi\rho(z)$.  We can see this by noting 
\[disc\ f(z)=\int_{-\infty}^\infty d\lambda \rho(\lambda){1\over z-\lambda+i\epsilon}-{1\over z-\lambda-i\epsilon}=\int_{-\infty}^\infty d\lambda \rho(\lambda)\left[-2\pi i\delta(z-\lambda)\right].\]

Using all this, and the fact that analytic functions are determined by their poles and cuts, we can write the propagator in the spectral form
\be\label{spectraldecomp}  {-i\over p^2+m\sqrt{p^2}}=\int_0^\infty ds {-i\over p^2+s}\rho(s),\ \ \ \rho(s)={m\over \pi \sqrt{s}(s+m^2)}>0.\ee
The spectral function is greater than zero, so this theory contains a continuum of ordinary (non-ghost, non-tachyon) gravitons, with masses ranging from $0$ to $\infty$.  This is what would be expected from dimensionally reducing a non-compact fifth dimension.  The Kaluza-Klein tower has collapsed down into a Kaluza-Klein continuum.

In the limit $m\rightarrow 0$, where the action becomes purely four dimensional, the spectral function reduces to a delta function,
\be \rho(s) \rightarrow 2\delta(s),\ee
and the propagator reduces to $-i/p^2$ representing a single massless graviton, vector and scalar, as can be seen from (\ref{gravstukelprops}) (the extra factor of two is taken care of by noting that the integral is from $0$ to $\infty$, so only half of the delta function actually gets counted).  This theory therefore contains a vDVZ discontinuity.

The potential of a point source of mass $M$ sourced by this resonance graviton displays an interesting crossover behavior. 
Looking back at (\ref{metric1masiv}) with the momentum space replacement $m^2\rightarrow m\sqrt{p^2}$, and using the relation $\phi={-h_{00}/M_4}$ for the newtonian potential, we have
\bea \phi(r)&=&{-2M\over 3M_4^2}\int{d^3\pb\over (2\pi)^3}e^{i\pb\cdot\xb}{-1\over   \pb^2+m |\pb|} \\ &=&{2\over 3}{M\over M_4^22\pi^2 r}\left[\sin\left(r\over r_0\right)\text{ci}\left(r\over r_0\right)+\half\cos\left(r\over r_0\right)\left(\pi-2\ \text{si}\left(r\over r_0\right)\right)\right],\nn \eea
where
\be \text{si}(x)\equiv\int_0^x{dt\over t}\sin t, \ \ \ \text{ci}(x)\equiv\gamma+\ln x+\int_0^x{dt\over t}\left(\cos t-1\right),\ee
$\gamma\approx 0.577\ldots$ is the Euler-Masceroni constant, and the length scale $r_0$ is
\be r_0\equiv{1\over m}= {M_4^2\over 2 M_5^3}.\ee
The potential interpolates between 4d $\sim 1/r$ and 5d $\sim 1/r^2$ behavior at the scale $r_0$,
\be V(r)=\begin{cases} {2\over 3}{M\over M_4^24\pi r}+{M\over 3\pi^2 M_4^2}{1\over r_0}\left[\gamma-1+\ln \left(r\over r_0\right)\right] +{\cal O}(r), & r\ll r_0, \\
{2\over 3}{M\over M_5^34\pi^2 r^2}+{\cal O}\left(1\over r^3\right), & r\gg r_0.\end{cases} \ee
Physically, we can think of gravity as being confined to the brane out to a distance $\sim r_0$, at which point it starts to weaken and leak off the brane, becoming five dimensional.  This is the behavior that is morally responsible for the self-accelerated solutions seen in DGP \cite{Deffayet:2001pu}.  It has been suggested that corrections to the newtonian potential for $r\ll r_0$ may be observable in lunar laser ranging experiments \cite{Lue:2002sw,Dvali:2002vf}.

The resonance massive graviton can also be generalized away from DGP, by replacing the mass term with an arbitrary function of the laplacian \cite{Gabadadze:2003ck,Dvali:2006su,Dvali:2007kt},
\be m^2\rightarrow m^2(\square).\ee
(See \cite{Dvali:2008em} for even further generalizations.) At large distances, where we want modifications to occur, the mass term will have a leading Taylor expansion,
\be \label{massalpha} m^2(\square)=L^{2(\alpha-1)}\square^{\alpha},\ee
with $L$ a length scale and $\alpha$ a constant.  In order to modify newtonian dynamics at large scales, $\partial\ll {1\over L}$, the mass term should dominate over the two derivative kinetic terms, so we should have $\alpha<1$.  Additionally, there is the constraint that the spectral function (\ref{spectraldecomp}) should be positive definite, so that there are no ghosts.  This puts a lower bound $\alpha\geq0$ \cite{Dvali:2006su}.  An analysis like that of Section \ref{filtersection} can be repeated with the more general mass term (\ref{massalpha}), and it turns out that degravitation can be made to work only for $\alpha<1/2$ \cite{Dvali:2007kt}.  DGP corresponds to $\alpha=1/2$, and so it just barely fails to degravitate, but by extending the DGP idea to higher co-dimension \cite{deRham:2007pz,Hassan:2010ys} or to multi-brane {cascading DGP} models \cite{deRham:2007xp,deRham:2008qx,deRham:2010rw}, $\alpha<1/2$ can be achieved and degravitation made to work \cite{deRham:2007rw}.  Some N-body simulations of de-gravitation and DGP have been done in \cite{Khoury:2009tk,Chan:2009ew,Schmidt:2009sv}.

\subsection{\label{auxdimsection}Massive gravity from an auxiliary extra dimension}

There is a way of writing a mass term by using an integral over a non-dynamical auxiliary extra dimension \cite{Gabadadze:2009ja,deRham:2009rm}, and which provides a re-summation of some particular massive gravity theories.  

To see what this entails, consider the following perverse way of writing the mass term for an ordinary scalar field, $\phi(x)$.  We let the scalar depend on an additional auxiliary parameter $u$, which takes values in the interval $u\in (0,1)$.  We then write the following action for the field $\phi(x,u)$,
\be S=\int d^4x\ \left[-\left.\half (\partial\phi)^2\right|_{u=0}-\half m^2\int_0^1 du (\partial_u\phi)^2\right].\ee
The first term is a normal kinetic term, with derivatives over the $x^\mu$ only, and the field there is $\phi(x)\equiv \phi(x,u=0)$.  The second term contains a $u$ derivative only, and an integral over all the $u$ values.  This can be thought of as an action for the infinite number of fields $\phi(x,u)$ parametrized by $u$, where only the $u=0$ field has a kinetic term.  

We can get a 4d effective action for this theory by integrating out the $u$ dimension, to obtain an equivalent theory for only $\phi(x)$.  To do this, we solve the equations of motion for all the $\phi(x,u)$ with $u\not=0$ in terms of $\phi(x)$, and then plug back into the action, i.e. we can think of all the $\phi(x,u)$ with $u\not=0$ as auxiliary fields and eliminate them via their equations of motion.  To get a unique solution, we must set boundary conditions at $u=1$, which we will choose to be a Dirichlet condition $\phi(x,1)=0$.  

The equations of motion for $u\not=0$ and the boundary conditions are then,
\be \partial^2_u\phi=0,\ \ \ \phi(x,0)=\phi(x),\ \ \ \phi(x,1)=0,\ee
with the solution
\be \phi(x,u)=(1-u)\phi(x).\ee
Plugging back into the action, we find
\be S=d^4x\ \left(-\left.\half (\partial\phi)^2-\half m^2\phi^2\right)\right|_{u=o}.\ee
The integral over $u$ has become an ordinary mass term for the scalar.

We now apply a similar construction to the Einstein-Hilbert action, writing
\be \label{auxgravityaction}S=d^4x\ {M_P\over 2}\left[\left.\(\sqrt{-g}R\)\right|_{u=0}-m^2\int_0^1 du\ \sqrt{-g}\(k_{\mu\nu}^2-k^2\)\right].\ee
Here the metric is $g_{\mu\nu}(x,u)$, it depends on the auxiliary parameter $u$.  $k_{\mu\nu}\equiv \half \partial_u g_{\mu\nu}$ is the extrinsic curvature into the auxiliary dimensions, and indices in the mass term are raised and lowered with $g_{\mu\nu}(x,u)$ and its inverse.  We choose boundary conditions so that the metric becomes flat at the end of the auxiliary dimension,
\be g_{\mu\nu}(x,0)=g_{\mu\nu}(x),\ \ \ g_{\mu\nu}(x,1)=\eta_{\mu\nu}.\ee
The lagrangian (\ref{auxgravityaction}) is invariant under 4d diffeomorphisms, but they are broken by the boundary conditions imposed at $u=1$.  This will be reflected in the effective 4d metric obtained from integrating out the extra dimension.  The boundary condition metric essentially plays the role of the fixed metric used for the purely 4d constructions.

We can now perturbatively solve the equations of motion in the auxiliary dimension and expand out the mass term.  We write the metric at $u=0$ as $g_{\mu\nu}=\eta_{\mu\nu}+h_{\mu\nu}$ and work in powers of $h_{\mu\nu}$.  The full $u$-dependent metric will have an expansion
\be\label{Hhexpansion} g_{\mu\nu}(x,u)=\eta_{\mu\nu}+H^{(1)}_{\mu\nu}(x,u)+H^{(2)}_{\mu\nu}(x,u)+\cdots,\ee
where $H^{(1)}_{\mu\nu}(x,u),\ H^{(2)}_{\mu\nu}(x,u),\ldots$ are the terms containing respective powers of $h_{\mu\nu}$.  The full equations of motion for $u\not=0$ derived from (\ref{auxgravityaction}) are
\be\label{gauxequation} \partial_u\left[\sqrt{-g}\(k^{\mu\nu}-k g^{\mu\nu}\)\right]=\half \sqrt{-g}g^{\mu\nu}\(k_{\alpha\beta}^2-k^2\)+2\sqrt{-g}\(kk^{\mu\nu}-k_{\alpha}^{\ \mu}k^{\nu\alpha}\).\ee
By plugging (\ref{Hhexpansion}) into (\ref{gauxequation}), we can collect like powers and solve order by order.  To lowest order we have simply $\partial_u\(H^{(1)}_{\mu\nu}-\eta_{\mu\nu}H^{(1)}\)=0$ with solution
\be H^{(1)}_{\mu\nu}(x,u)=(1-u)h_{\mu\nu}(x).\ee
Plugging back into (\ref{auxgravityaction}) we find exactly the Fierz-Pauli mass term at quadratic order, 
\be \label{auxgravityactionpert} S=d^4x\ {M_P\over 2}\left[\sqrt{-g}R-{m^2\over 4}\(h_{\mu\nu}^2-h^2+\mathcal{O}(h^3)\)\right].\ee
The order $h^3$ terms were calculated in \cite{deRham:2010gu}, and it is found that their coefficients are in the right combination for raising the cutoff to $\Lambda_3$, corresponding to $c_3=1/4$, and it seemed likely that this theory was a way of re-summing one of the $\Lambda_3$ theories.

Recently, an exact solution for the equation (\ref{gauxequation}) has been found, for an arbitrary boundary metric $f_{\mu\nu}(x)$ at $u=1$ \cite{Hassan:2011zr}.  This solution allows us to find the 4d lagrangian exactly,
\bea S=\int d^4x && {M_P\over 2}\bigg[\sqrt{-g}R \nn\\
&&+3m^2\sqrt{-g}\left(\left[\det L \right]^{-1/2}-2\left[\det L\right]^{-1/4}\cosh\left({1\over 2\sqrt{3}}\sqrt{\Tr\left[(\ln L)^2\right]-{1\over 4}\left(\Tr\ln L\right)^2}\right) +1\right)\bigg]. \nn\\
\eea
Here $L$ is the matrix $L^{\mu}_{\ \nu}=f^{\mu\lambda}g_{\lambda\nu}$, where $f^{\mu\nu}$ is the inverse of the boundary condition metric at $u=1$.  This lagrangian shows explicitly how the boundary condition metric becomes the fixed metric needed for the graviton mass term.  Taking $f_{\mu\nu}=\eta_{\mu\nu}$, then expanding to third order in $h_{\mu\nu}=g_{\mu\nu}-\eta_{\mu\nu}$ reproduces the results of \cite{deRham:2010gu}, but unfortunately the tuning necessary for the $\Lambda_3$ theory is violated at fourth order.  Thus this theory has a ghost, and a cutoff of $\Lambda_5$.  

It turns out, however, that a different choice for the boundary metric $f_{\mu\nu}$  can be made to restore the $\Lambda_3$ tuning order by order, though it requires $f_{\mu\nu}$ to depend on the metric at $u=0$ if flat space is to be a solution in 4d \cite{Berezhiani:2011nc}.   They also present an argument to explain why with $\eta_{\mu\nu}$ boundary conditions the third order tuning comes out right while the fourth order and higher do not.  Note that the combination $k_{\mu\nu}^2-k^2$ of extrinsic curvatures appearing in (\ref{auxgravityaction}) is the same combination that appears in the Gauss-Codazzi expansion of the five dimensional curvature in terms of four dimensional quantities (\ref{Radmdec}), which suggests that there may be some kind of hidden five dimensional symmetry at work in this model, one which enforces the tuning of coefficients in the Fierz-Pauli terms.  It is also possible to achieve the correct tuning by adding higher powers of $k_{\mu\nu}$ into the action (\ref{auxgravityaction}) with the right coefficients \cite{deRham:2009rm,Berezhiani:2011nc}.   Some non-flat cosmological solutions of the auxiliary extra dimension model are studied in \cite{Gabadadze:2009ja}.  Extensions to higher co-dimension radially symmetric models, and to models including additional bulk Gauss-Bonnet terms, are considered in \cite{Berezhiani:2011nc}.

\section{Massive gravity in three dimensions}

We have focused in this review on massive gravity in four dimensions.  Extending to higher dimensions is a straightforward exercise that does not reveal much which is conceptually new beyond what exists in four dimensions (with some exceptions \cite{Lu:2010sj}).   On the other hand, dropping down to three dimensions opens up new possibilities.  These possibilities are due to peculiarities of three dimensions, and they would take another review to fully cover (see \cite{Bergshoeff:2010ad} for a short review).  Here we will briefly mention the two main classes of fully interacting, covariant massive theories that have been studied in dimension three.  

\subsection{New massive gravity}

The first is \textit{new massive gravity} (NMG), proposed by Bergshoeff, Hohm, and Townsend \cite{Bergshoeff:2009hq}.  NMG relies on adding terms which carry four derivatives of the metric, such as $R^2$, $R_{\mu\nu}R^{\mu\nu}$, to the action in addition to the Einstein-Hilbert term.  Since these terms are higher derivative, degrees of freedom in addition to the graviton will generally propagate.  In four dimensions, adding such terms with generic coefficients will lead to the propagation around flat space of a massive spin 0 scalar and a massive spin 2 ghost, in addition to the massless graviton \cite{Stelle:1977ry}.  Judicious choices of coefficients exist which will project out the massive spin 2 ghost or the scalar, but no choice will eliminate the massless graviton.  There is no way to get an action that contains a healthy massive spin 2 and a healthy massless graviton.  If the sign of the action is flipped, the massive spin 2 can be made healthy, but then the massless graviton will be a ghost.  

In three dimensions, however, there is a loophole, because 3d Einstein gravity contains no propagating degrees of freedom (in three dimensions, the canonical analysis of Section \ref{canonicalanalysis} shows that there are 3 first class constraints on the 3 spatial components of the metric and their canonical momenta, leaving no degrees of freedom).  Thus, we can add curvature squared terms in a combination such that there is a massive spin 2 and no scalar, and then change the overall sign of the action so that the massive spin is healthy,
\be S={M_P\over 2}\int d^3x\ \sqrt{-g}\left[-R+{1\over m^2}\left(R_{\mu\nu}R^{\mu\nu}-{3\over 8}R^2\right)\right].\ee
 $M_P$ is the 3d Planck mass and $m$ is the mass of the graviton.  The specific factor of $3/8$ in the higher derivative terms kills the spin 0 mode.  
  Note the wrong sign Einstein-Hilbert term -- the massless spin 2 would be a ghost, but it does not matter because in three dimensions it carries no degrees of freedom anyway.  The result is a theory which propagates only a massive spin 2 around flat space, carrying two degrees of freedom\footnote{Recall that in three dimensions, spin for massive particles is like helicity for massless particles in four dimensions, because the little group is $so(2)$ in both cases.  Thus a parity conserving theory of a massive spin $s$ particle in three dimensions has two states, of $so(2)$ charge $\pm s$, which are related by parity.  A parity violating theory, on the other hand, will carry only one degree of freedom for some particle, the right or left handed state.  This will be the case for topologically massive gravity in Section \ref{topmasgrav}.}.  In contrast to the other massive gravity theories studied in this review, NMG requires no fixed metric in the action, so like Einstein gravity, the theory is free of prior geometry.
  
 This theory has been recently analyzed with the St\"ukelberg method \cite{deRham:2011ca}.  Like the $\Lambda_3$ theory in four dimensions, NMG theory carries a scale higher than that which would be generically expected (though in this case, it has been claimed that the theory is renormalizable \cite{Oda:2009ys}).  The de-coupling limit is ghost free, and the longitudinal mode interacts through a cubic galileon interaction.  In addition, the theory possesses no Boulware-Deser ghost to any order beyond the decoupling limit, and so it represents a completely consistent ghost free theory of a fully interacting massive graviton in three dimensions, one which is free of prior geometry.

\subsection{\label{topmasgrav}Topologically massive gravity}

The second type of massive gravity in three dimensions is \textit{topologically massive gravity} (TMG) \cite{Deser:1981wh}.  The action is
\be\label{topologicallymassive} S={M_P\over 2}\int d^3 x\sqrt{-g}\left[-R-{1\over 2\mu}\epsilon^{\lambda\mu\nu}\Gamma^\alpha_{\lambda\beta}\left(\partial_\mu\Gamma^\beta_{\alpha\nu}+{2\over 3}\Gamma^\beta_{\mu\gamma}\Gamma^\gamma_{\nu\alpha}\right)\right].
\ee
Here $\epsilon^{\lambda\mu\nu}$ is the 3d epsilon tensor, equal to $-{1\over \sqrt{-g}}\tilde\epsilon^{\lambda\mu\nu}$ where $\tilde\epsilon^{\lambda\mu\nu}$ is the alternating symbol with $\tilde\epsilon^{012}=+1$.  $\mu$ is a mass parameter, and the term proportional to ${1\over \mu}$ is a Chern-Simons form, or secondary characteristic class.  It is not a covariant tensor, but it changes into a total derivative under a diffeomorphism, so the action is indeed diffeomorphism invariant.  The equations of motion are 
\be \label{topmaseqom} R_{\mu\nu}-{1\over 2}Rg_{\mu\nu}+{1\over \mu}C_{\mu\nu}=0,\ee
where $C_{\mu\nu}=\epsilon_{\mu}^{\ \alpha\beta}\nabla_\alpha\left(R_{\beta\nu}-{1\over 4}g_{\beta\nu}R\right)$ is the Cotton tensor.  The Cotton tensor is symmetric, traceless, and covariantly conserved, and it vanishes if and only if the metric is conformally flat.

Linearizing the equations of motion (\ref{topmaseqom}) around flat space, and plugging in a $h_{\mu\nu}^{TT}$ which is transverse and traceless gives 
\be \left(\delta_\mu^\alpha \delta_\nu^\beta+{1\over \mu}\epsilon_\mu^{\ \lambda\alpha}\partial_\lambda\delta_\nu^\beta\right)\square h^{TT}_{\alpha\beta}=0.\ee
Acting on this with $\delta_\mu^\alpha \delta_\nu^\beta-{1\over \mu}\epsilon_\mu^{\ \lambda\alpha}\partial_\lambda\delta_\nu^\beta$, we find
\be\left(\square-\mu^2\right) \square h_{\mu\nu}^{TT}=0,\ee
suggesting that the linear curvature $R_{\mu\nu}^L\sim \square h_{\mu\nu}^{TT}$ propagates a particle with mass $\mu$.  The Chern-Simons term violates parity, and a more detailed analysis \cite{Deser:1981wh} shows that this model in fact propagates a single massive spin 2 degree of freedom, where the spin is left or right handed according to the sign of $\mu$.  This is in contrast to the parity conserving Fierz-Pauli mass term, which in three dimensions would propagate both a left and a right handed spin 2.  Finally, note the wrong sign Einstein-Hilbert term in (\ref{topologicallymassive}), which is necessary so that the spin 2 particle is not a ghost.

Adding a cosmological constant and putting this theory on $AdS_3$ yields an interesting system with links to AdS/CFT.  The $AdS_3$ isometry group is $SO(2,2)$ which locally decomposes to $SL(2,R) \times SL(2,R)$, the group of left-moving and right-moving symmetries in a dual boundary 2d conformal field theory.  This theory has ghosts or negative-energy modes for generic values of the parameters, but for a special choice, it was argued that half of the theory completely decouples, stability is restored, and the theory becomes chiral under only one of the $SL(2,R)$ factors \cite{Li:2008dq,Strominger:2008dp}.  For more on the subtleties, see \cite{Grumiller:2008qz,Skenderis:2009nt,Grumiller:2009mw,Gaberdiel:2010xv,Maloney:2009ck}.

\section{Conclusions and future directions}

Massive gravity remains an active research area, one which may provide a viable solution to the cosmological constant naturalness problem.  As we have seen, many interesting effects arise from the naive addition of a hard mass term to Einstein gravity.  There is a well defined effective field theory with a protected hierarchy between the cutoff and the graviton mass, and a screening mechanism which non-linearly hides the new degrees of freedom and restores continuity with GR in the massless limit.  

A massive graviton can screen a large cosmological constant, and a stable theory of massive gravity with a small protected mass offers a solution to the problem of quantum corrections to the cosmological constant.  It is a remarkable fact that the $\Lambda_3$ theories of Section \ref{lambda3section} exist and are ghost free, and that they are found simply by tuning some coefficients in the generic graviton potential. 

There are many interesting outstanding issues.  One is the nature of the $\Lambda_3$ theory.  Is there a symmetry or a topological construction that explains the tunings of the coefficients necessary to achieve the $\Lambda_3$ cutoff? Is there some construction free of prior geometry that would contain this theory?  Is there an extra dimensional construction?  

There are also many questions related to the quantum properties of these theories.  Apart from the order of magnitude estimates presented in this review and a few sporadic calculations, the detailed quantum properties of this theory and others remain relatively unexplored.  The same goes for non-perturbative quantum properties, such as how a massive graviton would modify black hole thermodynamics, Hawking radiation or holography \cite{Babak:2002uz,Katz:2005ir,Aharony:2006hz,Kiritsis:2006hy,Kiritsis:2008at,Niarchos:2009qb,Capela:2011mh}.

The cutoff  $\Lambda_3$ is still rather low, however, so at best this theory in its current perturbative expansion can only provide a partial solution to the cosmological constant naturalness problem.  Finding more natural constructions of these theories would go a long way towards solving the major issue, which is that of UV completion -- is it possible to find a standard UV completion for a massive graviton, analogous to what the Higgs mechanism provides for a massive vector?  Or is there some incontrovertible obstruction that forces any UV completion to violate Lorentz invariance, locality, or some other cherished property?  If so, there may be a non-standard UV completion, or it could be that massive gravity really is inconsistent, in the sense that there really is no way whatsoever to UV complete it.  There has been work on holography of massive gravitons in AdS/CFT \cite{Aharony:2006hz,Kiritsis:2006hy,Kiritsis:2008at,Niarchos:2009qb}, which would provide a UV completion for theories in AdS space containing massive gravity.

Even a partial UV completion, one that raises the cutoff to $M_P$, would be extremely important, as this is all that is required to offer a solution to the cosmological constant naturalness problem.  One possibility is that the scale $\Lambda_3$, while indicating a breakdown in perturbation theory, does not signal the activation of any new degrees of freedom, so that the theory is already self complete up to $M_P$.  Since there are multiple parameters in the theory, it is likely that there is some other expansion, such as a small $m$ expansion, which reorganizes the perturbation theory into one which yields perturbative access to scales above $\Lambda_3$.  If this is true, it is important that (as is the case) the $\Lambda_3$ theory is ghost free beyond the decoupling limit.

It should also be noted that massive gravitons already exist in nature, in the form of tensor mesons which carry spin 2.  There is a nonet of them, which at low energies can be described in chiral perturbation theory as a multiplet of massive gravitons \cite{Chow:1997sg}.  Here we know that these states find a UV completion in QCD, where they are simply excited states of bound quarks.  

We have focused in this review on theories with a vacuum that preserves Lorentz invariance, but there is a whole new world that opens up when ones allows for Lorentz violation.  There exist theories that explicitly break Lorentz invariance, and theories such as the ghost condensate \cite{ArkaniHamed:2003uy} which have Lorentz invariance spontaneously broken \cite{ArkaniHamed:2004ar} by some non-Lorentz invariant background.  In the former case, a systematic study of the possible mass terms and their degrees of freedom, generalizing the Fierz-Pauli analysis to the case where the mass term preserves only rotation invariance, is performed in \cite{Dubovsky:2004sg}.  For examples of the latter case, see \cite{Berezhiani:2007zf,Blas:2007zz}.  See also \cite{Gabadadze:2004iv,Rubakov:2004eb}, and \cite{Rubakov:2008nh,Bebronne:2009iy,Mironov:2009mx} for reviews.

There is still much to be learned about massive gravitons on curved spaces and cosmologies (see \cite{Babak:2002uz,Maeno:2008gj,Nair:2008yh,Grisa:2009yy,Berkhahn:2010hc} in addition to the references of Section \ref{massivecurvedspacesection}).  For instance, there are generalizations of the ghost-free higher cutoff $\Lambda_3$ theory, both for arbitrary curved backgrounds \cite{Hassan:2011tf}, and for bi-gravity \cite{Hassan:2011zd}.  Is there a consistent fully interacting theory of the partially massless theories on de Sitter space?  Are there consistent theories with cosmological backgrounds, and in particular can they non-linearly realize the screening of a large bare cosmological constant while maintaining consistency with solar system constraints?  

Finally, a topic worthy of a separate review is the observable signatures that would be characteristic of a massive graviton.  What would be the signatures of a cosmological constant screened by a graviton mass? For some examples of various proposed signatures, see \cite{Dubovsky:2004ud,Dubovsky:2009xk,Bessada:2009np,Bessada:2009qw,Bebronne:2009rq,Wyman:2011mp}.  One surprising feature is the absence of any non-trivial spatially flat FRW solutions \cite{DAmico:2011jj} (see also \cite{Gumrukcuoglu:2011ew}).  This means that the universe on scales larger than $\sim 1/m$ must be inhomogenous or anisotropic, with homogeneity and isotropy restored on smaller scales by the Vainshtein mechanism.  

We will end this review by quoting the tantalizing current experimental limits on the mass of the graviton (under some hypotheses, of course) $m\lesssim 7\times 10^{-32}$ eV \cite{Nakamura:2010zzi,Goldhaber:2008xy}, about an order of magnitude above the Hubble scale, the value needed to theoretically explain the cosmological constant naturalness problem.

\bigskip
{\bf Acknowledgements}: 
The author would like to thank Claudia de Rham, Gregory Gabadadze, Lam Hui, Justin Khoury, Janna Levin, Alberto Nicolis, Rachel Rosen and Claire Zukowski for discussions and for comments on the manuscript, and Mark Trodden for discussions, comments, and for encouraging the writing of this review. 
This work is supported in part by NSF grant PHY-0930521, and by funds provided by the University of Pennsylvania.

\appendix

\section{\label{totalDappendix}Total derivative combinations}
Define the matrix of second derivatives
\be \Pi_{\mu\nu}=\partial_\mu\partial_\nu\phi.\ee
At every order in $\phi$, there is a unique (up to overall constant) contraction of $\Pi$'s that reduces to a total derivative\footnote{The proof of this fact is the same as the proof showing the uniqueness of the galileons in \cite{Nicolis:2008in}.  See also \cite{Creminelli:2005qk}.},
\bea {\cal L}_1^{\rm TD}(\Pi) &=&[\Pi],\\
 {\cal L}_2^{\rm TD}(\Pi) &=&[\Pi]^2-[\Pi ^2], \\
{\cal L}_3^{\rm TD} (\Pi)&=& [\Pi]^3-3 [\Pi][\Pi ^2]+2[\Pi ^3] ,\\
{\cal L}_4^{\rm TD} (\Pi)&=&[\Pi]^4
-6[\Pi ^2][\Pi]^2+8[\Pi ^3][\Pi]+3[\Pi ^2]^2 -6[\Pi ^4] , \\
&\vdots& \nn
\eea
where the brackets are traces.
$ {\cal L}_2^{\rm TD} (h)$ is just the Fierz-Pauli term, and the others can be thought of as higher order generalizations of it.  They are characteristic polynomials, terms in the expansion of the determinant in powers of $H$,
\be \det(1+ \Pi)=1+{\cal L}_1^{\rm TD}(\Pi)+{1\over 2} {\cal L}_2^{\rm TD}(\Pi) +{1\over 3!} {\cal L}_3^{\rm TD}(\Pi)+{1\over 4!} {\cal L}_4^{\rm TD}(\Pi)+\cdots\ee
The term $ {\cal L}_n^{\rm TD}(\Pi)$ vanishes identically when $n>D$, with $D$ the spacetime dimension, so there are only $D$ non-trivial such combinations, those with $n=1,\cdots, D$.  

They can be written explicitly as
\be  {\cal L}_n^{\rm TD}(\Pi)=\sum_p\left(-1\right)^{p}\eta^{\mu_1p(\nu_1)}\eta^{\mu_2p(\nu_2)}\cdots\eta^{\mu_np(\nu_n)} \left(\Pi_{\mu_1\nu_1}\Pi_{\mu_2\nu_2}\cdots\Pi_{\mu_n\nu_n}\right).\ee
The sum is over all permutations of the $\nu$ indices, with $(-1)^p$ the sign of the permutation. 

They satisfy a recursion relation,
\be {\cal L}_n^{\rm TD}(\Pi)=-\sum_{m=1}^n(-1)^m{(n-1)!\over (n-m)!}\left[\Pi^m\right] {\cal L}_{n-m}^{\rm TD}(\Pi),\ee
with $ {\cal L}_0^{\rm TD}(\Pi)=1$.

In addition, there are tensors $X^{(n)}_{\mu\nu}$ that we construct out of $\Pi_{\mu\nu}$ as follows\footnote{Note that our definition of the $X^{(n)}_{\mu\nu}$ used here differs by a factor of 2 from that of \cite{deRham:2010kj}.},
\be X^{(n)}_{\mu\nu}={1\over n+1}{\delta \over \delta \Pi_{\mu\nu}} {\cal L}_{n+1}^{\rm TD}(\Pi).\ee
The first few are
\bea X^{(0)}_{\mu\nu}&=&\eta_{\mu\nu} \\
 X^{(1)}_{\mu\nu}&=&\left[\Pi\right]\eta_{\mu\nu}-\Pi_{\mu\nu} \\
  X^{(2)}_{\mu\nu}&=&\left(\left[\Pi\right]^2-\left[\Pi^2\right]\right)\eta_{\mu\nu}-2\left[\Pi\right]\Pi_{\mu\nu}+2\Pi^2_{\mu\nu} \\
   X^{(3)}_{\mu\nu}&=&\left(\left[\Pi\right]^3-3\left[\Pi\right]\left[\Pi^2\right]+2\left[\Pi^3\right]\right)\eta_{\mu\nu}-3\left(\left[\Pi\right]^2-\left[\Pi^2\right]\right)\Pi_{\mu\nu}+6\left[\Pi\right]\Pi^2_{\mu\nu}-6\Pi^3_{\mu\nu}\nn \\
   &\vdots&
   \eea

The following is an explicit expression,
\be X^{(n)}_{\mu\nu}=\sum_{m=0}^n(-1)^m{n!\over (n-m)!}\Pi^m_{\mu\nu}{\cal L}_{n-m}^{\rm TD}(\Pi).\ee
They satisfy the recursion relation
\be X^{(n)}_{\mu\nu}=-n\Pi_\mu^{\ \alpha}X^{(n-1)}_{\alpha\nu}+\Pi^{\alpha\beta}X^{(n-1)}_{\alpha\beta}\eta_{\mu\nu}.\ee

Since $ {\cal L}_n^{\rm TD}(\Pi)$ vanishes for $n>D$, $X^{(n)}_{\mu\nu}$ vanishes for $n\geq D$.

The $X^{(n)}_{\mu\nu}$ satisfy the following important properties:
\begin{itemize}
\item They are symmetric and identically conserved, and are the only combinations of $\Pi_{\mu\nu}$ at each order with these properties: 
\be \partial^\mu X^{(n)}_{\mu\nu}=0,\ee
\item For spatial indices $i,j$ and time index $0$,
\bea  &&X^{(n)}_{ij}\ \text{has at most two time derivatives,}\nn \\
&&X^{(n)}_{0i}\ \text{has at most one time derivative,}\nn \\
&&X^{(n)}_{00}\ \text{has no time derivatives.}\label{xtensorprops} 
\eea
\end{itemize}

Finally, we have the following relations involving the massless kinetic operator (\ref{Eoper}),
\bea &&{\cal E}_{\mu\nu}^{\ \ \alpha\beta}\left(\phi\eta_{\alpha\beta}\right)=-(D-2)X^{(1)}_{\mu\nu},\nn\\
&& {\cal E}_{\mu\nu}^{\ \ \alpha\beta}\left(\partial_\alpha\phi\partial_\beta\phi\right)=X^{(2)}_{\mu\nu}. \label{Eopergalrel}
\eea

\bibliographystyle{utphys}
\addcontentsline{toc}{section}{References}
\bibliography{massivegravity}

\end{document}